\documentclass[longauth]{aa}

\usepackage{amsmath}
\usepackage{graphicx}
\usepackage{natbib}
\usepackage{scalerel}
\usepackage{longtable}
\usepackage[normalem]{ulem}
\usepackage[dvipsnames]{xcolor}

\usepackage{txfonts}
\usepackage[pdfencoding=auto,psdextra]{hyperref}
\hypersetup{
    colorlinks=true,
    linkcolor=blue,
    filecolor=magenta,      
    urlcolor=blue,
    citecolor=blue
}
\makeatletter
\renewcommand*\aa@pageof{, page \thepage{} of \pageref*{LastPage}}
\makeatother

\usepackage[utf8]{inputenc}

\usepackage[switch, modulo]{lineno}
\usepackage{euclid}

\def\be{\begin{equation}}
\def\ee{\end{equation}}
\def\ba{\begin{eqnarray}}
\def\ea{\end{eqnarray}}
\def\eqi{\begin{equation}}
\def\eqf{\end{equation}}
\def\eqia{\begin{eqnarray}}
\def\eqfa{\end{eqnarray}}
\def\lcdm{$\Lambda$CDM\xspace}

\def\nn{\nonumber}
\usepackage[nameinlink,noabbrev]{cleveref}
\Crefname{equation}{Eq.}{Eqs.}
\Crefname{eqnarray}{Eq.}{Eqs.}
\Crefname{section}{Sect.}{Sects.}
\Crefname{figure}{Fig.}{Figs.}
\crefname{equation}{Equation}{Equations}
\crefname{section}{Section}{Sections}
\crefname{figure}{Figure}{Figures}

\defcitealias{IST:paper1}{EC20}
\usepackage{amstext}
\usepackage{soul}

\begin{document}

\title{\Euclid: Forecasts on \lcdm consistency tests with growth rate data \thanks{This paper is published on behalf of the Euclid Consortium.}}

\titlerunning{\Euclid: Forecast constraints on consistency tests of $\Lambda$CDM with growth data}
\authorrunning{I.~Ocampo et al.}

%%%%%%%%%%%%%%%%%%%%%%%%%%%%%%%

%%%% Version Wednesday 30th of July 2025 02:02:15 PM UT												
%%%% Please do not edit the author list -- contact ECEB Bureau for changes
\newcommand{\orcid}[1]{} %% if already defined in aa.cls: comment, or use renewcommand			   
\author{I.~Ocampo\orcid{0000-0001-7709-1930}\inst{\ref{aff1}}
\and D.~Sapone\orcid{0000-0001-7089-4503}\thanks{\email{dsapone@ing.uchile.cl}}\inst{\ref{aff2}}
\and S.~Nesseris\orcid{0000-0002-0567-0324}\inst{\ref{aff1}}
\and G.~Alestas\orcid{0000-0003-1790-4914}\inst{\ref{aff1}}
\and J.~Garc\'ia-Bellido\orcid{0000-0002-9370-8360}\inst{\ref{aff1}}
\and Z.~Sakr\orcid{0000-0002-4823-3757}\inst{\ref{aff3},\ref{aff4},\ref{aff5}}
\and C.~J.~A.~P.~Martins\orcid{0000-0002-4886-9261}\inst{\ref{aff6},\ref{aff7}}
\and J.~P.~Mimoso\orcid{0000-0002-9758-3366}\inst{\ref{aff8},\ref{aff9}}
\and A.~Carvalho\orcid{0000-0002-9301-262X}\inst{\ref{aff8},\ref{aff9}}
\and A.~Da~Silva\orcid{0000-0002-6385-1609}\inst{\ref{aff8},\ref{aff9}}
\and A.~Blanchard\orcid{0000-0001-8555-9003}\inst{\ref{aff4}}
\and S.~Casas\orcid{0000-0002-4751-5138}\inst{\ref{aff10}}
\and S.~Camera\orcid{0000-0003-3399-3574}\inst{\ref{aff11},\ref{aff12},\ref{aff13}}
\and M.~Martinelli\orcid{0000-0002-6943-7732}\inst{\ref{aff14},\ref{aff15}}
\and V.~Pettorino\orcid{0000-0002-4203-9320}\inst{\ref{aff16}}
\and A.~Amara\inst{\ref{aff17}}
\and S.~Andreon\orcid{0000-0002-2041-8784}\inst{\ref{aff18}}
\and N.~Auricchio\orcid{0000-0003-4444-8651}\inst{\ref{aff19}}
\and C.~Baccigalupi\orcid{0000-0002-8211-1630}\inst{\ref{aff20},\ref{aff21},\ref{aff22},\ref{aff23}}
\and M.~Baldi\orcid{0000-0003-4145-1943}\inst{\ref{aff24},\ref{aff19},\ref{aff25}}
\and A.~Balestra\orcid{0000-0002-6967-261X}\inst{\ref{aff26}}
\and S.~Bardelli\orcid{0000-0002-8900-0298}\inst{\ref{aff19}}
\and P.~Battaglia\orcid{0000-0002-7337-5909}\inst{\ref{aff19}}
\and F.~Bernardeau\orcid{0009-0007-3015-2581}\inst{\ref{aff27},\ref{aff28}}
\and A.~Biviano\orcid{0000-0002-0857-0732}\inst{\ref{aff21},\ref{aff20}}
\and E.~Branchini\orcid{0000-0002-0808-6908}\inst{\ref{aff29},\ref{aff30},\ref{aff18}}
\and M.~Brescia\orcid{0000-0001-9506-5680}\inst{\ref{aff31},\ref{aff32}}
\and G.~Ca\~nas-Herrera\orcid{0000-0003-2796-2149}\inst{\ref{aff16},\ref{aff33},\ref{aff34}}
\and V.~Capobianco\orcid{0000-0002-3309-7692}\inst{\ref{aff13}}
\and C.~Carbone\orcid{0000-0003-0125-3563}\inst{\ref{aff35}}
\and V.~F.~Cardone\inst{\ref{aff14},\ref{aff15}}
\and J.~Carretero\orcid{0000-0002-3130-0204}\inst{\ref{aff36},\ref{aff37}}
\and M.~Castellano\orcid{0000-0001-9875-8263}\inst{\ref{aff14}}
\and G.~Castignani\orcid{0000-0001-6831-0687}\inst{\ref{aff19}}
\and S.~Cavuoti\orcid{0000-0002-3787-4196}\inst{\ref{aff32},\ref{aff38}}
\and K.~C.~Chambers\orcid{0000-0001-6965-7789}\inst{\ref{aff39}}
\and A.~Cimatti\inst{\ref{aff40}}
\and C.~Colodro-Conde\inst{\ref{aff41}}
\and G.~Congedo\orcid{0000-0003-2508-0046}\inst{\ref{aff42}}
\and L.~Conversi\orcid{0000-0002-6710-8476}\inst{\ref{aff43},\ref{aff44}}
\and Y.~Copin\orcid{0000-0002-5317-7518}\inst{\ref{aff45}}
\and F.~Courbin\orcid{0000-0003-0758-6510}\inst{\ref{aff46},\ref{aff47}}
\and H.~M.~Courtois\orcid{0000-0003-0509-1776}\inst{\ref{aff48}}
\and H.~Degaudenzi\orcid{0000-0002-5887-6799}\inst{\ref{aff49}}
\and S.~de~la~Torre\inst{\ref{aff50}}
\and G.~De~Lucia\orcid{0000-0002-6220-9104}\inst{\ref{aff21}}
\and F.~Dubath\orcid{0000-0002-6533-2810}\inst{\ref{aff49}}
\and C.~A.~J.~Duncan\orcid{0009-0003-3573-0791}\inst{\ref{aff42},\ref{aff51}}
\and X.~Dupac\inst{\ref{aff44}}
\and S.~Dusini\orcid{0000-0002-1128-0664}\inst{\ref{aff52}}
\and S.~Escoffier\orcid{0000-0002-2847-7498}\inst{\ref{aff53}}
\and M.~Farina\orcid{0000-0002-3089-7846}\inst{\ref{aff54}}
\and R.~Farinelli\inst{\ref{aff19}}
\and S.~Farrens\orcid{0000-0002-9594-9387}\inst{\ref{aff55}}
\and F.~Faustini\orcid{0000-0001-6274-5145}\inst{\ref{aff14},\ref{aff56}}
\and S.~Ferriol\inst{\ref{aff45}}
\and F.~Finelli\orcid{0000-0002-6694-3269}\inst{\ref{aff19},\ref{aff57}}
\and P.~Fosalba\orcid{0000-0002-1510-5214}\inst{\ref{aff58},\ref{aff59}}
\and N.~Fourmanoit\orcid{0009-0005-6816-6925}\inst{\ref{aff53}}
\and M.~Frailis\orcid{0000-0002-7400-2135}\inst{\ref{aff21}}
\and E.~Franceschi\orcid{0000-0002-0585-6591}\inst{\ref{aff19}}
\and S.~Galeotta\orcid{0000-0002-3748-5115}\inst{\ref{aff21}}
\and K.~George\orcid{0000-0002-1734-8455}\inst{\ref{aff60}}
\and B.~Gillis\orcid{0000-0002-4478-1270}\inst{\ref{aff42}}
\and C.~Giocoli\orcid{0000-0002-9590-7961}\inst{\ref{aff19},\ref{aff25}}
\and J.~Gracia-Carpio\inst{\ref{aff61}}
\and A.~Grazian\orcid{0000-0002-5688-0663}\inst{\ref{aff26}}
\and F.~Grupp\inst{\ref{aff61},\ref{aff60}}
\and S.~V.~H.~Haugan\orcid{0000-0001-9648-7260}\inst{\ref{aff62}}
\and W.~Holmes\inst{\ref{aff63}}
\and F.~Hormuth\inst{\ref{aff64}}
\and A.~Hornstrup\orcid{0000-0002-3363-0936}\inst{\ref{aff65},\ref{aff66}}
\and K.~Jahnke\orcid{0000-0003-3804-2137}\inst{\ref{aff67}}
\and M.~Jhabvala\inst{\ref{aff68}}
\and B.~Joachimi\orcid{0000-0001-7494-1303}\inst{\ref{aff69}}
\and E.~Keih\"anen\orcid{0000-0003-1804-7715}\inst{\ref{aff70}}
\and S.~Kermiche\orcid{0000-0002-0302-5735}\inst{\ref{aff53}}
\and B.~Kubik\orcid{0009-0006-5823-4880}\inst{\ref{aff45}}
\and M.~Kunz\orcid{0000-0002-3052-7394}\inst{\ref{aff71}}
\and H.~Kurki-Suonio\orcid{0000-0002-4618-3063}\inst{\ref{aff72},\ref{aff73}}
\and A.~M.~C.~Le~Brun\orcid{0000-0002-0936-4594}\inst{\ref{aff74}}
\and S.~Ligori\orcid{0000-0003-4172-4606}\inst{\ref{aff13}}
\and P.~B.~Lilje\orcid{0000-0003-4324-7794}\inst{\ref{aff62}}
\and V.~Lindholm\orcid{0000-0003-2317-5471}\inst{\ref{aff72},\ref{aff73}}
\and I.~Lloro\orcid{0000-0001-5966-1434}\inst{\ref{aff75}}
\and G.~Mainetti\orcid{0000-0003-2384-2377}\inst{\ref{aff76}}
\and D.~Maino\inst{\ref{aff77},\ref{aff35},\ref{aff78}}
\and E.~Maiorano\orcid{0000-0003-2593-4355}\inst{\ref{aff19}}
\and O.~Mansutti\orcid{0000-0001-5758-4658}\inst{\ref{aff21}}
\and O.~Marggraf\orcid{0000-0001-7242-3852}\inst{\ref{aff79}}
\and K.~Markovic\orcid{0000-0001-6764-073X}\inst{\ref{aff63}}
\and N.~Martinet\orcid{0000-0003-2786-7790}\inst{\ref{aff50}}
\and F.~Marulli\orcid{0000-0002-8850-0303}\inst{\ref{aff80},\ref{aff19},\ref{aff25}}
\and R.~J.~Massey\orcid{0000-0002-6085-3780}\inst{\ref{aff81}}
\and E.~Medinaceli\orcid{0000-0002-4040-7783}\inst{\ref{aff19}}
\and S.~Mei\orcid{0000-0002-2849-559X}\inst{\ref{aff82},\ref{aff83}}
\and Y.~Mellier\inst{\ref{aff84},\ref{aff28}}
\and M.~Meneghetti\orcid{0000-0003-1225-7084}\inst{\ref{aff19},\ref{aff25}}
\and E.~Merlin\orcid{0000-0001-6870-8900}\inst{\ref{aff14}}
\and G.~Meylan\inst{\ref{aff85}}
\and A.~Mora\orcid{0000-0002-1922-8529}\inst{\ref{aff86}}
\and M.~Moresco\orcid{0000-0002-7616-7136}\inst{\ref{aff80},\ref{aff19}}
\and L.~Moscardini\orcid{0000-0002-3473-6716}\inst{\ref{aff80},\ref{aff19},\ref{aff25}}
\and C.~Neissner\orcid{0000-0001-8524-4968}\inst{\ref{aff87},\ref{aff37}}
\and S.-M.~Niemi\orcid{0009-0005-0247-0086}\inst{\ref{aff16}}
\and C.~Padilla\orcid{0000-0001-7951-0166}\inst{\ref{aff87}}
\and S.~Paltani\orcid{0000-0002-8108-9179}\inst{\ref{aff49}}
\and F.~Pasian\orcid{0000-0002-4869-3227}\inst{\ref{aff21}}
\and K.~Pedersen\inst{\ref{aff88}}
\and W.~J.~Percival\orcid{0000-0002-0644-5727}\inst{\ref{aff89},\ref{aff90},\ref{aff91}}
\and S.~Pires\orcid{0000-0002-0249-2104}\inst{\ref{aff55}}
\and G.~Polenta\orcid{0000-0003-4067-9196}\inst{\ref{aff56}}
\and M.~Poncet\inst{\ref{aff92}}
\and L.~A.~Popa\inst{\ref{aff93}}
\and F.~Raison\orcid{0000-0002-7819-6918}\inst{\ref{aff61}}
\and R.~Rebolo\orcid{0000-0003-3767-7085}\inst{\ref{aff41},\ref{aff94},\ref{aff95}}
\and A.~Renzi\orcid{0000-0001-9856-1970}\inst{\ref{aff96},\ref{aff52}}
\and J.~Rhodes\orcid{0000-0002-4485-8549}\inst{\ref{aff63}}
\and G.~Riccio\inst{\ref{aff32}}
\and E.~Romelli\orcid{0000-0003-3069-9222}\inst{\ref{aff21}}
\and M.~Roncarelli\orcid{0000-0001-9587-7822}\inst{\ref{aff19}}
\and C.~Rosset\orcid{0000-0003-0286-2192}\inst{\ref{aff82}}
\and R.~Saglia\orcid{0000-0003-0378-7032}\inst{\ref{aff60},\ref{aff61}}
\and B.~Sartoris\orcid{0000-0003-1337-5269}\inst{\ref{aff60},\ref{aff21}}
\and T.~Schrabback\orcid{0000-0002-6987-7834}\inst{\ref{aff97},\ref{aff79}}
\and A.~Secroun\orcid{0000-0003-0505-3710}\inst{\ref{aff53}}
\and E.~Sefusatti\orcid{0000-0003-0473-1567}\inst{\ref{aff21},\ref{aff20},\ref{aff22}}
\and G.~Seidel\orcid{0000-0003-2907-353X}\inst{\ref{aff67}}
\and M.~Seiffert\orcid{0000-0002-7536-9393}\inst{\ref{aff63}}
\and S.~Serrano\orcid{0000-0002-0211-2861}\inst{\ref{aff58},\ref{aff98},\ref{aff59}}
\and C.~Sirignano\orcid{0000-0002-0995-7146}\inst{\ref{aff96},\ref{aff52}}
\and G.~Sirri\orcid{0000-0003-2626-2853}\inst{\ref{aff25}}
\and A.~Spurio~Mancini\orcid{0000-0001-5698-0990}\inst{\ref{aff99}}
\and L.~Stanco\orcid{0000-0002-9706-5104}\inst{\ref{aff52}}
\and J.~Steinwagner\orcid{0000-0001-7443-1047}\inst{\ref{aff61}}
\and P.~Tallada-Cresp\'{i}\orcid{0000-0002-1336-8328}\inst{\ref{aff36},\ref{aff37}}
\and A.~N.~Taylor\inst{\ref{aff42}}
\and I.~Tereno\orcid{0000-0002-4537-6218}\inst{\ref{aff8},\ref{aff100}}
\and N.~Tessore\orcid{0000-0002-9696-7931}\inst{\ref{aff69}}
\and S.~Toft\orcid{0000-0003-3631-7176}\inst{\ref{aff101},\ref{aff102}}
\and R.~Toledo-Moreo\orcid{0000-0002-2997-4859}\inst{\ref{aff103}}
\and F.~Torradeflot\orcid{0000-0003-1160-1517}\inst{\ref{aff37},\ref{aff36}}
\and I.~Tutusaus\orcid{0000-0002-3199-0399}\inst{\ref{aff4}}
\and L.~Valenziano\orcid{0000-0002-1170-0104}\inst{\ref{aff19},\ref{aff57}}
\and J.~Valiviita\orcid{0000-0001-6225-3693}\inst{\ref{aff72},\ref{aff73}}
\and T.~Vassallo\orcid{0000-0001-6512-6358}\inst{\ref{aff60},\ref{aff21}}
\and G.~Verdoes~Kleijn\orcid{0000-0001-5803-2580}\inst{\ref{aff104}}
\and A.~Veropalumbo\orcid{0000-0003-2387-1194}\inst{\ref{aff18},\ref{aff30},\ref{aff29}}
\and Y.~Wang\orcid{0000-0002-4749-2984}\inst{\ref{aff105}}
\and J.~Weller\orcid{0000-0002-8282-2010}\inst{\ref{aff60},\ref{aff61}}
\and G.~Zamorani\orcid{0000-0002-2318-301X}\inst{\ref{aff19}}
\and F.~M.~Zerbi\inst{\ref{aff18}}
\and E.~Zucca\orcid{0000-0002-5845-8132}\inst{\ref{aff19}}
\and M.~Ballardini\orcid{0000-0003-4481-3559}\inst{\ref{aff106},\ref{aff107},\ref{aff19}}
\and C.~Burigana\orcid{0000-0002-3005-5796}\inst{\ref{aff108},\ref{aff57}}
\and L.~Gabarra\orcid{0000-0002-8486-8856}\inst{\ref{aff109}}
\and A.~Pezzotta\orcid{0000-0003-0726-2268}\inst{\ref{aff18}}
\and V.~Scottez\orcid{0009-0008-3864-940X}\inst{\ref{aff84},\ref{aff110}}
\and M.~Viel\orcid{0000-0002-2642-5707}\inst{\ref{aff20},\ref{aff21},\ref{aff23},\ref{aff22},\ref{aff111}}}
										   
%%%% please do not edit the affiliation list -- contact ECEB Bureau for changes
\institute{Instituto de F\'isica Te\'orica UAM-CSIC, Campus de Cantoblanco, 28049 Madrid, Spain\label{aff1}
\and
Departamento de F\'isica, FCFM, Universidad de Chile, Blanco Encalada 2008, Santiago, Chile\label{aff2}
\and
Institut f\"ur Theoretische Physik, University of Heidelberg, Philosophenweg 16, 69120 Heidelberg, Germany\label{aff3}
\and
Institut de Recherche en Astrophysique et Plan\'etologie (IRAP), Universit\'e de Toulouse, CNRS, UPS, CNES, 14 Av. Edouard Belin, 31400 Toulouse, France\label{aff4}
\and
Universit\'e St Joseph; Faculty of Sciences, Beirut, Lebanon\label{aff5}
\and
Centro de Astrof\'{\i}sica da Universidade do Porto, Rua das Estrelas, 4150-762 Porto, Portugal\label{aff6}
\and
Instituto de Astrof\'isica e Ci\^encias do Espa\c{c}o, Universidade do Porto, CAUP, Rua das Estrelas, PT4150-762 Porto, Portugal\label{aff7}
\and
Departamento de F\'isica, Faculdade de Ci\^encias, Universidade de Lisboa, Edif\'icio C8, Campo Grande, PT1749-016 Lisboa, Portugal\label{aff8}
\and
Instituto de Astrof\'isica e Ci\^encias do Espa\c{c}o, Faculdade de Ci\^encias, Universidade de Lisboa, Campo Grande, 1749-016 Lisboa, Portugal\label{aff9}
\and
Institute for Theoretical Particle Physics and Cosmology (TTK), RWTH Aachen University, 52056 Aachen, Germany\label{aff10}
\and
Dipartimento di Fisica, Universit\`a degli Studi di Torino, Via P. Giuria 1, 10125 Torino, Italy\label{aff11}
\and
INFN-Sezione di Torino, Via P. Giuria 1, 10125 Torino, Italy\label{aff12}
\and
INAF-Osservatorio Astrofisico di Torino, Via Osservatorio 20, 10025 Pino Torinese (TO), Italy\label{aff13}
\and
INAF-Osservatorio Astronomico di Roma, Via Frascati 33, 00078 Monteporzio Catone, Italy\label{aff14}
\and
INFN-Sezione di Roma, Piazzale Aldo Moro, 2 - c/o Dipartimento di Fisica, Edificio G. Marconi, 00185 Roma, Italy\label{aff15}
\and
European Space Agency/ESTEC, Keplerlaan 1, 2201 AZ Noordwijk, The Netherlands\label{aff16}
\and
School of Mathematics and Physics, University of Surrey, Guildford, Surrey, GU2 7XH, UK\label{aff17}
\and
INAF-Osservatorio Astronomico di Brera, Via Brera 28, 20122 Milano, Italy\label{aff18}
\and
INAF-Osservatorio di Astrofisica e Scienza dello Spazio di Bologna, Via Piero Gobetti 93/3, 40129 Bologna, Italy\label{aff19}
\and
IFPU, Institute for Fundamental Physics of the Universe, via Beirut 2, 34151 Trieste, Italy\label{aff20}
\and
INAF-Osservatorio Astronomico di Trieste, Via G. B. Tiepolo 11, 34143 Trieste, Italy\label{aff21}
\and
INFN, Sezione di Trieste, Via Valerio 2, 34127 Trieste TS, Italy\label{aff22}
\and
SISSA, International School for Advanced Studies, Via Bonomea 265, 34136 Trieste TS, Italy\label{aff23}
\and
Dipartimento di Fisica e Astronomia, Universit\`a di Bologna, Via Gobetti 93/2, 40129 Bologna, Italy\label{aff24}
\and
INFN-Sezione di Bologna, Viale Berti Pichat 6/2, 40127 Bologna, Italy\label{aff25}
\and
INAF-Osservatorio Astronomico di Padova, Via dell'Osservatorio 5, 35122 Padova, Italy\label{aff26}
\and
Institut de Physique Th\'eorique, CEA, CNRS, Universit\'e Paris-Saclay 91191 Gif-sur-Yvette Cedex, France\label{aff27}
\and
Institut d'Astrophysique de Paris, UMR 7095, CNRS, and Sorbonne Universit\'e, 98 bis boulevard Arago, 75014 Paris, France\label{aff28}
\and
Dipartimento di Fisica, Universit\`a di Genova, Via Dodecaneso 33, 16146, Genova, Italy\label{aff29}
\and
INFN-Sezione di Genova, Via Dodecaneso 33, 16146, Genova, Italy\label{aff30}
\and
Department of Physics "E. Pancini", University Federico II, Via Cinthia 6, 80126, Napoli, Italy\label{aff31}
\and
INAF-Osservatorio Astronomico di Capodimonte, Via Moiariello 16, 80131 Napoli, Italy\label{aff32}
\and
Institute Lorentz, Leiden University, Niels Bohrweg 2, 2333 CA Leiden, The Netherlands\label{aff33}
\and
Leiden Observatory, Leiden University, Einsteinweg 55, 2333 CC Leiden, The Netherlands\label{aff34}
\and
INAF-IASF Milano, Via Alfonso Corti 12, 20133 Milano, Italy\label{aff35}
\and
Centro de Investigaciones Energ\'eticas, Medioambientales y Tecnol\'ogicas (CIEMAT), Avenida Complutense 40, 28040 Madrid, Spain\label{aff36}
\and
Port d'Informaci\'{o} Cient\'{i}fica, Campus UAB, C. Albareda s/n, 08193 Bellaterra (Barcelona), Spain\label{aff37}
\and
INFN section of Naples, Via Cinthia 6, 80126, Napoli, Italy\label{aff38}
\and
Institute for Astronomy, University of Hawaii, 2680 Woodlawn Drive, Honolulu, HI 96822, USA\label{aff39}
\and
Dipartimento di Fisica e Astronomia "Augusto Righi" - Alma Mater Studiorum Universit\`a di Bologna, Viale Berti Pichat 6/2, 40127 Bologna, Italy\label{aff40}
\and
Instituto de Astrof\'{\i}sica de Canarias, V\'{\i}a L\'actea, 38205 La Laguna, Tenerife, Spain\label{aff41}
\and
Institute for Astronomy, University of Edinburgh, Royal Observatory, Blackford Hill, Edinburgh EH9 3HJ, UK\label{aff42}
\and
European Space Agency/ESRIN, Largo Galileo Galilei 1, 00044 Frascati, Roma, Italy\label{aff43}
\and
ESAC/ESA, Camino Bajo del Castillo, s/n., Urb. Villafranca del Castillo, 28692 Villanueva de la Ca\~nada, Madrid, Spain\label{aff44}
\and
Universit\'e Claude Bernard Lyon 1, CNRS/IN2P3, IP2I Lyon, UMR 5822, Villeurbanne, F-69100, France\label{aff45}
\and
Institut de Ci\`{e}ncies del Cosmos (ICCUB), Universitat de Barcelona (IEEC-UB), Mart\'{i} i Franqu\`{e}s 1, 08028 Barcelona, Spain\label{aff46}
\and
Instituci\'o Catalana de Recerca i Estudis Avan\c{c}ats (ICREA), Passeig de Llu\'{\i}s Companys 23, 08010 Barcelona, Spain\label{aff47}
\and
UCB Lyon 1, CNRS/IN2P3, IUF, IP2I Lyon, 4 rue Enrico Fermi, 69622 Villeurbanne, France\label{aff48}
\and
Department of Astronomy, University of Geneva, ch. d'Ecogia 16, 1290 Versoix, Switzerland\label{aff49}
\and
Aix-Marseille Universit\'e, CNRS, CNES, LAM, Marseille, France\label{aff50}
\and
Jodrell Bank Centre for Astrophysics, Department of Physics and Astronomy, University of Manchester, Oxford Road, Manchester M13 9PL, UK\label{aff51}
\and
INFN-Padova, Via Marzolo 8, 35131 Padova, Italy\label{aff52}
\and
Aix-Marseille Universit\'e, CNRS/IN2P3, CPPM, Marseille, France\label{aff53}
\and
INAF-Istituto di Astrofisica e Planetologia Spaziali, via del Fosso del Cavaliere, 100, 00100 Roma, Italy\label{aff54}
\and
Universit\'e Paris-Saclay, Universit\'e Paris Cit\'e, CEA, CNRS, AIM, 91191, Gif-sur-Yvette, France\label{aff55}
\and
Space Science Data Center, Italian Space Agency, via del Politecnico snc, 00133 Roma, Italy\label{aff56}
\and
INFN-Bologna, Via Irnerio 46, 40126 Bologna, Italy\label{aff57}
\and
Institut d'Estudis Espacials de Catalunya (IEEC),  Edifici RDIT, Campus UPC, 08860 Castelldefels, Barcelona, Spain\label{aff58}
\and
Institute of Space Sciences (ICE, CSIC), Campus UAB, Carrer de Can Magrans, s/n, 08193 Barcelona, Spain\label{aff59}
\and
Universit\"ats-Sternwarte M\"unchen, Fakult\"at f\"ur Physik, Ludwig-Maximilians-Universit\"at M\"unchen, Scheinerstrasse 1, 81679 M\"unchen, Germany\label{aff60}
\and
Max Planck Institute for Extraterrestrial Physics, Giessenbachstr. 1, 85748 Garching, Germany\label{aff61}
\and
Institute of Theoretical Astrophysics, University of Oslo, P.O. Box 1029 Blindern, 0315 Oslo, Norway\label{aff62}
\and
Jet Propulsion Laboratory, California Institute of Technology, 4800 Oak Grove Drive, Pasadena, CA, 91109, USA\label{aff63}
\and
Felix Hormuth Engineering, Goethestr. 17, 69181 Leimen, Germany\label{aff64}
\and
Technical University of Denmark, Elektrovej 327, 2800 Kgs. Lyngby, Denmark\label{aff65}
\and
Cosmic Dawn Center (DAWN), Denmark\label{aff66}
\and
Max-Planck-Institut f\"ur Astronomie, K\"onigstuhl 17, 69117 Heidelberg, Germany\label{aff67}
\and
NASA Goddard Space Flight Center, Greenbelt, MD 20771, USA\label{aff68}
\and
Department of Physics and Astronomy, University College London, Gower Street, London WC1E 6BT, UK\label{aff69}
\and
Department of Physics and Helsinki Institute of Physics, Gustaf H\"allstr\"omin katu 2, 00014 University of Helsinki, Finland\label{aff70}
\and
Universit\'e de Gen\`eve, D\'epartement de Physique Th\'eorique and Centre for Astroparticle Physics, 24 quai Ernest-Ansermet, CH-1211 Gen\`eve 4, Switzerland\label{aff71}
\and
Department of Physics, P.O. Box 64, 00014 University of Helsinki, Finland\label{aff72}
\and
Helsinki Institute of Physics, Gustaf H{\"a}llstr{\"o}min katu 2, University of Helsinki, Helsinki, Finland\label{aff73}
\and
Laboratoire d'etude de l'Univers et des phenomenes eXtremes, Observatoire de Paris, Universit\'e PSL, Sorbonne Universit\'e, CNRS, 92190 Meudon, France\label{aff74}
\and
SKA Observatory, Jodrell Bank, Lower Withington, Macclesfield, Cheshire SK11 9FT, UK\label{aff75}
\and
Centre de Calcul de l'IN2P3/CNRS, 21 avenue Pierre de Coubertin 69627 Villeurbanne Cedex, France\label{aff76}
\and
Dipartimento di Fisica "Aldo Pontremoli", Universit\`a degli Studi di Milano, Via Celoria 16, 20133 Milano, Italy\label{aff77}
\and
INFN-Sezione di Milano, Via Celoria 16, 20133 Milano, Italy\label{aff78}
\and
Universit\"at Bonn, Argelander-Institut f\"ur Astronomie, Auf dem H\"ugel 71, 53121 Bonn, Germany\label{aff79}
\and
Dipartimento di Fisica e Astronomia "Augusto Righi" - Alma Mater Studiorum Universit\`a di Bologna, via Piero Gobetti 93/2, 40129 Bologna, Italy\label{aff80}
\and
Department of Physics, Institute for Computational Cosmology, Durham University, South Road, Durham, DH1 3LE, UK\label{aff81}
\and
Universit\'e Paris Cit\'e, CNRS, Astroparticule et Cosmologie, 75013 Paris, France\label{aff82}
\and
CNRS-UCB International Research Laboratory, Centre Pierre Bin\'etruy, IRL2007, CPB-IN2P3, Berkeley, USA\label{aff83}
\and
Institut d'Astrophysique de Paris, 98bis Boulevard Arago, 75014, Paris, France\label{aff84}
\and
Institute of Physics, Laboratory of Astrophysics, Ecole Polytechnique F\'ed\'erale de Lausanne (EPFL), Observatoire de Sauverny, 1290 Versoix, Switzerland\label{aff85}
\and
Telespazio UK S.L. for European Space Agency (ESA), Camino bajo del Castillo, s/n, Urbanizacion Villafranca del Castillo, Villanueva de la Ca\~nada, 28692 Madrid, Spain\label{aff86}
\and
Institut de F\'{i}sica d'Altes Energies (IFAE), The Barcelona Institute of Science and Technology, Campus UAB, 08193 Bellaterra (Barcelona), Spain\label{aff87}
\and
DARK, Niels Bohr Institute, University of Copenhagen, Jagtvej 155, 2200 Copenhagen, Denmark\label{aff88}
\and
Waterloo Centre for Astrophysics, University of Waterloo, Waterloo, Ontario N2L 3G1, Canada\label{aff89}
\and
Department of Physics and Astronomy, University of Waterloo, Waterloo, Ontario N2L 3G1, Canada\label{aff90}
\and
Perimeter Institute for Theoretical Physics, Waterloo, Ontario N2L 2Y5, Canada\label{aff91}
\and
Centre National d'Etudes Spatiales -- Centre spatial de Toulouse, 18 avenue Edouard Belin, 31401 Toulouse Cedex 9, France\label{aff92}
\and
Institute of Space Science, Str. Atomistilor, nr. 409 M\u{a}gurele, Ilfov, 077125, Romania\label{aff93}
\and
Consejo Superior de Investigaciones Cientificas, Calle Serrano 117, 28006 Madrid, Spain\label{aff94}
\and
Universidad de La Laguna, Departamento de Astrof\'{\i}sica, 38206 La Laguna, Tenerife, Spain\label{aff95}
\and
Dipartimento di Fisica e Astronomia "G. Galilei", Universit\`a di Padova, Via Marzolo 8, 35131 Padova, Italy\label{aff96}
\and
Universit\"at Innsbruck, Institut f\"ur Astro- und Teilchenphysik, Technikerstr. 25/8, 6020 Innsbruck, Austria\label{aff97}
\and
Satlantis, University Science Park, Sede Bld 48940, Leioa-Bilbao, Spain\label{aff98}
\and
Department of Physics, Royal Holloway, University of London, TW20 0EX, UK\label{aff99}
\and
Instituto de Astrof\'isica e Ci\^encias do Espa\c{c}o, Faculdade de Ci\^encias, Universidade de Lisboa, Tapada da Ajuda, 1349-018 Lisboa, Portugal\label{aff100}
\and
Cosmic Dawn Center (DAWN)\label{aff101}
\and
Niels Bohr Institute, University of Copenhagen, Jagtvej 128, 2200 Copenhagen, Denmark\label{aff102}
\and
Universidad Polit\'ecnica de Cartagena, Departamento de Electr\'onica y Tecnolog\'ia de Computadoras,  Plaza del Hospital 1, 30202 Cartagena, Spain\label{aff103}
\and
Kapteyn Astronomical Institute, University of Groningen, PO Box 800, 9700 AV Groningen, The Netherlands\label{aff104}
\and
Infrared Processing and Analysis Center, California Institute of Technology, Pasadena, CA 91125, USA\label{aff105}
\and
Dipartimento di Fisica e Scienze della Terra, Universit\`a degli Studi di Ferrara, Via Giuseppe Saragat 1, 44122 Ferrara, Italy\label{aff106}
\and
Istituto Nazionale di Fisica Nucleare, Sezione di Ferrara, Via Giuseppe Saragat 1, 44122 Ferrara, Italy\label{aff107}
\and
INAF, Istituto di Radioastronomia, Via Piero Gobetti 101, 40129 Bologna, Italy\label{aff108}
\and
Department of Physics, Oxford University, Keble Road, Oxford OX1 3RH, UK\label{aff109}
\and
ICL, Junia, Universit\'e Catholique de Lille, LITL, 59000 Lille, France\label{aff110}
\and
ICSC - Centro Nazionale di Ricerca in High Performance Computing, Big Data e Quantum Computing, Via Magnanelli 2, Bologna, Italy\label{aff111}}    

\abstract{
The large-scale structure (LSS) of the Universe is an important probe for deviations from the canonical cosmological constant $\Lambda$ and cold dark matter ($\Lambda$CDM) model. A statistically significant detection of any deviations would signify the presence of new physics or the breakdown of any number of the underlying assumptions of the standard cosmological model or possible systematic errors in the data.
In this paper, we quantify the ability of the LSS data products of the spectroscopic survey of the \Euclid mission, together with other contemporary surveys, to improve the constraints on deviations from \lcdm in the redshift range $0<z<1.75$.
We consider both currently available growth rate data and simulated data with specifications from \Euclid and external surveys, based on \lcdm and a modified gravity (MoG) model with an evolving Newton's constant (denoted $\mu$CDM), and carry out a binning method and a machine learning reconstruction, based on genetic algorithms (GAs), of several LSS null tests.
Using the forecast \Euclid growth data from the spectroscopic survey in the range $0.95<z<1.75$, we find that in combination with external data products (covering the range $0<z<0.95$), \Euclid will be able to improve on current constraints of null tests of the LSS on average by a factor of eight when using a binning method and a factor of six when using the GAs. 
Our work highlights the need for synergies between \Euclid and other surveys, but also the usefulness of statistical analyses, such as GAs, in order to disentangle any degeneracies in the cosmological parameters. Both are necessary to provide tight constraints over an extended redshift range and to probe for deviations from the \lcdm model.
}


\maketitle
%
%-----------------------------------------------

\section{Introduction \label{sec:intro}}
Modern cosmology has evolved through a blend of theoretical insights, observational data, and simulations~\citep{2020coce.book.....P}. An example of this integration is the cosmological principle, combining the observed local isotropy (no preferred sky direction) and the Copernican principle (we are not in a unique position in the Universe). These concepts together suggest that the Universe, at its largest scales, is both homogeneous (lacking a special place) and isotropic (lacking a special direction), leading to its representation via the Friedmann--Lema\^itre--Robertson--Walker (FLRW) metric.

Modelling the cosmic expansion history within the FLRW framework requires not only considerations of symmetries but also the specification of dynamical prescriptions, drawing insights from both theoretical physics and observations. Traditionally, general relativity (GR) describes the interplay between matter and space-time geometry. Here we focus on the late stages of cosmic evolution, when pressure-less matter and dark energy predominantly govern the Universe's energy budget. The recent accelerated expansion is typically ascribed to Einstein's cosmological constant, denoted by $\Lambda$, which can be conceptualised as a negative-pressure fluid described by an equation-of-state parameter $w_\mathrm{de}=p_\mathrm{de}\,\rho^{-1}_\mathrm{de}\,c^{-2}=-1$, and is one of the many possible forms of dark energy, see for example  \citet{Joyce:2016vqv} for a recent review. Additionally, the homogeneity hypersurfaces exhibit vanishing curvature.

The standard cosmological model, known as \lcdm, incorporates the cosmological constant $\Lambda$, cold dark matter (CDM), and initial conditions as dictated by the inflationary paradigm. To date, \lcdm stands in excellent agreement with a myriad of cosmological observations~\citep{Aghanim:2018eyx}. However, ongoing debates within the scientific community, such as the $H_0$ tension~\citep{DiValentino:2021izs}, the $\sigma_8$ tension~\citep{Sakr:2018new}, and others \citep{Perivolaropoulos:2021jda} have motivated the exploration of alternatives to the standard \lcdm model. Among the many proposed extensions, a model featuring a Taylor expansion in terms of the scale factor around the present day value of the dark energy equation-of-state parameter $w_\mathrm{de}$, has emerged as a valid competing model that is statistically preferred over $\Lambda$CDM \citep{DESI2025}.

Several methods have been employed to test for such extensions of the \lcdm model, utilizing diverse cosmological probes such as luminosity distances from standard candles \citep{Chiang:2017yrq}, galaxy age estimates from fossil records \citep{Heavens:2011mr}, quasar samples \citep{Laurent:2016eqo}, spectroscopic or photometric surveys \citep{Scrimgeour:2012wt, Alonso:2014xca}, measurements from the kinetic Sunyaev--Zeldovich effect, and peculiar velocity measurements \citep{Nadolny:2021hti}. Theoretical tests have been proposed, which include considering light ray propagation in an inhomogeneous universe using numerical relativity and direct distance measurements to constrain spatial curvature \citep{Giblin:2016mjp, Clarkson:2007pz,Clarkson:2012bg,Valkenburg:2012td}. Additionally, consistency tests based on dynamical probes, growth rate data on its own \citep{Nesseris:2014mfa,Nesseris:2014qca} or in conjunction with other probes \citep{Andrade:2021njl, Achitouv:2016mbn,2021A&A...655A..11R,Viljoen:2020efi,Capozziello:2004jy}, and a linear model formalism have been conducted to comprehensively assess deviations in cosmic structure homogeneity \citep{Marra:2017pst}.

Current and forthcoming surveys are poised to rigorously test $\Lambda$CDM and its underlying principles, as well as possible extensions beyond GR, offering unprecedented data quality and volume. In this work, we particularly focus on \Euclid, which is an M-class space mission of the European Space Agency (ESA), see \citet{Racca:2016qpi}, launched in July 2023 and is already operational. The near-infrared spectrometer and photometer \citep{NISP_paper_2022}
and the visible imager \citep{VIS_paper} are carried aboard the spacecraft and will conduct spectroscopic and photometric galaxy surveys of about 14\,000 deg$^2$ of the sky \citep{2024arXiv240513491E}. The aim is to map the geometry of the observable Universe and measure the growth of structures up to $z\sim 2$~\citep{Laureijs:2011gra, 2024arXiv240513491E}. 

The key observables for probing cosmology through \Euclid involve studying galaxy clustering in both the spectroscopic and photometric surveys. The spectroscopic survey offers detailed radial precision, whereas the photometric survey covers a broader range of observables, although with somewhat larger redshift uncertainties. With the high spectroscopic accuracy of \Euclid, we expect exceptionally precise measurements of galaxy clustering. In this study, we generate simulated growth rate data using the spectroscopic specifications of \Euclid, modelled via the Fisher matrix method, as employed in \citet[][hereafter EC20]{IST:paper1}.

Moreover, we highlight the synergies resulting from the intersection of \Euclid with other large-scale structure (LSS) surveys, specifically the one conducted by the Dark Energy Spectroscopic Instrument (DESI), see \citet{DESI2016}. Both surveys are designed to investigate the expansion history and LSS of the Universe. Notably, their redshift ranges complement each other, thereby substantially broadening the overall redshift range of our observational constraints. This collaborative approach enhances the collective capacity to explore and understand the cosmic evolution over a broader range of redshifts.

The LSS of the Universe offers a powerful probe for exploring modifications of gravity beyond GR, since the latter not only affects the expansion history but also the growth rate of matter density perturbations, thus leading to potentially smoking-gun signatures for deviations from the \lcdm model \citep{Wang:2007ht,Guzzo:2008ac}. Since \Euclid will produce measurements of the growth rate of unprecedented precision, it is expected to provide stringent constraints of the order of 3$\%$ on parameters of covariant modifications of GR (Modified gravity - MoG), according to recent forecasts \citep[][]{Euclid:2023tqw,Euclid:2023rjj}. However, the aforementioned analyses considered extremely specific modifications of GR: \citet[][]{Euclid:2023tqw} assumed the Hu--Sawicki $f(R)$ model, while \citet[][]{Euclid:2023rjj} assumed the Jordan--Brans--Dicke, Dvali--Gabadadze--Porrati, and $k$-mouflage models. Similarly, \citet{ReviewDoc} had presented forecast constraints on deviations from GR by performing a Fisher analysis of the Euclid galaxy survey \citep{Laureijs:2011gra} using similar parameterisations for the MoG models.

In this work we aim to present broader constraints, thus we will consider a generic time-dependent parameterisation of the Newton's constant. This approach can effectively capture deviations from GR at late times and can mimic several MoG models \citep{Nesseris:2017vor}, see also \citet{Ishak:2018his} for a comparison of different phenomenological expressions for Newton's constant at late times. For this, we will also rely on the recent \Euclid specifications \citepalias[see][]{IST:paper1} and also explore synergies with other surveys. Although the currently foreseen specifications of the Euclid Wide Survey (EWS) differ from that assumed in \citetalias{IST:paper1}, for example in terms of the survey area, we use their results to allow for comparison with earlier forecasts.

For our analysis, we will use the \lcdm model and a MoG model based on an evolving Newton's constant at late times. The second part of our analysis is dedicated to the assessment of the so-called `null-tests', which are specifically crafted to identify any deviations from the predicted history of cosmic expansion as dictated by the standard cosmological model. They are agnostic in the sense that they do not require any specific alternative model; instead, the data is used directly to test consistency relations of the \lcdm model.

If any of the null tests in this context finds statistically significant evidence against \lcdm, they can help in identifying which fundamental assumptions of \lcdm are being violated. This might raise questions about the large-scale homogeneity or anisotropy of the Universe, the nature of dark energy relative to $\Lambda$, the interaction between dark matter and dark energy, the spatial curvature of the Universe, among other relevant considerations. For instance, forecasts on \Euclid's ability to constrain deviations from the duality relation were performed in \citet{Martinelli:2020hud}, tests of scalar fields coupling to the electromagnetic sector in \citet{Euclid:2021cfn}, while forecasts on a plethora of null tests of the spatial homogeneity and isotropy of the Universe were performed in \citet{Euclid:2021frk}. In this work, we focus on null tests of \lcdm using mock growth rate data from the Euclid spectroscopic survey, an aspect not explored in the mentioned work, where different null tests were reconstructed, using other observables.

The outline of our paper is as follows: in \Cref{sec:theory} we review the theoretical background of the fundamental assumptions of the standard cosmological model and various ways in which they can be broken. In \Cref{sec:tests} we summarise the null tests employed in our analysis, while in \Cref{sec:method} we describe our methodology for testing the fundamental assumptions of the standard cosmological model via machine learning reconstructions of null tests of the \lcdm model, using mock \Euclid growth rate data.  
Specifically,  in \Cref{sec:mock_method}  we present how we generated mock data based on two different cosmologies: the vanilla \lcdm model, which we used as our null hypothesis, and an evolving with time parameterisation for Newton's constant that at late times can capture deviations from GR and mimic a large class of MoG models in order to examine the response of our null tests to a different cosmology. In \Cref{sec:GA} we review the machine learning approach used for the reconstruction of the null tests. Finally, the results of our analysis are discussed in \Cref{sec:results} for the currently available data and for the mock data, while in \Cref{sec:conclusions} we draw our conclusions.

%--------------------------------------------------------------------
\section{Theoretical background \label{sec:theory}}
In this section, we provide an overview of the theoretical framework for the treatment of the matter density perturbations at linear order, under the fundamental assumption of spatial homogeneity and isotropy of the Universe. In order to describe the LSS of the Universe, we need to consider the perturbed FLRW metric, which at linear order and in the conformal Newtonian gauge is given by,  
\be
\diff s^2=a^2(\eta)\,\Bigg\{-\left[1+\frac{2\,\Psi_\mathrm{N}(\vec{x},\eta)}{c^2}\right]\,c^2\,\diff \eta^2+\left[1-\frac{2\,\Phi_\mathrm{N}(\vec{x},\eta)}{c^2}\right]\,\diff\vec{x}^2\Bigg\}\,,
\label{eq:FRWpert}
\ee
written in terms of the conformal time $\eta$ defined via $\diff\eta=a^{-1}(t)\,\diff t$, where $a(t)$ is the scale factor and $t$ the cosmic time. Furthermore, $\Psi_\mathrm{N}(\vec{x},\eta)$ and $\Phi_\mathrm{N}(\vec{x},\eta)$ are the Newtonian potentials, following \citet{Ma:1995ey}.\footnote{Our conventions are: ($-$+++) for the metric signature, the Riemann and Ricci tensors are given by $V_{b;cd}-V_{b;dc}=V_a\,R^a_{bcd}$ and $R_{ab}=R^s_{asb}$. The Einstein equations are $G_{\mu\nu}=+\kappa\,T_{\mu\nu}$ for $\kappa=8\pi\,G_\mathrm{N}\,c^{-4}$ and $G_\mathrm{N}$ is the bare Newton's constant.}

We can then track the evolution of the matter perturbations via the variable $\delta_\mathrm{m}:= \delta \rho_\mathrm{m}/\rho_\mathrm{m}$, where $\rho_\mathrm{m}$ is the average matter density and $\delta\rho_\mathrm{m}$ represents its linear order perturbation. To do so, it is convenient to assume flatness, neglect neutrinos and radiation at late times, and apply the sub-horizon and quasi-static approximations, where we only consider terms of order $k^2 \gg a^2\,H^2\,c^{-2}$, $k$ is the wavenumber of the Fourier mode of the perturbations, and we neglect terms with time derivatives in the potentials respectively, see for example \citet{Tsujikawa:2007gd}. The neglect of neutrinos and radiation is well justified at late times, where their contributions to the background evolution and structure formation are subdominant. However, for analyses extending to very high redshifts, these effects may become relevant. From the other side, the sub-horizon and quasi-static approximations allow for a simplified treatment of perturbations by focusing on scales well within the horizon, where modifications to gravity often have the most significant impact, and by neglecting the time derivatives of the potentials, respectively. The latter approximation is, in fact, valid also below sound-horizon scales.

In most modified gravity models Newton's gravitational constant can vary with both time and spatial scale, giving rise to an effective coupling $G_\mathrm{eff}(a,k)$, compared to the usual constant $G_\mathrm{N}$. It is then convenient to define a quantity $\mu_\mathrm{mg}(a,k):= G_\mathrm{eff}(a,k)/G_\mathrm{N}$ normalised to unity, and that can be shown to affect the evolution of the matter perturbations \citep{Tsujikawa:2007gd, Amendola:2007rr, Nesseris:2008mq, Nesseris:2009jf}. However, current and forthcoming growth rate data are not sensitive to a possible scale dependence of the matter density perturbations, so we will ignore it in this analysis and focus on the time evolution only, that is $\mu_\mathrm{mg}(a,k)=\mu_\mathrm{mg}(a)$. However, in general this is not the case when DE perturbations are not subdominant \citep{Sapone:2010uy, Sapone:2012nh, Nesseris:2015fqa}.

Then, we can assume that $\delta_\mathrm{m} (\vec{x},a)=g_{+}(a)\,\delta_\mathrm{m}(\vec{x},1)$, where $g_{+}(a)$ is the space-independent linear growth, which be shown to satisfy the following differential equation in many MoG theories:
\begin{equation}
g_{+}^{\prime \prime}(a)+\left[\frac{3}{a}+\frac{H^{\prime}(a)}{H(a)}\right]\,g_{+}^{\prime}(a)-\frac{3}{2}\,\frac{\Omega_\mathrm{m}\,\mu_\mathrm{mg}(a)}{a^{5}\,H^2(a) / H_{0}^{2}}\,g_{+}(a)=0\,, \label{eq:growth}
\end{equation}
where the primes indicate differentiation with respect to the scale factor $a$. Finally, in \Cref{eq:growth} we have set $H(a):=\dot{a}/a$ as the Hubble parameter, where the dot denotes a derivative with respect to cosmic time $t$, and $\Omega_\mathrm{m}$ is the current fractional matter density parameter. In principle, using \Cref{eq:growth} one can only measure the combination $\Omega_\mathrm{m}\, \mu_\mathrm{mg}(a)$, unless some other data or an external measurement of $\Omega_\mathrm{m}$ is used to break the degeneracy between the two; for a more extended discussion of this important point we refer the interested reader to \citet{Zheng:2023yco}.

The MoG fiducial cosmology that we use in our analysis is described by \citep{Nesseris:2017vor}, 
\begin{align}
\mu_\mathrm{mg}(a)=1+g_n\,(1-a)^n-g_n\,(1-a)^{2 n},\label{eq:mu}
\end{align}
where $g_n$ is a constant related to the time derivative of $G_\mathrm{eff}$ and $n\geq2$ due to Solar System constraints that require the first derivative at present, that is at time $t=t_0$, to be effectively zero \citep{Pitjeva:2021hnc}, 
\begin{align}
\frac{\dot{G}_{\mathrm{eff}}}{G_\mathrm{N}}\,\Bigg|_{t=t_0}=\left(0.0_{\,-2.9}^{\,+4.6}\,\times\,10^{-14}\right)\, \mathrm{yr}^{-1}.\label{eq:Geff}
\end{align}
Constraints with currently available growth rate data are much less stringent and seem to indicate that $g_n= -1.16\pm 0.34$ for $n=2$ \citep{Nesseris:2017vor}, which corresponds roughly to a value of $|f_{R0}|\simeq 5\,\times\,10^{-5}$ \citep{kazantzidis2021sigma} in the context of the Hu--Sawicki model, while similar constraints were also obtained in \citet{Ballardini:2019tho, Ballardini:2021evv}. 

In the context of the parametrization described in Eq.~\eqref{eq:mu} one can operate under the assumption that the value of the effective Newton's constant $G_{\mathrm{eff}}$ is matter-density and environment-independent in sub-horizon scales. While this is expected to hold true for the case of MoG models where in the physical frame there is no direct coupling of the scalar degree of freedom to matter density, the same cannot be said for Chameleon-type, scalar-tensor field models. For the latter, we could even consider $n\geq0$ values since the Solar System constraints can be broken. In our analysis, we consider $n=2$ since it applies to both cases. Moreover, this parameterisation has the advantage that can mimic many different MoG models at late times, thus keeping our analysis as general as possible, but at early times is consistent with GR, in agreement with the Big Bang nucleosynthesis constraints.

On the other hand, in the limit when $\mu_\mathrm{mg}(a)=1$, we return to the framework of GR. Then, assuming flatness, a constant dark energy equation-of-state parameter $w_\mathrm{de}$ and neglecting neutrinos and radiation at late times, the Hubble parameter is simply given by
\be
\frac{H^2(a)}{H^2_0}=\Omega_{\rm m}\;a^{-3}+(1-\Omega_{\rm m})\,a^{-3\,(1+w_\mathrm{de})}\,,\label{eq:HlcdmDE}
\ee
while the solution to \Cref{eq:growth} in this case can be found to be \citep{Silveira:1994yq,Percival:2005vm}
\begin{equation}
g_{+}(a)=a\,{}_{2} F_{1}\left[-\frac{1}{3 w_\mathrm{de}}, \frac{1}{2}-\frac{1}{2 w_\mathrm{de}} ; 1-\frac{5}{6 w_\mathrm{de}} ; a^{-3 w_\mathrm{de}}\,\left(1-\Omega_\mathrm{m}^{-1}\right)\right]\,,\label{eq:hyper}
\end{equation}
where ${}_{2} F_{1}(a, b ; c ; z)$ is a hypergeometric function, see \citet{Abramowitz:1974} for more details.

At this point, we can also define the two particular models we will consider in our analysis. They are:
\begin{enumerate}
    \item \lcdm, with $w_\mathrm{de}=-1$, the expansion history given by~\Cref{eq:HlcdmDE}, and the evolution of the perturbations via~\Cref{eq:growth} for $\mu_\mathrm{mg}(a)=1$, leading to the analytic solution for the growth given by~\Cref{eq:hyper}.
    \item $\mu$CDM, which we take to be a designer model with the same background expansion history as in \lcdm, but with $\mu_\mathrm{mg}(a)$ given by~\Cref{eq:mu}. In this case, the growth is calculated by numerically solving~\Cref{eq:growth}.
\end{enumerate}
Finally, we can define the normalised growth $\Delta(a)$ as
\ba
\Delta(a)&:=&\frac{g_{+}(a)}{g_{+}(1)}\,;
\label{eq:sigma}
\ea
however, what is measurable by LSS surveys at linear scales is not exactly the variable $\Delta(a)$, but rather the combination
\ba
f\sigma_8(a)&:=& f_\mathrm{g}(a)\; \sigma_8(a)\nn\\
&=&\frac{\sigma_{8}}{g_{+}(1)}~a~g_{+}'(a) = \sigma_{8}\,a\,\Delta'(a)\,,\label{eq:fsigma8}
\ea
where
\ba
f_\mathrm{g}(a)=\frac{\diff \ln g_{+}(a)}{\diff \ln a}
\ea
is the logarithmic derivative of the growth, called the growth rate, and $\sigma_8(a)=\sigma_{8}\,\Delta(a)$ is the redshift-dependent root mean square (RMS) fluctuations of the linear density field at $R=8\, h^{-1}\, \mathrm{Mpc}$, while the parameter $\sigma_{8}$ is its value today. The value of $f\sigma_8(a)$ can be obtained from the ratio of the monopole to the quadrupole of the redshift-space power spectrum, which depends on the parameter $\beta=f_\mathrm{g}\, b_\mathrm{g}^{-1}$, where $b_\mathrm{g}$ is the galaxy bias. The combination of $f\sigma_8(a)$ is bias-free since both $f_\mathrm{g}(a)$ and $\sigma_8(a)$ have a dependence on bias which is the inverse of the other, thus cancelling out, and it has been shown to be a good discriminator of DE models \citep{Song:2008qt}. 

Performing direct manipulations on the definition of $f\sigma_8$ given by Eq.~\eqref{eq:fsigma8} and the differential equation of the growth given by Eq.~\eqref{eq:growth}, one can prove several useful relations which are valid for GR ($\mu_\mathrm{mg}=1$), see also \cite{Nesseris:2011pc}, for example 
\ba
\Delta(a)&=&\frac{1}{\sigma_{8}}\,\int_0^a \diff x\,\frac{f\sigma_8(x)}{x}\,, \label{eq:Delta_m} \\
f_\mathrm{g}(a) &=& \frac{f\sigma_8(a)}{\int_0^a\diff x\,\frac{f\sigma_8(x)}{x}}\,,\label{eq:f(a)}\\
H^{2}(a)/H_0^2&=&\frac{3\,\Omega_\mathrm{m}}{a^4\,f\sigma_8^{\,2}(a)}\,\int_0^a \diff x\,f\sigma_8(x)\,\int_0^x \diff y \,\frac{f\sigma_8(y)}{y}\,,\label{eq:H}\\
\sigma_{8}&=&\int_0^1 \diff x\,\frac{f\sigma_8(x)}{x}\,, \label{eq:sigma80} \\
\Omega_\mathrm{m}&=&\frac{1}{3\int_0^1 \diff x\,\frac{f\sigma_8(x)}{f\sigma_8(1)}\,\int_0^x \diff y \,\frac1{y}\,\frac{f\sigma_8(y)}{f\sigma_8(1)}}\,.\label{eq:Om}
\ea
Next, we also present the variables related to the matter density perturbations, in particular the growth index $\gamma_\mathrm{g}(a)$, which is defined via \citep{Wang:1998gt},
\be
f_\mathrm{g}(a)=\Omega_\textrm{m}^{\,\gamma_\mathrm{g}(a)}(a)\,,
\ee
where the matter fractional density is given by\footnote{The current value of the fractional matter density parameter is defined as $\Omega_\textrm{m}=\rho_\textrm{m}(t_0)\,\rho_\mathrm{cr}^{-1}$ while at a different epoch it is given by $\Omega_\textrm{m}(z)=\rho_\textrm{m}(z)\,\rho_\mathrm{cr}^{-1}(z)$, and similarly for $\sigma_8$.} 
\be
\Omega_\mathrm{m}(a)=\frac{\Omega_\mathrm{m}\,a^{-3}}{H^{2}(a)/H_0^2}\,.
\ee
Solving for the growth index $\gamma_\mathrm{g}$ we find that it can be expressed via
\be
\gamma_\mathrm{g}(a)=\frac{\ln f_\mathrm{g}(a)}{\ln\Omega_\textrm{m}(a)}
=\frac{\ln f_\mathrm{g}(a)}{\ln\left[\frac{\Omega_\textrm{m}\,a^{-3}}{H^{\,2}(a)/H_0^2}\right]}\,. \label{eq:gamma1}
\ee
In the \lcdm model the growth index has the asymptotic value at early times of $\gamma_\mathrm{g}(z\gg0)\longrightarrow\gamma_{\infty} = 6/11$, which can be quite different from the value today $\gamma_\mathrm{g}(z=0)$.

We note, however, that $f\sigma_8$ has some weak model dependence on the underlying fiducial cosmology via the parameter $\Omega_\mathrm{m}$ used by the surveys to extract $f\sigma_8$. This may be corrected and a robust growth data set can be compiled, see for example Sect. 3 in \cite{Sagredo:2018ahx} where corrections for the Alcock--Paczynski (AP) effect are used to minimise the mismatch in the underlying cosmologies. On the other hand, there is the issue of how to model the nonlinear effects properly, which requires higher-order perturbation theory or costly numerical simulations, such that one can marginalise those nonlinearities and extract the linear growth. Thus, while the proper way to analyse the data is to first model the nonlinearities and then extract the growth, our approach is still useful given the current data and their overall uncertainties. 

For future high-precision surveys like \Euclid, where nonlinear scales will be probed with greater accuracy, a more detailed treatment of nonlinear effects may become necessary. However, as long as the dominant information in our analysis comes from linear scales, our approach remains a reasonable approximation. Additionally, methods to mitigate nonlinear contamination, such as scale cuts or empirical calibrations using simulations, could be employed to refine our results without requiring a full nonlinear modelling.

\section{Consistency tests \label{sec:tests}}
The consistency tests for the growth of matter perturbations we consider in this work are motivated by the properties and symmetries of the \lcdm model. We thus probe for deviations from this baseline model that range from simple modifications of the expansion history to models that are extensions of GR at late times.

In this section, we now introduce these null tests for LSS  assuming the \lcdm model and using mock data from \Euclid and other contemporary surveys. Our approach encompasses a diverse range of null tests targeting both the \lcdm model and the FLRW metric, as each test is designed to be sensitive to different observational probes within the redshift range that \Euclid covers. 
\subsection{The $\gamma_\mathrm{g}(\alpha)$ null test\label{sec:test1}}
Assuming the \lcdm model ($\mu_\mathrm{mg}=1$) and combining Eqs.~\eqref{eq:f(a)}, \eqref{eq:Om}, and \eqref{eq:gamma1}, we obtain our main result for the growth index:
\be
\gamma_\mathrm{g}(a)=\frac{\gamma_A(a)}{\gamma_B(a)}\,,\label{eq:gamma2}
\ee
where
\ba
\gamma_A(a)&=&\ln\left[\frac{f\sigma_8(a)}{\int_0^a\,\diff x\,\frac{1}{x}\,f\sigma_8(x)}\right]\,,\\
\gamma_B(a)&=&\ln\left[\frac{a\,f\sigma_8^2(a)}{3\int_0^a\,\diff x\,f\sigma_8(x)\,\int_0^x \diff y\,\frac{1}{y}\,f\sigma_8(y)}\right]\,.
\ea
The main advantage of Eq.~\eqref{eq:gamma2} is that it only requires knowledge of $f\sigma_8(a)$, and does not explicitly depend on $\Omega_\mathrm{m}$ or $H(a)$, $\sigma_{8}$ or any other parameter. Furthermore, the growth index $\gamma_\mathrm{g}$ has been shown to be sensitive to modifications of GR that have a time-dependent Newton's constant \citep{Linder:2005in}.

In the case of Eq.~\eqref{eq:gamma2}, since the growth index only depends on $f\sigma_8(a)$ we can directly propagate the error numerically by using the value of the growth rate at the $1\sigma$ boundaries as
\be 
f\sigma_{8,1\sigma}(a)=f\sigma_8(a)\pm\delta f\sigma_8(a),
\ee 
where $\delta f\sigma_8(a)$ is the growth rate uncertainty. Then, the value of the growth index at the $1\sigma$ boundaries will be
\be 
\gamma_{1\sigma}(a)=\gamma\left[f\sigma_{8,1\sigma}(a)\right]=\gamma\left[f\sigma_8(a)\pm\delta f\sigma_8(a)\right],
\ee
which is a valid approximation since the uncertainty on $f\sigma_8$ is small.

\subsection{The $\mathcal{O}(a)$ null test\label{sec:test2}}
Exploiting the Noether symmetries of Eq.~\eqref{eq:growth} within the context of GR ($\mu_\mathrm{mg}=1$) we can define a conserved charge that has to be constant at all redshifts, thus is an ideal null test. Following this procedure, it was shown by \cite{Nesseris:2014mfa} that a null test for the growth can be written as
\ba
\mathcal{O}(a)&=& a^2\, E(a)\,\frac{f\sigma_8(a)}{f\sigma_8(1)}\,\mathrm{e}^{\,I(a)}\,, \\
I(a)&=&-\frac{3}{2}\,\Omega_\mathrm{m}\int_{1}^a\,\diff x\,\frac{\sigma_{8}+\int_{1}^x\,\diff y\,\frac{f\sigma_8(y)}{y}}{x^4\,E^2(x)\,f\sigma_8(x)}\,,~~
\label{eq:nullf}
\ea
where for simplicity we have set $E(a):= H(a)/H_0$. As before, using Eqs.~\eqref{eq:H}, \eqref{eq:sigma80}, and \eqref{eq:Om} we can rewrite the previous expressions solely in terms of $f\sigma_8$ and $E$. Doing so we find,
\ba 
\mathcal{O}^2(a)&=& G(a)\,\exp\left[-\frac{\int_{1}^a\diff x\, F(x)/G(x)}{\int_{0}^1\diff x\, F(x)}\right]\,,\label{eq:nullf1}
\ea
where we have set
\ba
F(a)&:=&\frac{f\sigma_8(a)}{f\sigma_8(1)}\,\int_0^a \diff y\,\frac{1}{y}\,\frac{f\sigma_8(y)}{f\sigma_8(1)},\\
G(a)&:=& a^4\, E^2(a)\,\left[\frac{f\sigma_8(a)}{f\sigma_8(1)}\right]^2.
\ea 
Clearly, not only the $\mathcal{O}(a)$ test has to be constant for all redshifts $z$, but more strongly
\be 
\mathcal{O}(a)\longrightarrow 1~~\mathrm{implies}~~\mathrm{GR}.
\ee 
As mentioned earlier, the differential equation for the growth, given by Eq.~\eqref{eq:growth}, assumes the validity of linear perturbation theory in a flat FLRW metric, that radiation and neutrinos can be neglected at late times, but also the validity of the sub-horizon and quasi-static approximations. Since the $\mathcal{O}(a)$ test relies on Eq.~\eqref{eq:growth}, then any deviations of the null test from unity also imply that one or more of the aforementioned assumptions are no longer valid. Finally, the $68.3\%$ uncertainty of $\mathcal{O}$ was calculated by adding the uncertainties of the various terms in quadrature, assuming standard Gaussian error propagation.

\subsection{The $\mathrm{Om}_{f{\sigma_8}}(a,\Delta)$ null test\label{sec:test3}}
For our final null test we follow \citet{Arjona:2021mzf} and use the fact that in \lcdm the normalised growth $\Delta$, as can be seen by inspecting Eqs.~\eqref{eq:hyper} and \eqref{eq:sigma}, depends only on the scale factor $a$ and the matter density parameter $\Omega_\mathrm{m}$, meaning that we have
\be
\Delta=\mathcal{D}(a, \Omega_\mathrm{m}),\label{eq:delta_D}
\ee
where $\mathcal{D}$ is given in terms of hypergeometric functions. To derive a quantity suitable to be a null test, we notice that in \lcdm the matter density parameter $\Omega_\mathrm{m}$ is, by definition, a constant at all redshifts, thus we invert Eq.~\eqref{eq:delta_D} and solve for $\Omega_\mathrm{m}$.

Specifically, we first perform a series expansion on Eq.~\eqref{eq:delta_D} around $\Omega^{-1}_\mathrm{m}=1$, since we found it is more numerically stable and robust, and we then keep the first $15$ terms of the expansion. Next, we apply the Lagrange inversion theorem to invert the aforementioned series expansion and write the inverse matter density $\Omega^{-1}_\mathrm{m}$ as a function of $\Delta$.\footnote{The Lagrange inversion theorem asserts that given an analytic function, one can estimate the Taylor series expansion of the inverse function. In other words, given the function $y=f(x)$, where $f$ is analytic at a point $p$ and $f'(p)\neq 0$, the theorem allows us to solve the equation for $x$ and write it as a power series $x=g(y)$, see \cite{Abramowitz:1974}.}
Doing so, we can then define a quantity $\mathrm{Om}_{f{\sigma_8}}(a,\Delta)$ that is equal to the formal inversion of Eq.~\eqref{eq:delta_D}, so that it reduces to a constant $\Omega_{m0}$ only if $\Delta$ corresponds to \lcdm, or schematically
\ba
\mathrm{Om}_{f{\sigma_8}}(a,\Delta)&:=&\mathcal{D}^{-1}\left(a,\Delta\right)\nn\\
&\longrightarrow& \Omega_\mathrm{m}~~\mathrm{(constant)}~~\mathrm{implies}~~\Lambda\mathrm{CDM}.\label{eq:nulltest}
\ea
Following this procedure we find that the first few terms of our null test are given by
\be 
\mathrm{Om}_{f{\sigma_8}}(a,\Delta)=\left[1+\frac{11\left(\Delta-a\right)}{2a\left(1-a^3\right)}+\cdots\right]^{-1}.
\ee 
As in previous tests, $\mathrm{Om}_{f{\sigma_8}}$ has to be constant at all redshifts and any deviation from $\Omega_\mathrm{m}$ could be due to the change in the expansion history, similarly to the Om statistic of \citet{Shafieloo:2009ti}. 

\section{Methodology \label{sec:method}}
In this section we present our methodology and, in particular, our recipe for making mock realisations of data, based on \Euclid's specifications.\footnote{As this and the following sections deal with observational data, for convenience we will now switch to using the redshift $z$
as a time variable, instead of the scale factor $a$ used in the previous section. As the two variables are simply related to each other via $a=1/(1+z)$, it is straightforward to interchange between the two, thus by writing $\mathcal{O}(z)$ we will in fact mean the $\mathcal{O}(a)$ test given by Eq.~\eqref{eq:nullf1}, and similarly for the rest of the tests.}
\begin{table}
\begin{center}
\setlength{\tabcolsep}{4pt}
\renewcommand{\arraystretch}{1.35}
\caption{The parameter values of the fiducial models we used for the mocks. The values for the \lcdm follow the fiducial of \citetalias{IST:paper1}; in particular, spatial flatness is assumed. Also, $H_0$ is shown in units of km s$^{-1}$~Mpc$^{-1}$. \label{tab:fiducials}}
\begin{tabular}{lcccccccc}
\hline
\hline
model  & $\Omega_{\rm m}$ & $\Omega_{\rm b}\,h^2$ & $H_0$ & $w_0$ & $w_a$ & $\sigma_{8}$ & $n_\mathrm{s}$ & $g_n$\\ \\
 \hline
\lcdm & $0.32$ & $0.02225$ & $67$ & $-1$ & $0$ &0.816 & 0.96 &0 \\
$\mu$CDM & $0.32$ & $0.02225$ & $67$ & $-1$ & $0$ & 0.816 & 0.96 &$-1$\\
\hline
\hline
\end{tabular}\\
\end{center}
\end{table}
\subsection{Current Data\label{sec:current_method}}
The RSD data are characterised by a phenomenon occurring at both large and small scales during observation. In essence, due to the peculiar velocities of galaxies, an overdense region appears compressed in redshift space at large scales, while at small scales, the same region is compressed along the line of sight. This effect influences the 2-point correlation function, leading to an anisotropic power spectrum. Additionally, on large scales, part of the observed anisotropy in the power spectrum may be attributed to the use of an incorrect fiducial cosmology, specifically the Hubble parameter, necessitating consideration when analysing growth rate data, which can be corrected for when taking into account the AP effect.

On the other hand, the CC data correspond to another quite useful cosmological probe, as they can directly constrain the Hubble rate at different redshifts by using the differential age method for passively evolving galaxies. Here we mainly use the compilation of 39 CC data discussed in \citet{Alestas:2022gcg}, see \Cref{sec:dataappdx}. The data set used in this work is also similar to that of \citet{Moresco:2022phi} with the exception of a few data points. The total covariance matrix $C_{ij}^\mathrm{tot}$ for the CC data is defined as the combination of its statistical and systematic uncertainties
\be
C_{ij}^\mathrm{tot} = C^{\,\mathrm{stat}}_{ij}+C^{\,\mathrm{syst}}_{ij}\,, \label{eq:CCcovmat}
\ee
where $C^{\,\mathrm{syst}}_{ij}$ is attributed to the physical properties of the observed galaxies (like stellar metallicity, galaxy population age etc.). In this work however, we do not use the covariance matrix of the cosmic chronometer data. The main effect of its inclusion would be to increase the uncertainties in the data points and complicate the overall analysis. 

Furthermore, there are a few observational effects that could potentially introduce biases in the CC data analysis. One such effect is the so-called progenitor bias which entails the fact that there could be a substantial enough difference in the selection criteria of early-type galaxies when dealing with high versus low-redshift ones. Specifically, galaxies with high redshifts meeting the sample criteria could potentially be both older and more massive compared to those chosen at lower redshifts, essentially representing the progenitor population of the low-redshift sample \citep{vanDokkum:2000rs, Moresco:2022phi}.

\subsection{Mock Data\label{sec:mock_method}}
In order to assess \Euclid's sensitivity to deviations from the \lcdm model using the null tests discussed in \Cref{sec:tests}, mock data are utilised based on \Euclid's specifications outlined in \citetalias[][]{IST:paper1}. Specifically, simulated data sets for $f\sigma_8$ measurements are generated using both the \lcdm and $\mu$CDM models, with parameter values sourced from \Cref{tab:fiducials}. It is worth noting that the \lcdm fiducial aligns with that employed in \citetalias{IST:paper1}. 

Upon establishing the two primary cosmological models under scrutiny, we proceeded to derive the fiducial values for the growth rate $f\sigma_8$ and the Hubble parameter across multiple redshifts $z$. Subsequently, our mock data sets were generated, adhering to the specifications of forthcoming surveys, as discussed in the subsequent sections.

While current and forthcoming surveys, such as \Euclid, will deliver highly precise data, validating their accuracy is of paramount importance. This necessitates addressing potential observational systematic uncertainties that could impact the survey analyses. For example, \citet{Paykari2020} undertook a comprehensive examination of the observational systematic effects associated with the \Euclid VIS instrument, delving into considerations related to the modelling of the point spread function and charge transfer inefficiency. 

Similarly, accurate calibration of the halo mass function is essential for mitigating theoretical systematics that impact galaxy-halo connection models, which are key for interpreting galaxy clustering measurements \citep{Euclid:2022dbc}.
Although the specifications may in fact change, here we assume a comprehensive understanding of the systematic effects outlined will be attained by the time data collection commences. Consequently, we have incorporated all pertinent astrophysical systematic effects, including galaxy bias, into our analysis, as detailed in what follows.

Our calculation of covariance matrices assumes a fiducial \lcdm cosmology. This means that uncertainties due to using a different cosmological model have not been considered in our analysis. Computing covariance matrices for non-\lcdm models is still a complex and unresolved challenge for future surveys, as discussed in \citet{Harnois-Deraps:2019rsd} and \citet{Friedrich:2020dqo}. Therefore, in our work, the mock data were generated using the Fisher matrix approach outlined in \citetalias{IST:paper1} for the spectroscopic survey.

Our analysis is focused on the spectroscopic part of \Euclid, with the primary goal of attaining precise measurements of the Hubble parameter $H(z)$, the angular diameter distance $D_\mathrm{A}(z)$ and the growth rate $f\sigma_8(z)$. In particular, \Euclid is anticipated to explore the galaxy power spectrum within the redshift range $z\in [0.95,1.75]$, with a particular focus on ${\rm H}_{\alpha}$ emitters, as indicated in \citetalias{IST:paper1}. In this context, \Euclid is projected to accumulate approximately 30 million spectroscopic redshifts with an uncertainty of $\sigma_z=0.001\,(1+z)$, following the findings of \citet{Pozzetti:2016cch}. The primary observational target is the galaxy power spectrum, offering insights into the AP effect, residual shot noise, redshift uncertainties, and anisotropies arising from redshift-space distortions (RSD) and the galaxy bias. Additionally, we consider nonlinear effects impacting the power spectrum's shape, such as the nonlinear smearing of the baryon acoustic oscillations (BAO) feature, as elucidated in \citet{Wang:2012bx}. Potential considerations also encompass a nonlinear, scale-dependent galaxy bias, as explored in \citet{delaTorre:2012dg}.

In this part, our binning strategy is similar to the one outlined in \citet{Martinelli:2020hud}, departing slightly from the approach employed in \citetalias{IST:paper1}. Specifically, we assumed nine redshift bins, each with a width of $\Delta z= 0.1$, a departure from the four bins previously employed. By following this binning structure, we recalibrated the specifications for the comoving galaxy number density $n(z)$ in units of Mpc$^{-3}$ and the galaxy bias $b_\mathrm{g}(z)$, expressed as follows:
\begin{align}
n(z)&= \left\{2.04, 2.08, 1.78, 1.58, 1.39, 1.15, 0.97, 0.7, 0.6\right\}\times 10^{-4},\,\nonumber \\
b_\mathrm{g}(z)&= \left\{1.42, 1.5, 1.57, 1.64, 1.71, 1.78, 1.84, 1.90, 1.96\right\} \,.\nonumber
\end{align}
This binning approach offers a greater number of data points and substantially enhances the efficacy of the machine learning analysis, as shown later on. However, it is important to note that we subjected this specific choice to a comparative analysis against that of \citetalias{IST:paper1} in \citet{Martinelli:2020hud}, revealing no statistically significant difference. 

To derive the Fisher matrix for all of the cosmological parameters, employed to estimate the parameter covariance matrix and propagate the uncertainties of the parameters, we followed the methodology detailed in \citetalias{IST:paper1}. Our analysis contains five redshift-dependent parameters: $\ln \left(D_{\rm A}\,\mathrm{Mpc}^{-1}\right)$, $\ln \left(H\,\mathrm{km}^{-1}\,\mathrm{s}\,\mathrm{Mpc}\right)$, $\ln f\sigma_8$, $\ln b\sigma_8$, and $P_{\rm s}$, evaluated in each redshift bin. Here, $b\sigma_8= b_\mathrm{g}(z)\,\sigma_8(z)$, and $P_{\rm s}$ characterise the galaxy bias and the shot noise, respectively (see \citetalias{IST:paper1}). 

As we are considering the optimistic scenario of \citetalias{IST:paper1}, the four shape parameters $\{\omega_{\rm m}=h^2\,\Omega_{\rm m}$, $h$, $\omega_{\rm b}=h^2\,\Omega_{\rm b}$, and $n_{\rm s}\}$, and the two nonlinear parameters $\{\sigma_{\rm p}\text{ and }\sigma_{\rm v}\}$ have been kept fixed, the latter to the values given by Eqs.~(85)-(86) of \citetalias{IST:paper1}, while for the maximum wavenumber, we considered the value $k_{\max } =0.30 \, h\, \mathrm{Mpc}^{-1}$. Finally, our Fisher forecast code also includes the AP effects, as discussed in detail in \citetalias{IST:paper1}. 

Using this methodology, we can compute the expected uncertainties for the Hubble parameter $H(z)$ and the growth rate $f\sigma_8(z)$ in each of the nine redshift bins. Although the final Fisher matrix might theoretically depend on the specific fiducial cosmology, for this analysis we assumed that any such dependence is at most weak.

Considering the restricted redshift span, $z \in [0.95,1.75]$, covered by the Euclid spectroscopic survey, our access to BAO data is inherently constrained. To overcome this limitation, we augmented our analysis by incorporating data from the DESI survey, ensuring comprehensive coverage of null tests across a broader redshift range. DESI commenced survey operations in 2021, and is poised to acquire optical spectra for tens of millions of quasars and galaxies up to $z\sim 4$, facilitating BAO and RSD analyses.

For the Hubble parameter $H(z)$ and growth rate $f\sigma_8(z)$, we adopted the official DESI forecasts outlined in \citet{DESI2016}. These forecasts, derived using a Fisher matrix approach detailed in \citet{2014JCAP...05..023F}, leverage the full anisotropic galaxy power spectrum. This spectrum incorporates measurements of the matter power spectrum relative to the line of sight, accounting for variations across redshift and wavenumber. Analogous to the \Euclid forecasts, this methodology includes all available information from the 2-point correlation function, not merely the position of the BAO peak.

In particular, our analysis centred on the baseline DESI survey, covering 14\,000\,deg$^2$ and targeting various populations, including bright galaxies (BGs), luminous red galaxies (LRGs), emission-line galaxies (ELGs), and quasars in the redshift range $z\in [0.05,3.55]$. The precision of the survey explicitly depends on the target population. For instance, BGs span the redshift range $z\in [0.05,0.45]$ in five equally spaced redshift bins, while the subsequent targets, LRGs and ELGs, cover the range $z\in [0.65,1.85]$ in $13$ equally spaced redshift bins. Notably, in our analysis, we selectively incorporated DESI data at late times, ensuring no overlap with \Euclid points to avoid spurious correlations between the two surveys. In order to perform a consistent and fair analysis with both surveys, we also use the same specifications as the ones used for \textit{Euclid}, namely the same $k_{\max }$ and the same parameter values for the fiducial models used in our analysis (also shown in \Cref{tab:fiducials}).

\subsection{Reconstruction methods\label{sec:GA}}
In this work, we use two approaches for the reconstruction of the null tests, first a binning method and second, a machine learning technique known as the genetic algorithms (GAs).

For the binning method, we performed the growth index reconstruction using Eq.~\eqref{eq:gamma2}, the $\mathcal{O}(z)$ test by Eq.~\eqref{eq:nullf1}, and the $\mathrm{Om}_{f{\sigma_8}}(z,\Delta)$ test by Eq.~\eqref{eq:nulltest}. For our analysis, we require both Hubble parameter $H(z)$ and growth rate $f\sigma_8(z)$ data. Given the limited number of $f\sigma_8$ data available and their distribution, predominantly denser at lower redshifts, the choice of the bins was determined by the $f\sigma_8$ data. Our goal was to create bins that are as equally spaced as possible, while also ensuring each bin contains a similar number of data points. We adopted this approach of equal spacing in order to produce accurate results while performing the trapezoidal integration.

On the other hand, GAs represent a collection of stochastic optimisation techniques frequently employed for non-parametric data reconstructions using analytic functions. Drawing inspiration from grammatical evolution, these symbolic regression algorithms incorporate genetic operations such as mutation (a random change in an individual) and crossover (the combination of different individuals to generate offspring). These operators function in the same way as their counterparts in evolutionary biology, mirroring the principles of natural selection and enabling the adaptability of the best-fit functions that ultimately describe the data.

The likelihood of an individual function within the population created in one iteration of the GA to successfully pass on its features to the next iteration's population is typically assumed to be directly proportional to the functions' fitness. Fitness, in turn, is a measure of how well each function describes the provided data, quantified through a standard $\chi^2$ statistic. The GAs have found diverse applications, including exploring extensions beyond the standard model \citep[]{Akrami:2009hp}, investigating deviations from $\Lambda$CDM at both the background and perturbation levels in linear order \citep[]{Nesseris:2012tt, Arjona:2020kco, Arjona:2019fwb, Alestas:2022gcg}, and reconstructing various cosmological data sets such as supernovae data \citep[]{Bogdanos:2009ib, Arjona:2020doi}. Moreover, GAs have been instrumental in performing reconstructions of null tests, such as the Om statistic,  the curvature test, and others  \citep[]{Nesseris:2010ep, Nesseris:2013bia, Sapone:2014nna, Euclid:2021frk}, but also as analytical emulators of the matter power spectrum \citep{Kammerer:2025dbi, Sui:2024wob, Bartlett:2023cyr, Bartlett:2024jes, Orjuela-Quintana:2024hha}.

In the initialisation phase of the algorithm, a set of functions is randomly selected from an orthogonal basis constituted by various functions such as orthogonal polynomials, exponentials, logarithms, and other elementary functions. Additionally, it encompasses operators like $+$, $-$, $\times$, $\div$, and $\wedge$, where $\wedge$ represents exponentiation. For example, for two functions $f(x)$ and $g(x)$, $f\wedge g= f^g=\exp\left[g\,\ln(f)\right]$. This selection forms the grammar of our algorithm, guiding the GA to generate the best-fitting functions. The output functions are not constrained in terms of their properties, except for the requirement to fit the data exceptionally well, as is defined by the $\chi^2$ statistic. We intentionally avoid imposing a prior on the functional space. However, to ensure realistic results, two main considerations were implemented. First, we insisted that all functions be smooth, continuous, and differentiable across the entire redshift range covered by the data, a condition automatically enforced in our GA code implementation. Secondly, additional physical priors were assumed to further validate the realism of the derived functions.

Once the initial population of functions is established, we proceeded to evaluate the fitness of each member by employing a $\chi^2$ statistic. This involved utilising the available data and their respective covariance matrices simultaneously as input, allowing for a comprehensive assessment of the overall fitness of the set. Using the tournament selection method as described by \citet{Bogdanos:2009ib}, a random subset of the best-fitting functions from each generation is chosen. Subsequently, we applied the crossover and mutation operators to the selected functions. To ensure the algorithm's convergence, we iterated this process thousands of times. Lastly, as a precaution against any potential bias arising from the selection of a specific seed, we performed multiple runs with different random seeds.

We determined the uncertainties associated with the best-fit functions by employing a path integral approach introduced by \cite{Nesseris:2012tt} and \cite{Nesseris:2013bia} within the framework of GAs. In essence, the process involved integrating the likelihood function across the entire functional space explored by the GA, providing a comprehensive assessment of uncertainties. The  reliability of this approach has been extensively verified by \citet{Nesseris:2012tt}, demonstrating consistency with uncertainty estimates derived from a bootstrap Monte Carlo method. We performed the GAs numerical analysis using the open source code \texttt{Genetic Algorithms}.\footnote{\url{https://github.com/snesseris/Genetic-Algorithms}}

\section{Results\label{sec:results}}
In this section, we provide the results from the null tests that were discussed in \Cref{sec:tests}, calculated using the currently available data, mock \Euclid and DESI data for both the \lcdm and the $\mu$CDM fiducial cosmologies. Finally, by using the currently available data we quantify the improvement that \Euclid will bring.

\subsection{GA analysis of the currently available data\label{sec:real_results}}
In this part of our analysis, we use compilations of the most up-to-date RSD and cosmic chronometer (CC) data that are currently available (see \Cref{sec:dataappdx}). Using these data sets we perform the GA reconstructions of all three functions that constitute the null test discussed in \Cref{sec:tests}, and subsequently compare their forms to those predicted in the context of \lcdm.
\begin{figure*}[!t]
\centering
\includegraphics[width = 0.475\textwidth]{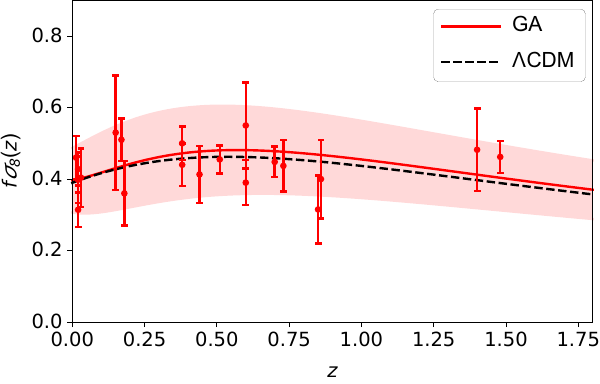}
\includegraphics[width = 0.49\textwidth]{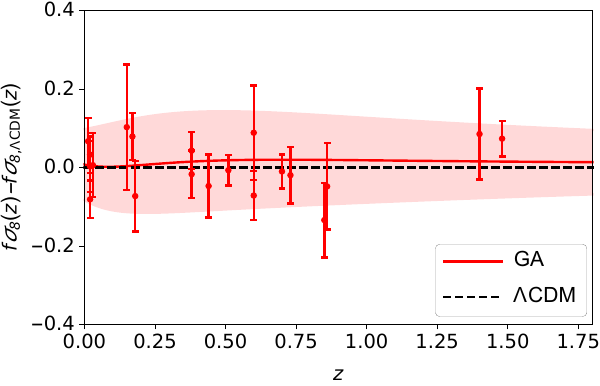}
\hspace*{-0.3cm}\includegraphics[width = 0.49\textwidth]{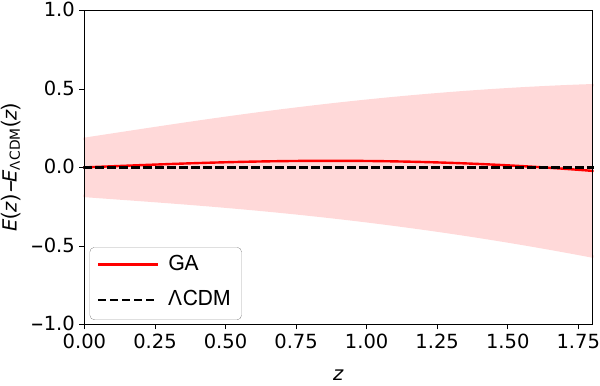}
\hspace*{0.0cm}\includegraphics[width = 0.48\textwidth]{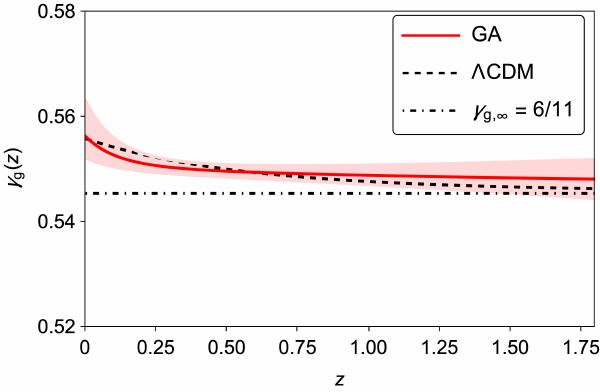}
\includegraphics[width = 0.475\textwidth]{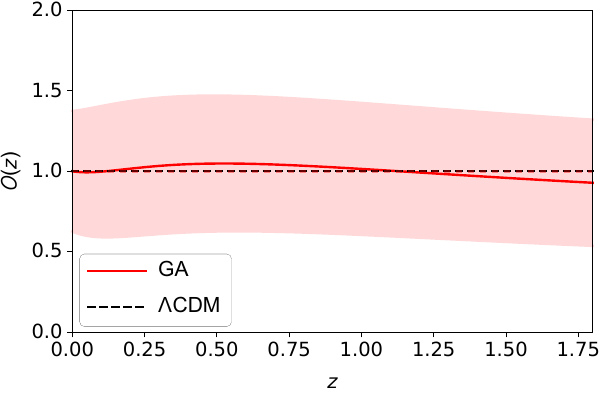}
\hspace*{0.2cm}\includegraphics[width = 0.47\textwidth]{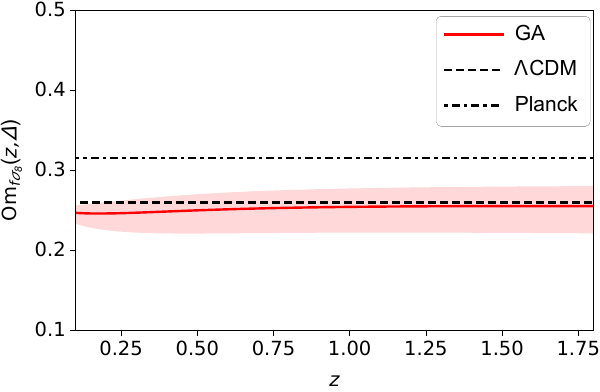}
\caption{Plots of the GA and \lcdm reconstructions with the currently available data. In all cases, we show the best-fit GA (red line), its $68.3\%$ uncertainty region (light red shaded area), and \lcdm (dashed black line). \emph{Top}: The growth rate $f\sigma_8$ as a function of redshift (left) and the difference between the GA and \lcdm $f\sigma_8$ best-fits (right). \emph{Centre}: The difference between the GA and \lcdm best-fits of the normalised Hubble parameter (left) and the GA growth index reconstruction (right) using Eq.~\eqref{eq:gamma2}. \emph{Bottom}: The $\mathcal{O}(z)$ test given by Eq.~\eqref{eq:nullf1}, left panel, and the $\mathrm{Om}_{f{\sigma_8}}(z,\Delta)$ test given by Eq.~\eqref{eq:nulltest}, right panel.\label{fig:plots_real}}
\end{figure*}

\begin{figure*}[!t]
\centering
\includegraphics[width = 0.48\textwidth]{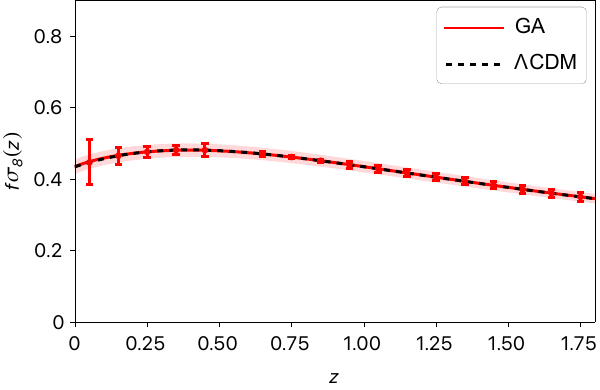}
\includegraphics[width = 0.49\textwidth]{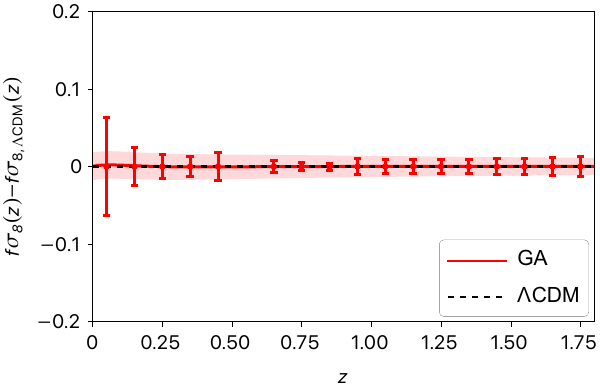}
\hspace*{-0.5cm}\includegraphics[width = 0.499\textwidth]{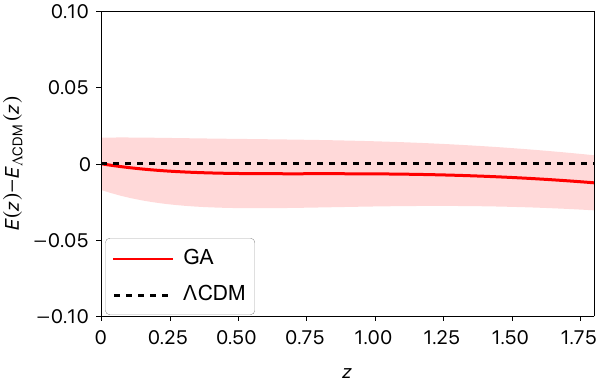}
\includegraphics[width = 0.49\textwidth]{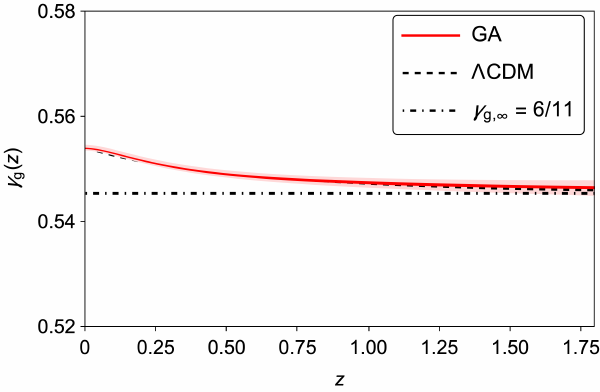}
\hspace*{-0.2cm}\includegraphics[width = 0.49\textwidth]{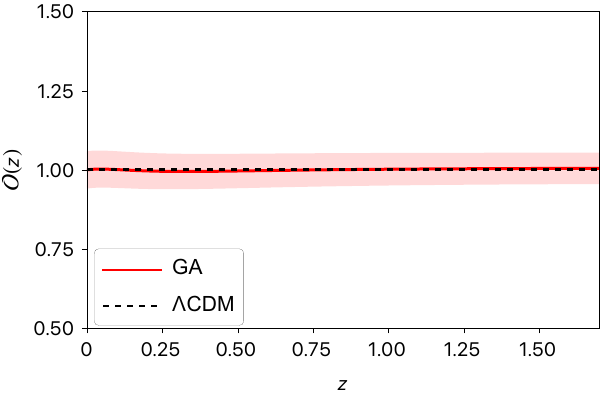}
\includegraphics[width = 0.49\textwidth]{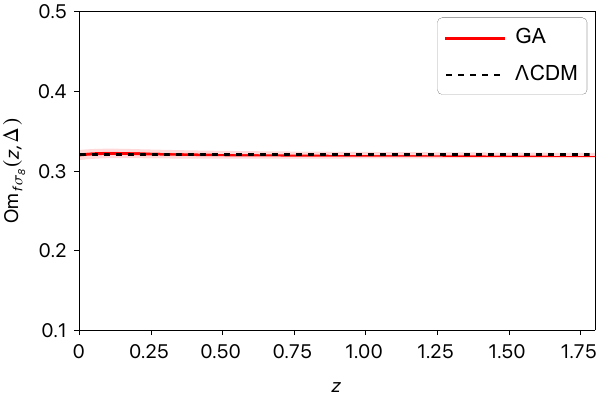}
\caption{The same as in Fig.~\ref{fig:plots_real}, but using the \Euclid + DESI \lcdm mock data. \label{fig:plots_mock_LCDM}}
\end{figure*}

In the top left panel of Fig.~\ref{fig:plots_real} we show the GA best-fit (red solid line) for $f\sigma_8$ and its $68.3\%$ uncertainty region (light red shaded region), the \lcdm best-fit (dashed black line) along with the growth rate data (red points), while in the top right panel, we show the difference between the GA best-fit and the \lcdm one. Similarly, in the centre left panel we show the difference of the GA best-fit normalised Hubble parameter $E(z)$ against that of the best-fit \lcdm one. Then, in the centre right panel we show the GA reconstruction of the growth index $\gamma_\mathrm{g}(z)$ (red line) using Eq.~\eqref{eq:gamma2} and the RSD data, while the \lcdm reconstruction is shown in the black dashed line and the asymptotic value of $6/11$ of the growth index at infinity in the black dot-dashed line. 

Similarly, for the case of the second null test described by Eq.~\eqref{eq:nullf1}, we see in the bottom left panel of Fig.~\ref{fig:plots_real} that its value given via the GA pipeline (red line), using both RSD and CC data, is in agreement within a $68.3\%$ confidence level with \lcdm (red dashed line). For the third-in-line null test described in \Cref{sec:tests} we see in the bottom right panel of Fig.~\ref{fig:plots_real} that the best-fit function of Eq.~\eqref{eq:nulltest}, as it is reconstructed via the GA pipeline and using the RSD data compilation, is again in agreement at the $68.3\%$ confidence level with the best-fit value of the $\Omega_\mathrm{m}$ given by a standard $\chi^{2}$ minimisation process (black dashed line). However, since we considered only RSD data the best-fit value that we obtained is at a $95.5\%$ confidence level distance away from the Planck best-fit value attributed to \lcdm (black dashed line), see \citet{Aghanim:2018eyx}.

In summary, we observe on average over the redshift range covered by the currently available data, constraints on $\gamma_\mathrm{g}(z)$ at the level of $0.5\%$, on $\mathcal{O}(z)$ at the level of $40\%$ and $\mathrm{Om}_{f{\sigma_8}}(z,\Delta)$ at the $8\%$ level. 
\begin{figure*}[!t]
\centering
\includegraphics[width = 0.47\textwidth]{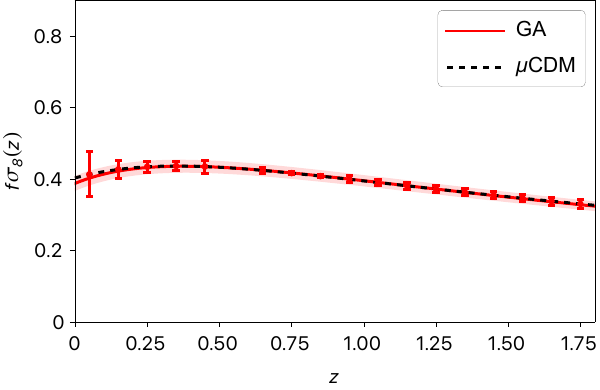}
\includegraphics[width = 0.48\textwidth]{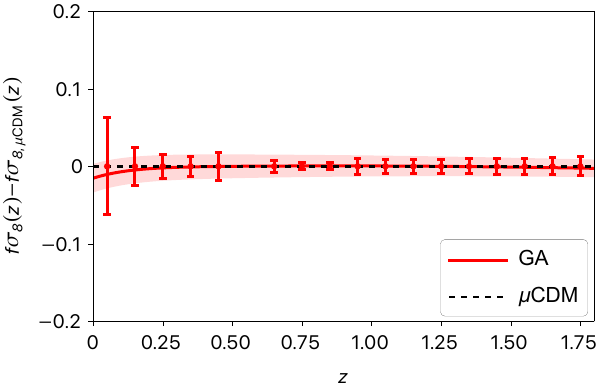}
\hspace*{-0.5cm}\includegraphics[width = 0.495\textwidth]{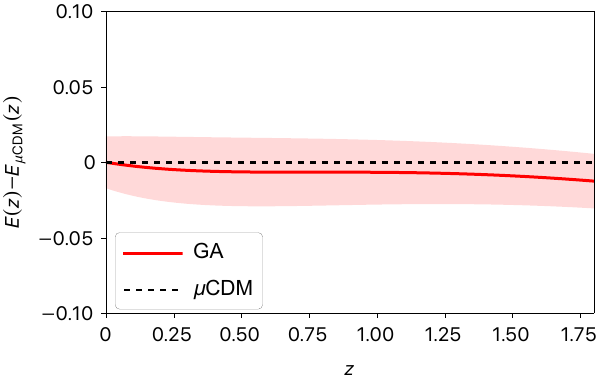}
\includegraphics[width = 0.48\textwidth]{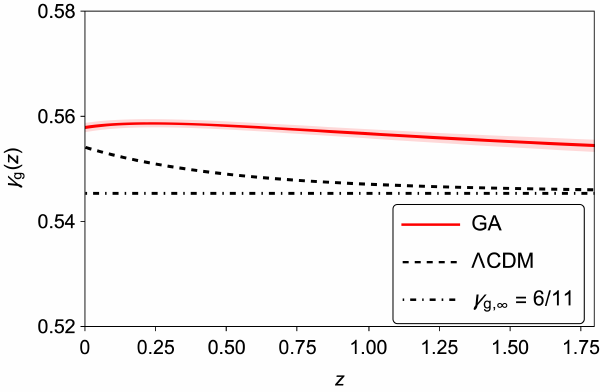}
\includegraphics[width = 0.48\textwidth]{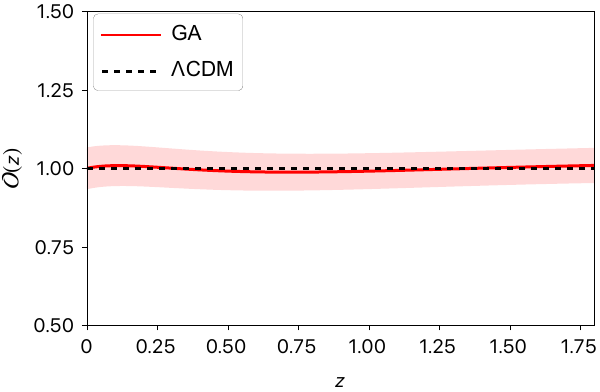}
\includegraphics[width = 0.48\textwidth]{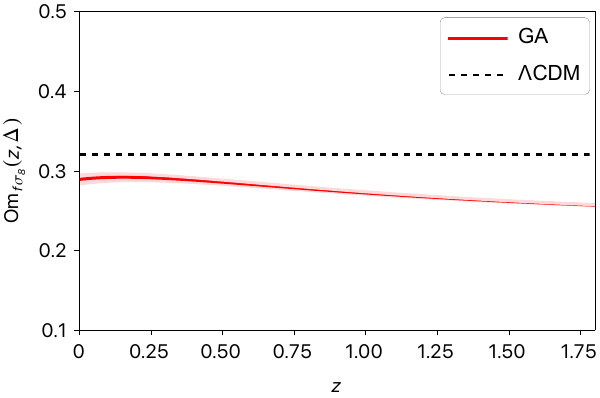}
\caption{The same as in Fig.~\ref{fig:plots_real}, but using the \Euclid  + DESI $\mu$CDM mock data.\label{fig:plots_mock_muCDM}}
\end{figure*}

\subsection{GA analysis of the mock data\label{sec:mock_results}}
We repeat the analysis using mock \Euclid and DESI-like data created in the context of two fiducial cosmologies as discussed in \Cref{sec:method}. In the first case we consider the standard \lcdm model as our background model, while for the second one, we used a simple MoG parametrisation given by Eq.~\eqref{eq:mu} as our fiducial cosmology.

\subsubsection{\lcdm fiducial cosmology}
Assuming the standard \lcdm cosmology we construct the two different sets of mock data in the manner that is described, in detail, in \Cref{sec:mock_method}. Using the pipeline that we have tested using the currently available data, we repeat the analysis described in \Cref{sec:real_results} obtaining the GA reconstructed best-fit functions of the three null tests in question.

These results along with those for the $f\sigma_8(z)$ and $E(z)$ reconstructed functions are shown in Fig.~\ref{fig:plots_mock_LCDM}, similar to Fig.~\ref{fig:plots_real}. We see that for all of the null tests the GA reconstructed functions are in very good agreement with the predicted \lcdm curve, to within the $68.3\%$ confidence level. We also observe that the uncertainty bands of the reconstructed functions that correspond to the null tests derived using the mock data are substantially narrower compared to those derived using the currently available ones. This is because the mock data were created using the \Euclid specifications and therefore are of higher quality than the currently available data, the same argument can be made about the DESI-like mock data as well.

Overall, we observe on average over the redshift range covered by the data, constraints on $\gamma_\mathrm{g}(z)$ at the level of $0.1\%$, on $\mathcal{O}(z)$ at the level of $6\%$ and $\mathrm{Om}_{f{\sigma_8}}(z,\Delta)$  also at the $1\%$ level. This implies an expected improvement brought forth by \Euclid of a factor of five for the growth index $\gamma_\mathrm{g}(z)$, but a factor of seven to eight for both the $\mathcal{O}(z)$ and $\mathrm{Om}_{f{\sigma_8}}(z,\Delta)$ tests, with respect to the currently available data. On average across all tests, this implies an improvement of around a factor of six with respect to the currently available data.
\begin{figure*}[!h]
\centering
\includegraphics[width = 0.495\textwidth]{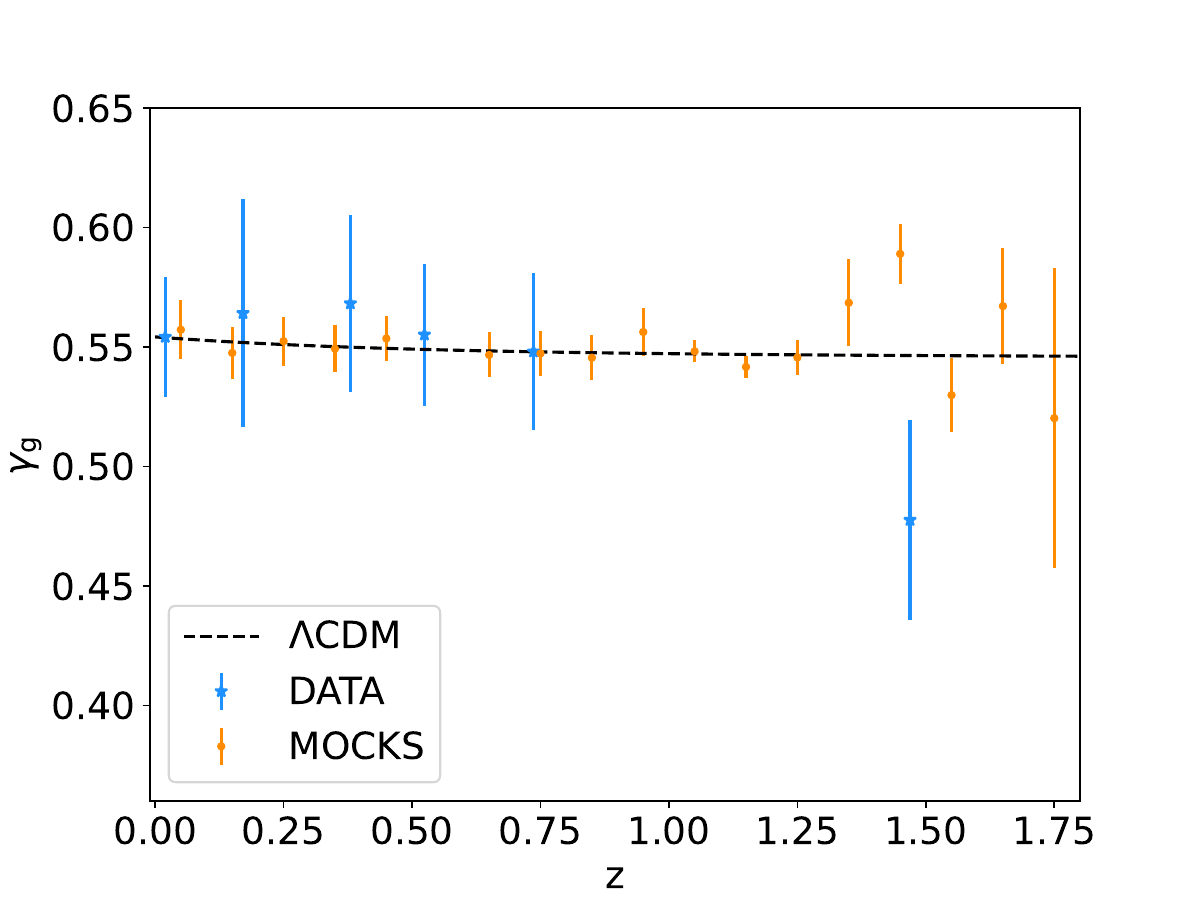}
\includegraphics[width = 0.495\textwidth]{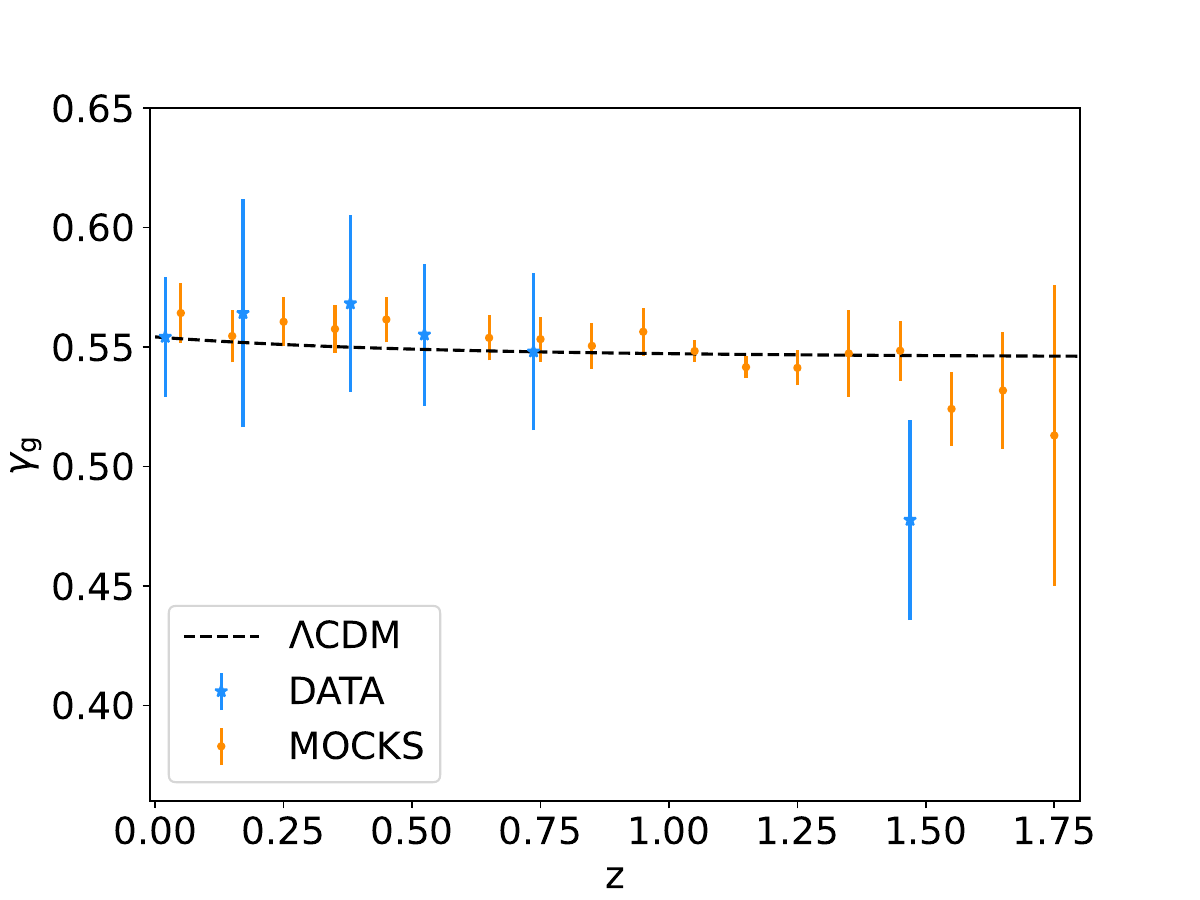}
\includegraphics[width = 0.495\textwidth]{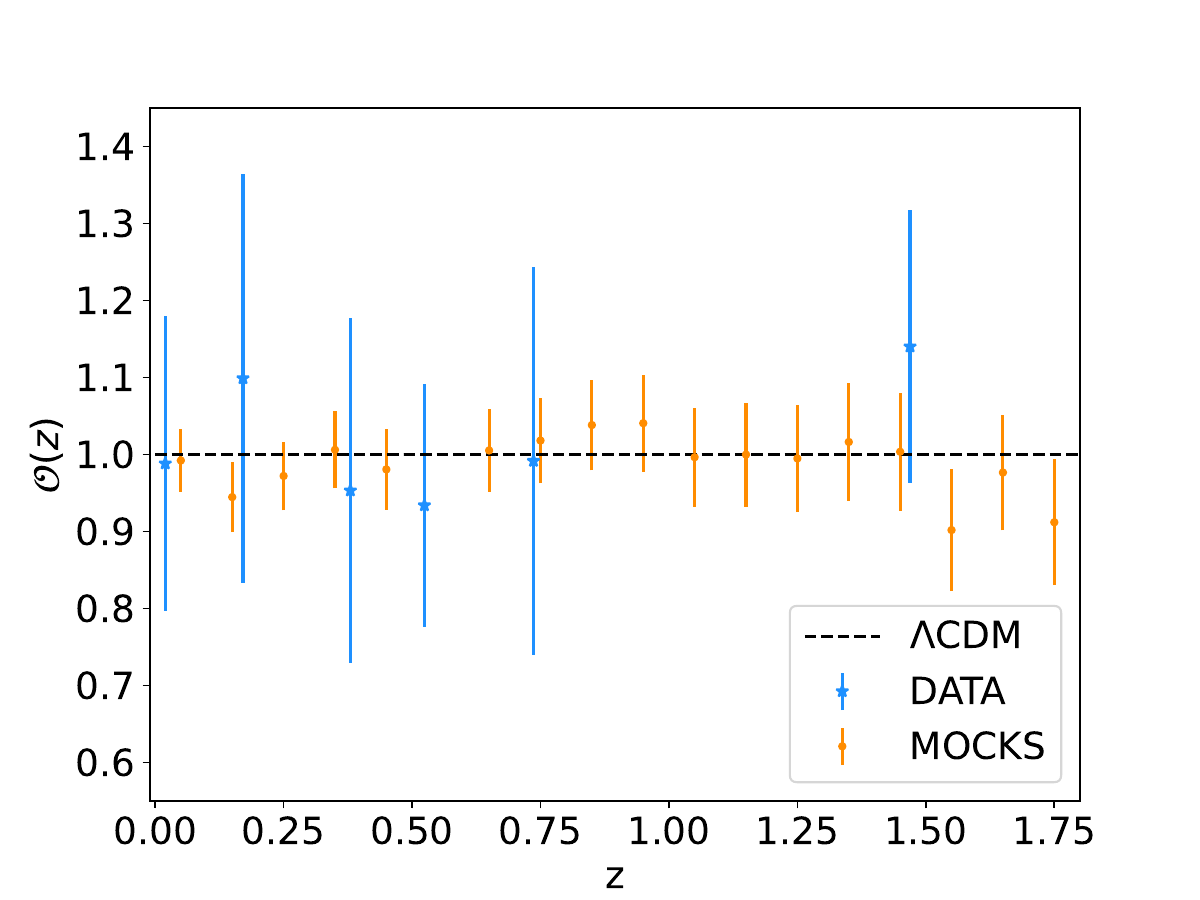}
\includegraphics[width = 0.495\textwidth]{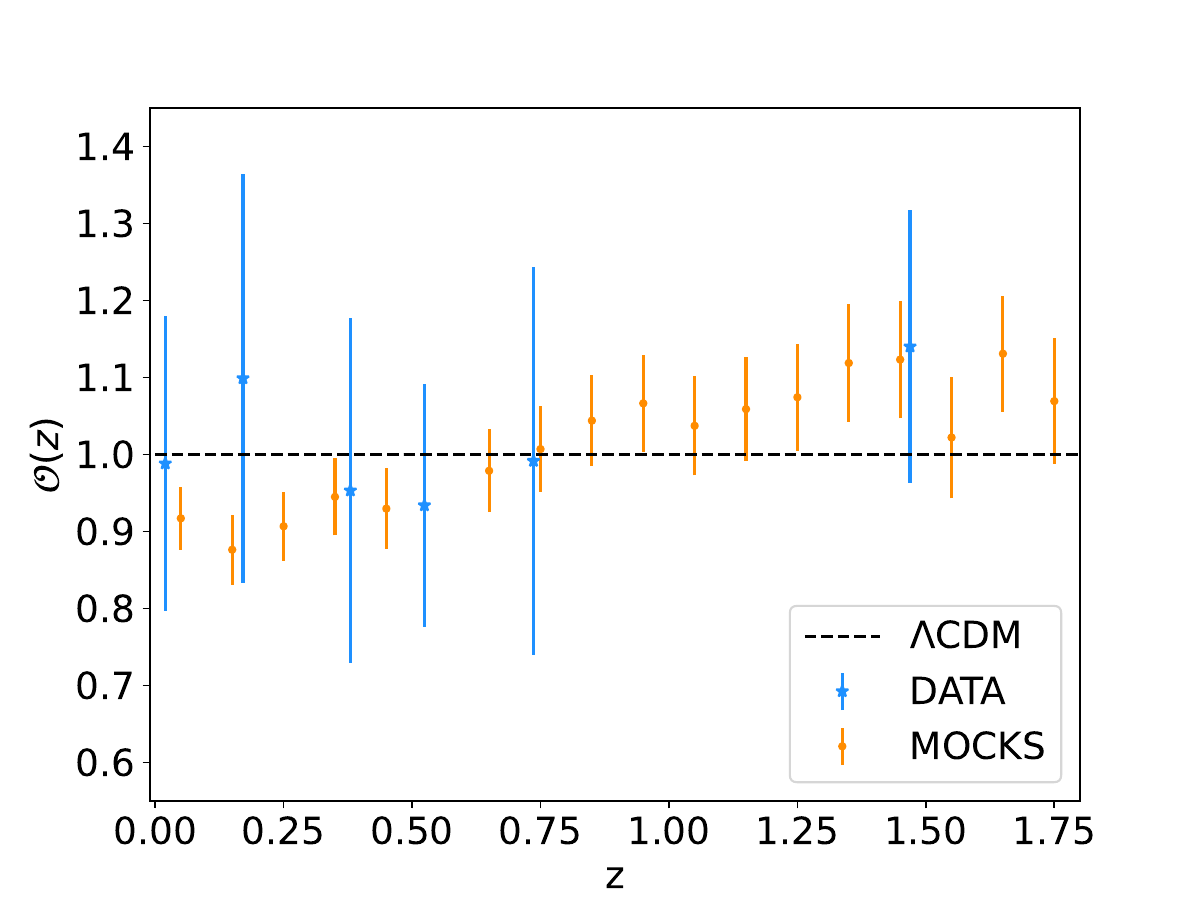}
\includegraphics[width = 0.495\textwidth]{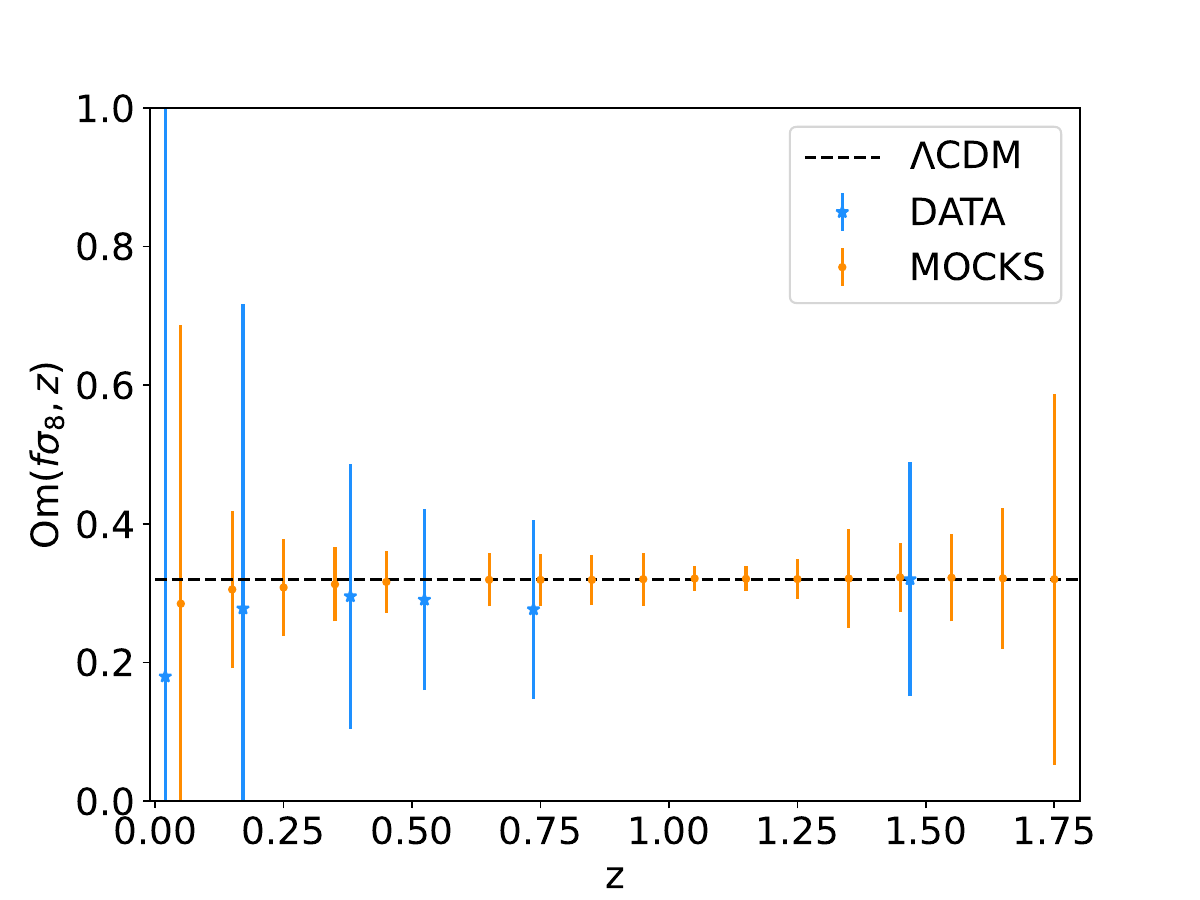}
\includegraphics[width = 0.495\textwidth]{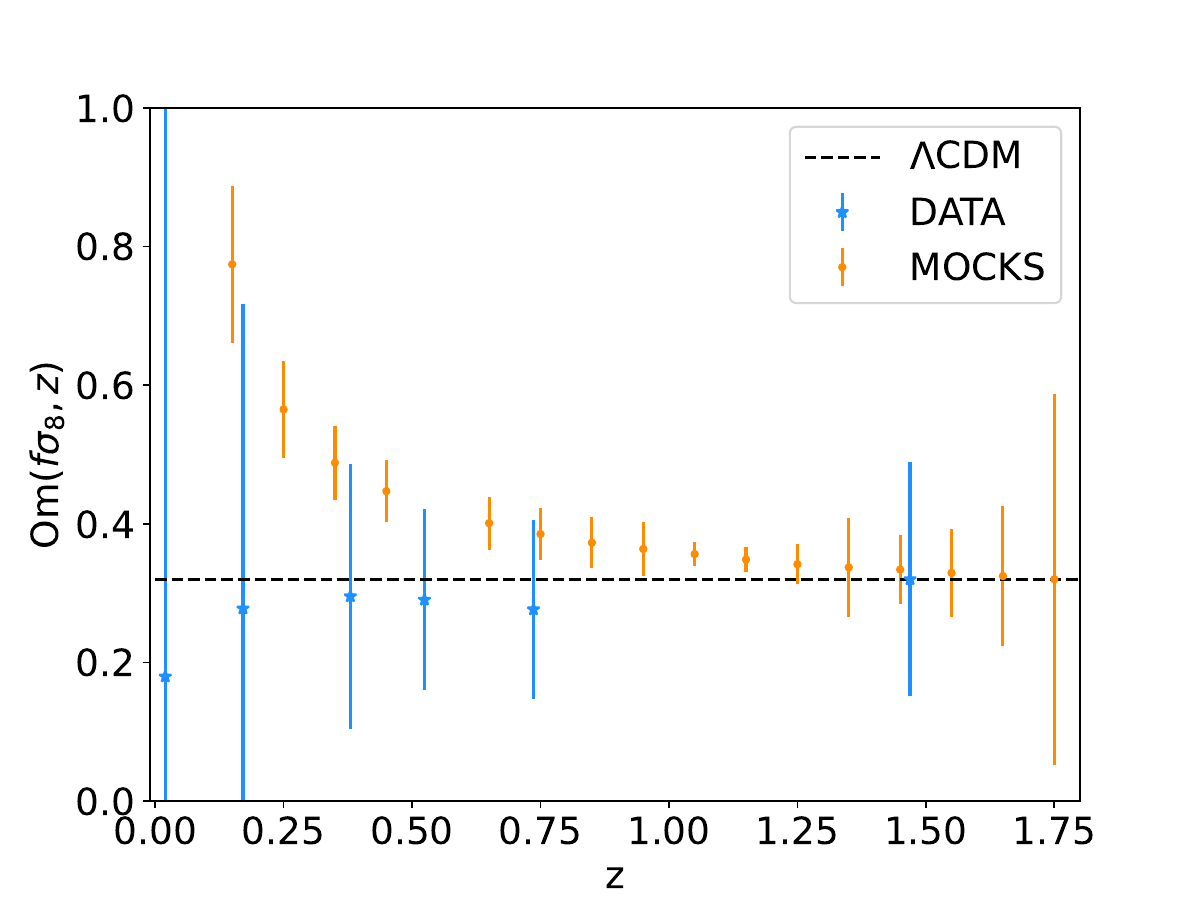}
\caption{Binned tests for the data: \lcdm model (left panels) and $\mu$CDM model (right panels). The blue points correspond to the currently available data, the orange points to the mock data, and the dashed line to the \lcdm fiducial. Specifically, we performed the growth index reconstruction using Eq.~\eqref{eq:gamma2}, the $\mathcal{O}(z)$ test by Eq.~\eqref{eq:nullf1}, and the $\mathrm{Om}_{f{\sigma_8}}(z,\Delta)$ test by Eq.~\eqref{eq:nulltest}. \label{fig:tests_binned}}
\end{figure*}

\subsubsection{$\mu$CDM fiducial cosmology}
Using our pipeline we construct the GA best-fits of the same null tests, this time in the context of $f\sigma_{8}(z)$ and $H(z)$ mock data created using the $\mu$CDM model described via Eq.~\eqref{eq:mu}. As we observe in Fig.~\ref{fig:plots_mock_muCDM}, there is a substantial deviation between the values of the GA reconstructed functions of the null tests and the one corresponding to \lcdm. This deviation is especially pronounced in the case of the $\mathrm{Om}_{f{\sigma_8}}(z,\Delta)$ and $\gamma_\mathrm{g}(z)$ null tests reaching more than the $95.5\%$ confidence level. However, this difference is expected since the reconstructed functions were created using the MoG fiducial cosmology and not \lcdm, making these tests ideal for testing for deviations from GR. 

\subsection{Binning of the data\label{sec:binning}}
Finally, we show here the results of the reconstruction of the null tests using a binning approach. In the left panels of Fig.\ref{fig:tests_binned} we show the results of the reconstructions of the null tests, after binning the currently available data (blue points) and the \lcdm mock data (orange points), along with the \lcdm fiducial. 

We found that the optimal number of bins is six, centred in $z =  \{0.020,\,0.171,\,0.380,\,0.524,\, 0.736,\, 1.469\}$. 
For the $H(z)$ data, we adopted the following methodology initially, we created the redshift bins as defined by the $f\sigma_8(z)$ binned data. We then filtered the $H(z)$ data to include only the data points within that bin's range and evaluated the weighted mean and uncertainty for each bin. The only exception is for the first bin where we anchored the Hubble data to the lowest redshift, for which there is a mismatch in redshift of $5\%$ between the two data sets. This procedure ensures that all binned data have the same redshifts. 

As can be seen, \Euclid with other contemporary surveys, will be able to improve upon current constraints by a factor of three on average for the $\gamma_\mathrm{g}(z)$ and $\mathcal{O}(z)$ tests, and by a factor of $10$ for the $\mathrm{Om}_{f{\sigma_8}}(z,\Delta)$ test. Furthermore, it will also provide more consistent constraints and better sampling across the whole redshift range of approximately $z\in [0,2]$. 

Performing the analysis as before, we find that the constraining power of the Euclid survey is, on average,  twice as constraining than the DESI survey for the tests considered in this work, as for the $\gamma_\mathrm{g}(z)$ test \Euclid provides a factor of three to four improvement, while DESI only a factor of two. Similarly for the other two tests the improvements were a factor of four and 15 for \Euclid, while a factor of three and 13 for DESI for the $\mathcal{O}(z)$ test, and a factor of seven and 15 for \Euclid, while a factor of two and 11 for DESI for the $\mathcal{O}(z)$ and $\mathrm{Om}_{f{\sigma_8}}(z,\Delta)$ tests respectively. On average and taking into account all the tests we observe an overall improvement of a factor of six over the current data with the binning method.

Finally, in the right panels of Fig.~\ref{fig:tests_binned} we show the plots for the $\mu$CDM model mocks. As can be seen, in this case, the yellow points show a clear deviation from the \lcdm expectation for the null tests, which is more pronounced for the $\mathrm{Om}_{f{\sigma_8}}(z,\Delta)$ test. 

\section{Conclusions \label{sec:conclusions}}
In this work we constrained deviations from the \lcdm model using three complementary consistency tests of the LSS, expressed in terms of the growth rate $f\sigma_8(z)$. We used both currently available and forecast \Euclid growth rate data, while for the latter we also explored synergies with other contemporary surveys.

In \Cref{sec:theory} we presented the theoretical background for the treatment of matter density perturbations, but also the two theoretical models we will consider in our analysis, namely the \lcdm and $\mu$CDM models. The latter is a designer model that exhibits an evolving Newton's constant at late times and can imitate several MoG models. Thus, using the $\mu$CDM model we can test a wide variety of beyond-GR models without having to specify a specific theory. 

Then, in \Cref{sec:tests}, we provided the explicit expressions for the consistency tests we considered in our analysis, namely the growth index $\gamma_\mathrm{g}(z)$ described by Eq.~\eqref{eq:gamma2}, the $\mathcal{O}(z)$ test given by Eq.~\eqref{eq:nullf1}, and the $\mathrm{Om}_{f{\sigma_8}}(z,\Delta)$ test given by Eq.~\eqref{eq:nulltest}. All of the former tests come directly via direct manipulations on the definition of $f\sigma_8(z)$ and its evolution equation in the \lcdm model, thus directly test the fundamental assumptions in the model.

In \Cref{sec:method} we presented the methodology we followed in our work, namely the compilation of the currently available growth rate and CC data in \Cref{sec:current_method}, the forecast Euclid and DESI data in \Cref{sec:mock_method}. We also presented our two reconstruction methods, based on a machine learning and binning approach in \Cref{sec:GA}. Both the GA and the binning approach are non-parametric and can capture features in the data, thus we found them to be ideal for our analysis, as we preferred to be as agnostic as possible.

Then, in \Cref{sec:results} we presented the findings of our analysis. Specifically, in \Cref{sec:real_results} we discussed the results of the GA analysis of the currently available data, which in \Cref{sec:mock_results} we used as a benchmark to quantify the improvement that \Euclid and other contemporary surveys will bring. By inspecting Figs.~\ref{fig:plots_mock_LCDM} and \ref{fig:plots_mock_muCDM} we observe an overall factor of improvement of approximately six, over the currently available data.

Finally, in \Cref{sec:binning} we performed a binning of both the currently available data and the mocks. As can be seen, there is an overall improvement upon current constraints by a factor of three on average for the $\gamma_\mathrm{g}(z)$ and $\mathcal{O}(z)$ tests, whereas by a factor of $14$ for the $\mathrm{Om}_{f{\sigma_8}}(z,\Delta)$ test. We also find more consistent constraints and better sampling across the whole redshift range of approximately $z\in [0,2]$ and an overall improvement on all of the tests by approximately a factor of eight.

In summary, our work highlights the advantages of using standard and machine learning analyses, exploiting synergies between \Euclid and other contemporary LSS surveys for the growth rate data, in order to constrain deviations from \lcdm and GR. However, with precision cosmology, coming soon with the forthcoming exquisite data from \Euclid, it will be possible to probe for very subtle deviations from \lcdm and GR. While this might prove to be challenging for standard analyses,  machine learning may offer innovative methods to distinguish these deviations from \lcdm and GR, especially when effectively interpreting outcomes \citep{Ocampo:2024fvx}.

\begin{acknowledgements}
The authors would like to thank M.~de los Rios, J.~L.~Tastet and E.~Donini for useful discussions. IO, SN, GA, and JGB acknowledge support from the research project PID2021-123012NB-C43 and the Spanish Research Agency (Agencia Estatal de Investigaci\'on) through the Grant IFT Centro de Excelencia Severo Ochoa No CEX2020-001007-S, funded by MCIN/AEI/10.13039/501100011033. IO is supported by the fellowship LCF/BQ/DI22/11940033 from ``la Caixa” Foundation (ID 100010434) and by a Graduate Fellowship at Residencia de Estudiantes supported by Madrid City Council (Spain), 2022-2023. DS acknowledges financial support from the Fondecyt Regular project N.~1251339. ZS acknowledges funding from DFG project 456622116 and support from the IRAP and IN2P3 Lyon computing centres. This work was financed by Portuguese funds through FCT - Funda\c c\~ao para a Ci\^encia e a Tecnologia (FCT) in the framework of the projects 2022.04048.PTDC (Phi in the Sky, DOI 10.54499/2022.04048.PTDC) UIDB/04434/2020, UIDP/04434/2020, EXPL/FIS-AST/1368/2021 (ML\_CLUSTER, DOI 10.54499/EXPL/FIS-AST/1368/2021) and PTDC/FIS-AST/0054/2021. CJAPM also acknowledges FCT and POCH/FSE (EC) support through Investigador FCT Contract 2021.01214.CEECIND/CP1658/CT0001. AC acknowledges support form the FCT research grant 2020.06644.BD.\\

\AckEC
\end{acknowledgements}

\bibliographystyle{aa}
\bibliography{references} 

\providecommand*\hyphen{-}
\begin{thebibliography}{112}
\expandafter\ifx\csname natexlab\endcsname\relax\def\natexlab#1{#1}\fi

\bibitem[{Abramowitz(1974)}]{Abramowitz:1974}
Abramowitz, M. 1974, Handbook of Mathematical Functions, With Formulas, Graphs, and Mathematical Tables, (USA: Dover Publications, Inc.)

\bibitem[{Achitouv {et~al.}(2017)Achitouv, Blake, Carter, Koda, \& Beutler}]{Achitouv:2016mbn}
Achitouv, I., Blake, C., Carter, P., Koda, J., \& Beutler, F. 2017, Phys. Rev. D, 95, 083502

\bibitem[{{Akrami} {et~al.}(2010){Akrami}, {Scott}, {Edsj{\"o}}, {Conrad}, \& {Bergstr{\"o}m}}]{Akrami:2009hp}
{Akrami}, Y., {Scott}, P., {Edsj{\"o}}, J., {Conrad}, J., \& {Bergstr{\"o}m}, L. 2010, JHEP, 04, 57

\bibitem[{{Alam} {et~al.}(2021){Alam}, {Aubert}, {Avila}, {Balland}, {Bautista}, {Bershady}, {Bizyaev}, {Blanton}, {Bolton}, {Bovy}, {Brinkmann}, {Brownstein}, {Burtin}, {Chabanier}, {Chapman}, {Choi}, {Chuang}, {Comparat}, {Cousinou}, {Cuceu}, {Dawson}, {de la Torre}, {de Mattia}, {Agathe}, {des Bourboux}, {Escoffier}, {Etourneau}, {Farr}, {Font-Ribera}, {Frinchaboy}, {Fromenteau}, {Gil-Mar{\'\i}n}, {Le Goff}, {Gonzalez-Morales}, {Gonzalez-Perez}, {Grabowski}, {Guy}, {Hawken}, {Hou}, {Kong}, {Parker}, {Klaene}, {Kneib}, {Lin}, {Long}, {Lyke}, {de la Macorra}, {Martini}, {Masters}, {Mohammad}, {Moon}, {Mueller}, {Mu{\~n}oz-Guti{\'e}rrez}, {Myers}, {Nadathur}, {Neveux}, {Newman}, {Noterdaeme}, {Oravetz}, {Oravetz}, {Palanque-Delabrouille}, {Pan}, {Paviot}, {Percival}, {P{\'e}rez-R{\`a}fols}, {Petitjean}, {Pieri}, {Prakash}, {Raichoor}, {Ravoux}, {Rezaie}, {Rich}, {Ross}, {Rossi}, {Ruggeri}, {Ruhlmann-Kleider}, {S{\'a}nchez}, {S{\'a}nchez}, {S{\'a}nchez-Gallego}, {Sayres}, {Schneider}, {Seo}, {Shafieloo},
  {Slosar}, {Smith}, {Stermer}, {Tamone}, {Tinker}, {Tojeiro}, {Vargas-Maga{\~n}a}, {Variu}, {Wang}, {Weaver}, {Weijmans}, {Y{\`e}che}, {Zarrouk}, {Zhao}, {Zhao}, \& {Zheng}}]{eBOSS:2020yzd}
{Alam}, S., {Aubert}, M., {Avila}, S., {et~al.} 2021, \prd, 103, 083533

\bibitem[{Alestas {et~al.}(2022)Alestas, Kazantzidis, \& Nesseris}]{Alestas:2022gcg}
Alestas, G., Kazantzidis, L., \& Nesseris, S. 2022, Phys. Rev. D, 106, 103519

\bibitem[{Alonso {et~al.}(2015)Alonso, Salvador, S\'anchez, Bilicki, Garc\'\i{}a-Bellido, \& S\'anchez}]{Alonso:2014xca}
Alonso, D., Salvador, A.~I., S\'anchez, F.~J., {et~al.} 2015, MNRAS, 449, 670

\bibitem[{{Amendola} {et~al.}(2018){Amendola}, {Appleby}, {Avgoustidis}, {Bacon}, {Baker}, {Baldi}, {Bartolo}, {Blanchard}, {Bonvin}, {Borgani}, {Branchini}, {Burrage}, {Camera}, {Carbone}, {Casarini}, {Cropper}, {de Rham}, {Dietrich}, {Di Porto}, {Durrer}, {Ealet}, {Ferreira}, {Finelli}, {Garc{\'\i}a-Bellido}, {Giannantonio}, {Guzzo}, {Heavens}, {Heisenberg}, {Heymans}, {Hoekstra}, {Hollenstein}, {Holmes}, {Hwang}, {Jahnke}, {Kitching}, {Koivisto}, {Kunz}, {La Vacca}, {Linder}, {March}, {Marra}, {Martins}, {Majerotto}, {Markovic}, {Marsh}, {Marulli}, {Massey}, {Mellier}, {Montanari}, {Mota}, {Nunes}, {Percival}, {Pettorino}, {Porciani}, {Quercellini}, {Read}, {Rinaldi}, {Sapone}, {Sawicki}, {Scaramella}, {Skordis}, {Simpson}, {Taylor}, {Thomas}, {Trotta}, {Verde}, {Vernizzi}, {Vollmer}, {Wang}, {Weller}, \& {Zlosnik}}]{ReviewDoc}
{Amendola}, L., {Appleby}, S., {Avgoustidis}, A., {et~al.} 2018, Living Rev. Rel., 21, 2

\bibitem[{Amendola {et~al.}(2008)Amendola, Kunz, \& Sapone}]{Amendola:2007rr}
Amendola, L., Kunz, M., \& Sapone, D. 2008, JCAP, 04, 013

\bibitem[{{Anderson} {et~al.}(2014){Anderson}, {Aubourg}, {Bailey}, {Beutler}, {Bhardwaj}, {Blanton}, {Bolton}, {Brinkmann}, {Brownstein}, {Burden}, {Chuang}, {Cuesta}, {Dawson}, {Eisenstein}, {Escoffier}, {Gunn}, {Guo}, {Ho}, {Honscheid}, {Howlett}, {Kirkby}, {Lupton}, {Manera}, {Maraston}, {McBride}, {Mena}, {Montesano}, {Nichol}, {Nuza}, {Olmstead}, {Padmanabhan}, {Palanque-Delabrouille}, {Parejko}, {Percival}, {Petitjean}, {Prada}, {Price-Whelan}, {Reid}, {Roe}, {Ross}, {Ross}, {Sabiu}, {Saito}, {Samushia}, {S{\'a}nchez}, {Schlegel}, {Schneider}, {Scoccola}, {Seo}, {Skibba}, {Strauss}, {Swanson}, {Thomas}, {Tinker}, {Tojeiro}, {Maga{\~n}a}, {Verde}, {Wake}, {Weaver}, {Weinberg}, {White}, {Xu}, {Y{\`e}che}, {Zehavi}, \& {Zhao}}]{BOSS:2013rlg}
{Anderson}, L., {Aubourg}, {\'E}., {Bailey}, S., {et~al.} 2014, \mnras, 441, 24

\bibitem[{Andrade {et~al.}(2021)Andrade, Anbajagane, von Marttens, Huterer, \& Alcaniz}]{Andrade:2021njl}
Andrade, U., Anbajagane, D., von Marttens, R., Huterer, D., \& Alcaniz, J. 2021, JCAP, 11, 014

\bibitem[{Arjona(2020)}]{Arjona:2020doi}
Arjona, R. 2020, JCAP, 08, 009

\bibitem[{Arjona {et~al.}(2022)Arjona, Melchiorri, \& Nesseris}]{Arjona:2021mzf}
Arjona, R., Melchiorri, A., \& Nesseris, S. 2022, JCAP, 05, 047

\bibitem[{Arjona \& Nesseris(2020{\natexlab{a}})}]{Arjona:2020kco}
Arjona, R. \& Nesseris, S. 2020{\natexlab{a}}, JCAP, 11, 042

\bibitem[{Arjona \& Nesseris(2020{\natexlab{b}})}]{Arjona:2019fwb}
Arjona, R. \& Nesseris, S. 2020{\natexlab{b}}, Phys. Rev. D, 101, 123525

\bibitem[{Avila {et~al.}(2021)Avila, Bernui, de~Carvalho, \& Novaes}]{Avila:2021dqv}
Avila, F., Bernui, A., de~Carvalho, E., \& Novaes, C.~P. 2021, MNRAS, 505, 3404

\bibitem[{Ballardini {et~al.}(2022)Ballardini, Finelli, \& Sapone}]{Ballardini:2021evv}
Ballardini, M., Finelli, F., \& Sapone, D. 2022, JCAP, 06, 004

\bibitem[{Ballardini {et~al.}(2019)Ballardini, Sapone, Umilt\`a, Finelli, \& Paoletti}]{Ballardini:2019tho}
Ballardini, M., Sapone, D., Umilt\`a, C., Finelli, F., \& Paoletti, D. 2019, JCAP, 05, 049

\bibitem[{Bartlett {et~al.}(2024{\natexlab{a}})Bartlett, Kammerer, Kronberger, Desmond, Ferreira, Wandelt, Burlacu, Alonso, \& Zennaro}]{Bartlett:2023cyr}
Bartlett, D.~J., Kammerer, L., Kronberger, G., {et~al.} 2024{\natexlab{a}}, A\&A, 686, A209

\bibitem[{Bartlett {et~al.}(2024{\natexlab{b}})Bartlett, Wandelt, Zennaro, Ferreira, \& Desmond}]{Bartlett:2024jes}
Bartlett, D.~J., Wandelt, B.~D., Zennaro, M., Ferreira, P.~G., \& Desmond, H. 2024{\natexlab{b}}, A\&A, 686, A150

\bibitem[{{Blake} {et~al.}(2013){Blake}, {Baldry}, {Bland-Hawthorn}, {Christodoulou}, {Colless}, {Conselice}, {Driver}, {Hopkins}, {Liske}, {Loveday}, {Norberg}, {Peacock}, {Poole}, \& {Robotham}}]{Blake:2013nif}
{Blake}, C., {Baldry}, I.~K., {Bland-Hawthorn}, J., {et~al.} 2013, MNRAS, 436, 3089

\bibitem[{{Blake} {et~al.}(2012){Blake}, {Brough}, {Colless}, {Contreras}, {Couch}, {Croom}, {Croton}, {Davis}, {Drinkwater}, {Forster}, {Gilbank}, {Gladders}, {Glazebrook}, {Jelliffe}, {Jurek}, {Li}, {Madore}, {Martin}, {Pimbblet}, {Poole}, {Pracy}, {Sharp}, {Wisnioski}, {Woods}, {Wyder}, \& {Yee}}]{Blake:2012pj}
{Blake}, C., {Brough}, S., {Colless}, M., {et~al.} 2012, \mnras, 425, 405

\bibitem[{Bogdanos \& Nesseris(2009)}]{Bogdanos:2009ib}
Bogdanos, C. \& Nesseris, S. 2009, JCAP, 05, 006

\bibitem[{Borghi {et~al.}(2022)Borghi, Moresco, \& Cimatti}]{Borghi:2021rft}
Borghi, N., Moresco, M., \& Cimatti, A. 2022, ApJ, 928, L4

\bibitem[{Capozziello {et~al.}(2004)Capozziello, Cardone, Funaro, \& Andreon}]{Capozziello:2004jy}
Capozziello, S., Cardone, V.~F., Funaro, M., \& Andreon, S. 2004, Phys. Rev. D, 70, 123501

\bibitem[{{Casas} {et~al.}(2023){Casas}, {Cardone}, {Sapone}, {Frusciante}, {Pace}, {Parimbelli}, {Archidiacono}, {Koyama}, {Tutusaus}, {Camera}, {Martinelli}, {Pettorino}, {Sakr}, {Lombriser}, {Silvestri}, {Pietroni}, {Vernizzi}, {Kunz}, {Kitching}, {Pourtsidou}, {Lacasa}, {Carbone}, {Garcia-Bellido}, {Aghanim}, {Altieri}, {Amara}, {Auricchio}, {Baldi}, {Bodendorf}, {Branchini}, {Brescia}, {Brinchmann}, {Capobianco}, {Carretero}, {Castellano}, {Cavuoti}, {Cimatti}, {Cledassou}, {Congedo}, {Conselice}, {Conversi}, {Copin}, {Corcione}, {Courbin}, {Courtois}, {DaSilva}, {Degaudenzi}, {Dubath}, {Duncan}, {Dupac}, {Dusini}, {Farrens}, {Ferriol}, {Fosalba}, {Frailis}, {Franceschi}, {Fumana}, {Galeotta}, {Garilli}, {Gillard}, {Gillis}, {Giocoli}, {Grazian}, {Grupp}, {Guzzo}, {Haugan}, {Hormuth}, {Hornstrup}, {Hudelot}, {Jahnke}, {Kermiche}, {Kiessling}, {Kilbinger}, {Kurki-Suonio}, {Ligori}, {Lilje}, {Lloro}, {Maiorano}, {Mansutti}, {Marggraf}, {Marulli}, {Massey}, {Medinaceli}, {Mellier}, {Meneghetti}, {Merlin},
  {Meylan}, {Moresco}, {Moscardini}, {Munari}, {Niemi}, {Padilla}, {Paltani}, {Pasian}, {Pedersen}, {Percival}, {Pires}, {Polenta}, {Poncet}, {Popa}, {Raison}, {Renzi}, {Rhodes}, {Riccio}, {Romelli}, {Roncarelli}, {Rossetti}, {Saglia}, {Sartoris}, {Scottez}, {Secroun}, {Seidel}, {Serrano}, {Sirignano}, {Sirri}, {Stanco}, {Starck}, {Surace}, {Tallada-Cresp{\'\i}}, {Taylor}, {Tereno}, {Toledo-Moreo}, {Torradeflot}, {Valentijn}, {Valenziano}, {Vassallo}, {Wang}, {Weller}, \& {Zoubian}}]{Euclid:2023tqw}
{Casas}, S., {Cardone}, V.~F., {Sapone}, D., {et~al.} 2023, arXiv:2306.11053

\bibitem[{Castro {et~al.}(2023)}]{Euclid:2022dbc}
Castro, T. {et~al.} 2023, A\&A, 671, A100

\bibitem[{Chiang {et~al.}(2019)Chiang, Romano, Nugier, \& Chen}]{Chiang:2017yrq}
Chiang, H.~W., Romano, A.~E., Nugier, F., \& Chen, P. 2019, JCAP, 11, 016

\bibitem[{Chuang \& Wang(2013)}]{Chuang:2012qt}
Chuang, C.-H. \& Wang, Y. 2013, MNRAS, 435, 255

\bibitem[{Clarkson(2012)}]{Clarkson:2012bg}
Clarkson, C. 2012, Comptes Rendus Physique, 13, 682

\bibitem[{Clarkson {et~al.}(2008)Clarkson, Bassett, \& Lu}]{Clarkson:2007pz}
Clarkson, C., Bassett, B., \& Lu, T. H.-C. 2008, Phys. Rev. Lett., 101, 011301

\bibitem[{Cropper {et~al.}(2018)Cropper, Pottinger, Azzollini, Szafraniec, Awan, Mellier, Berthé, Martignac, Cara, Giorgio, Sciortino, Bozzo, Genolet, Philippon, Hailey, Hunt, Swindells, Holland, Gow, Murray, Hall, Skottfelt, Amiaux, Laureijs, Racca, Salvignol, Short, Alvarez, Kitching, Hoekstra, Galli, Willis, Hu, Candini, Boucher, Bahlawan, Chaudery, de~Lacy, Pendem, Smit, Dubois, Horeau, Carty, Fontignie, Doumayrou, Larcheveque, Castelli, Cole, Niemi, Denniston, Massey, Kohley, Ferrando, \& Conversi}]{VIS_paper}
Cropper, M., Pottinger, S., Azzollini, R., {et~al.} 2018, in Space Telescopes and Instrumentation 2018: Optical, Infrared, and Millimeter Wave, ed. M.~Lystrup, H.~A. MacEwen, G.~G. Fazio, N.~Batalha, N.~Siegler, \& E.~C. Tong, Vol. 10698, International Society for Optics and Photonics (SPIE), 709 -- 729

\bibitem[{Davis {et~al.}(2011)Davis, Nusser, Masters, Springob, Huchra, \& Lemson}]{Davis:2010sw}
Davis, M., Nusser, A., Masters, K., {et~al.} 2011, MNRAS, 413, 2906

\bibitem[{de~la Torre \& Guzzo(2012)}]{delaTorre:2012dg}
de~la Torre, S. \& Guzzo, L. 2012, MNRAS, 427, 327

\bibitem[{{Delubac} {et~al.}(2015){Delubac}, {Bautista}, {Busca}, {Rich}, {Kirkby}, {Bailey}, {Font-Ribera}, {Slosar}, {Lee}, {Pieri}, {Hamilton}, {Aubourg}, {Blomqvist}, {Bovy}, {Brinkmann}, {Carithers}, {Dawson}, {Eisenstein}, {Gontcho}, {Kneib}, {Le Goff}, {Margala}, {Miralda-Escud{\'e}}, {Myers}, {Nichol}, {Noterdaeme}, {O'Connell}, {Olmstead}, {Palanque-Delabrouille}, {P{\^a}ris}, {Petitjean}, {Ross}, {Rossi}, {Schlegel}, {Schneider}, {Weinberg}, {Y{\`e}che}, \& {York}}]{BOSS:2014hwf}
{Delubac}, T., {Bautista}, J.~E., {Busca}, N.~G., {et~al.} 2015, \aap, 574, A59

\bibitem[{{DESI Collaboration: Abdul-Karim} {et~al.}(2025){DESI Collaboration: Abdul-Karim}, {Aguilar}, {Ahlen}, {Alam}, {Allen}, {Allende Prieto}, {Alves}, {Anand}, {Andrade}, {Armengaud}, {Aviles}, {Bailey}, {Baltay}, {Bansal}, {Bault}, {Behera}, {BenZvi}, {Bianchi}, {Blake}, \& {Brieden}}]{DESI2025}
{DESI Collaboration: Abdul-Karim}, M., {Aguilar}, J., {Ahlen}, S., {et~al.} 2025, arXiv:2503.14738

\bibitem[{{DESI Collaboration: Aghamousa} {et~al.}(2016){DESI Collaboration: Aghamousa}, {Aguilar}, {Ahlen}, {Alam}, {Allen}, {Allende Prieto}, {Annis}, {Bailey}, {Balland}, {Ballester}, {Baltay}, {Beaufore}, {Bebek}, {Beers}, {Bell}, {Bernal}, {Besuner}, {Beutler}, {Blake}, {Bleuler}, {Blomqvist}, {Blum}, {Bolton}, {Briceno}, {Brooks}, {Brownstein}, {Buckley-Geer}, {Burden}, {Burtin}, {Busca}, {Cahn}, {Cai}, {Cardiel-Sas}, {Carlberg}, {Carton}, {Casas}, {Castand er}, {Cervantes-Cota}, {Claybaugh}, {Close}, {Coker}, {Cole}, {Comparat}, {Cooper}, {Cousinou}, {Crocce}, {Cuby}, {Cunningham}, {Davis}, {Dawson}, {de la Macorra}, {De Vicente}, {Delubac}, {Derwent}, {Dey}, {Dhungana}, {Ding}, {Doel}, {Duan}, {Ealet}, {Edelstein}, {Eftekharzadeh}, {Eisenstein}, {Elliott}, {Escoffier}, {Evatt}, {Fagrelius}, {Fan}, {Fanning}, {Farahi}, {Farihi}, {Favole}, {Feng}, {Fernandez}, {Findlay}, {Finkbeiner}, {Fitzpatrick}, {Flaugher}, {Flender}, {Font-Ribera}, {Forero-Romero}, {Fosalba}, {Frenk}, {Fumagalli}, {Gaensicke},
  {Gallo}, {Garc\'ia-Bellido}, {Gaztanaga}, {Pietro Gentile Fusillo}, {Gerard}, {Gershkovich}, {Giannantonio}, {Gillet}, {Gonzalez-de-Rivera}, {Gonzalez-Perez}, {Gott}, {Graur}, {Gutierrez}, {Guy}, {Habib}, {Heetderks}, {Heetderks}, {Heitmann}, {Hellwing}, {Herrera}, {Ho}, {Holland}, {Honscheid}, {Huff}, {Hutchinson}, {Huterer}, {Hwang}, {Illa Laguna}, {Ishikawa}, {Jacobs}, {Jeffrey}, {Jelinsky}, {Jennings}, {Jiang}, {Jimenez}, {Johnson}, {Joyce}, {Jullo}, {Juneau}, {Kama}, {Karcher}, {Karkar}, {Kehoe}, {Kennamer}, {Kent}, {Kilbinger}, {Kim}, {Kirkby}, {Kisner}, {Kitanidis}, {Kneib}, {Koposov}, {Kovacs}, {Koyama}, {Kremin}, {Kron}, {Kronig}, {Kueter-Young}, {Lacey}, {Lafever}, {Lahav}, {Lambert}, {Lampton}, {Land riau}, {Lang}, {Lauer}, {Le Goff}, {Le Guillou}, {Le Van Suu}, {Lee}, {Lee}, {Leitner}, {Lesser}, {Levi}, {L'Huillier}, {Li}, {Liang}, {Lin}, {Linder}, {Loebman}, {Luki{\'c}}, {Ma}, {MacCrann}, {Magneville}, {Makarem}, {Manera}, {Manser}, {Marshall}, {Martini}, {Massey}, {Matheson}, {McCauley},
  {McDonald}, {McGreer}, {Meisner}, {Metcalfe}, {Miller}, {Miquel}, {Moustakas}, {Myers}, {Naik}, {Newman}, {Nichol}, {Nicola}, {Nicolati da Costa}, {Nie}, {Niz}, {Norberg}, {Nord}, {Norman}, {Nugent}, {O'Brien}, {Oh}, {Olsen}, {Padilla}, {Padmanabhan}, {Padmanabhan}, {Palanque-Delabrouille}, {Palmese}, {Pappalardo}, {P{\^a}ris}, {Park}, {Patej}, {Peacock}, {Peiris}, {Peng}, {Percival}, {Perruchot}, {Pieri}, {Pogge}, {Pollack}, {Poppett}, {Prada}, {Prakash}, {Probst}, {Rabinowitz}, {Raichoor}, {Ree}, {Refregier}, {Regal}, {Reid}, {Reil}, {Rezaie}, {Rockosi}, {Roe}, {Ronayette}, {Roodman}, {Ross}, {Ross}, {Rossi}, {Rozo}, {Ruhlmann-Kleider}, {Rykoff}, {Sabiu}, {Samushia}, {Sanchez}, {Sanchez}, {Schlegel}, {Schneider}, {Schubnell}, {Secroun}, {Seljak}, {Seo}, {Serrano}, {Shafieloo}, {Shan}, {Sharples}, {Sholl}, {Shourt}, {Silber}, {Silva}, {Sirk}, {Slosar}, {Smith}, {Smoot}, {Som}, {Song}, {Sprayberry}, {Staten}, {Stefanik}, {Tarle}, {Sien Tie}, {Tinker}, {Tojeiro}, {Valdes}, {Valenzuela}, {Valluri},
  {Vargas-Magana}, {Verde}, {Walker}, {Wang}, {Wang}, {Weaver}, {Weaverdyck}, {Wechsler}, {Weinberg}, {White}, {Yang}, {Yeche}, {Zhang}, {Zhao}, {Zheng}, {Zhou}, {Zhou}, {Zhu}, {Zou}, \& {Zu}}]{DESI2016}
{DESI Collaboration: Aghamousa}, A., {Aguilar}, J., {Ahlen}, S., {et~al.} 2016, arXiv:1611.00036

\bibitem[{Di~Valentino {et~al.}(2021)Di~Valentino, Mena, Pan, Visinelli, Yang, Melchiorri, Mota, Riess, \& Silk}]{DiValentino:2021izs}
Di~Valentino, E., Mena, O., Pan, S., {et~al.} 2021, Class. Quant. Grav., 38, 153001

\bibitem[{{Euclid Collaboration: Blanchard} {et~al.}(2020){Euclid Collaboration: Blanchard}, {Camera}, {Carbone}, {Cardone}, {Casas}, {Clesse}, {Ili{\'c}}, {Kilbinger}, {Kitching}, {Kunz}, {Lacasa}, {Linder}, {Majerotto}, {Markovi{\v{c}}}, {Martinelli}, {Pettorino}, {Pourtsidou}, {Sakr}, {S{\'a}nchez}, {Sapone}, {Tutusaus}, {Yahia-Cherif}, {Yankelevich}, {Andreon}, {Aussel}, {Balaguera-Antol{\'\i}nez}, {Baldi}, {Bardelli}, {Bender}, {Biviano}, {Bonino}, {Boucaud}, {Bozzo}, {Branchini}, {Brau-Nogue}, {Brescia}, {Brinchmann}, {Burigana}, {Cabanac}, {Capobianco}, {Cappi}, {Carretero}, {Carvalho}, {Casas}, {Castander}, {Castellano}, {Cavuoti}, {Cimatti}, {Cledassou}, {Colodro-Conde}, {Congedo}, {Conselice}, {Conversi}, {Copin}, {Corcione}, {Coupon}, {Courtois}, {Cropper}, {Da Silva}, {de la Torre}, {Di Ferdinando}, {Dubath}, {Ducret}, {Duncan}, {Dupac}, {Dusini}, {Fabbian}, {Fabricius}, {Farrens}, {Fosalba}, {Fotopoulou}, {Fourmanoit}, {Frailis}, {Franceschi}, {Franzetti}, {Fumana}, {Galeotta}, {Gillard},
  {Gillis}, {Giocoli}, {G{\'o}mez-Alvarez}, {Graci{\'a}-Carpio}, {Grupp}, {Guzzo}, {Hoekstra}, {Hormuth}, {Israel}, {Jahnke}, {Keihanen}, {Kermiche}, {Kirkpatrick}, {Kohley}, {Kubik}, {Kurki-Suonio}, {Ligori}, {Lilje}, {Lloro}, {Maino}, {Maiorano}, {Marggraf}, {Martinet}, {Marulli}, {Massey}, {Medinaceli}, {Mei}, {Mellier}, {Metcalf}, {Metge}, {Meylan}, {Moresco}, {Moscardini}, {Munari}, {Nichol}, {Niemi}, {Nucita}, {Padilla}, {Paltani}, {Pasian}, {Percival}, {Pires}, {Polenta}, {Poncet}, {Pozzetti}, {Racca}, {Raison}, {Renzi}, {Rhodes}, {Romelli}, {Roncarelli}, {Rossetti}, {Saglia}, {Schneider}, {Scottez}, {Secroun}, {Sirri}, {Stanco}, {Starck}, {Sureau}, {Tallada-Cresp{\'\i}}, {Tavagnacco}, {Taylor}, {Tenti}, {Tereno}, {Toledo-Moreo}, {Torradeflot}, {Valenziano}, {Vassallo}, {Verdoes Kleijn}, {Viel}, {Wang}, {Zacchei}, {Zoubian}, \& {Zucca}}]{IST:paper1}
{Euclid Collaboration: Blanchard}, A., {Camera}, S., {Carbone}, C., {et~al.} 2020, \aap, 642, A191

\bibitem[{{Euclid Collaboration: Mellier} {et~al.}(2025){Euclid Collaboration: Mellier}, {Abdurro\'uf}, {Acevedo Barroso}, {Ach{\'u}carro}, {Adamek}, {Adam}, {Addison}, {Aghanim}, {Aguena}, {Ajani}, {Akrami}, {Al-Bahlawan}, {Alavi}, {Albuquerque}, {Alestas}, {Alguero}, {Allaoui}, {Allen}, {Allevato}, {Alonso-Tetilla}, {Altieri}, {Alvarez-Candal}, {Alvi}, {Amara}, {Amendola}, {Amiaux}, {Andika}, {Andreon}, {Andrews}, {Angora}, {Angulo}, {Annibali}, {Anselmi}, {Anselmi}, {Arcari}, {Archidiacono}, {Aric{\`o}}, {Arnaud}, {Arnouts}, {Asgari}, {Asorey}, {Atayde}, {Atek}, {Atrio-Barandela}, {Aubert}, {Aubourg}, {Auphan}, {Auricchio}, {Aussel}, {Aussel}, {Avelino}, {Avgoustidis}, {Avila}, {Awan}, {Azzollini}, {Baccigalupi}, {Bachelet}, {Bacon}, {Baes}, {Bagley}, {Bahr-Kalus}, {Balaguera-Antolinez}, {Balbinot}, {Balcells}, {Baldi}, {Baldry}, {Balestra}, {Ballardini}, {Ballester}, {Balogh}, {Ba{\~n}ados}, {Barbier}, {Bardelli}, {Baron}, {Barreiro}, {Barrena}, {Barriere}, {Barros}, {Barthelemy}, {Bartolo}, {Basset},
  {Battaglia}, {Battisti}, {Baugh}, {Baumont}, {Bazzanini}, {Beaulieu}, {Beckmann}, {Belikov}, {Bel}, {Bellagamba}, {Bella}, {Bellini}, {Benabed}, {Bender}, {Benevento}, {Bennett}, {Benson}, {Bergamini}, {Bermejo-Climent}, {Bernardeau}, {Bertacca}, {Berthe}, {Berthier}, {Bethermin}, {Beutler}, {Bevillon}, {Bhargava}, {Bhatawdekar}, {Bianchi}, {Bisigello}, {Biviano}, {Blake}, {Blanchard}, {Blazek}, {Blot}, {Bosco}, {Bodendorf}, {Boenke}, {B{\"o}hringer}, {Boldrini}, {Bolzonella}, {Bonchi}, {Bonici}, {Bonino}, {Bonino}, {Bonvin}, {Bon}, {Booth}, {Borgani}, {Borlaff}, {Borsato}, {Bosco}, {Bose}, {Botticella}, {Boucaud}, {Bouche}, {Boucher}, {Boutigny}, {Bouvard}, {Bouwens}, {Bouy}, {Bowler}, {Bozza}, {Bozzo}, {Branchini}, {Brando}, {Brau-Nogue}, {Brekke}, {Bremer}, {Brescia}, {Breton}, {Brinchmann}, {Brinckmann}, {Brockley-Blatt}, {Brodwin}, {Brouard}, {Brown}, {Bruton}, {Bucko}, {Buddelmeijer}, {Buenadicha}, {Buitrago}, {Burger}, {Burigana}, {Busillo}, {Busonero}, {Cabanac}, {Cabayol-Garcia}, {Cagliari},
  {Caillat}, {Caillat}, {Calabrese}, {Calabro}, {Calderone}, {Calura}, {Camacho Quevedo}, {Camera}, {Campos}, {Canas-Herrera}, {Candini}, {Cantiello}, {Capobianco}, {Cappellaro}, {Cappelluti}, {Cappi}, {Caputi}, {Cara}, {Carbone}, {Cardone}, {Carella}, {Carlberg}, {Carle}, {Carminati}, {Caro}, {Carrasco}, {Carretero}, {Carrilho}, \& {Carron Duque}}]{2024arXiv240513491E}
{Euclid Collaboration: Mellier}, Y., {Abdurro\'uf}, {Acevedo Barroso}, J.~A., {et~al.} 2025, A\&A, 697, A1

\bibitem[{{Euclid Collaboration: Paykari} {et~al.}(2020){Euclid Collaboration: Paykari}, {Kitching}, {Hoekstra}, {Azzollini}, {Cardone}, {Cropper}, {Duncan}, {Kannawadi}, {Miller}, {Aussel}, {Conti}, {Auricchio}, {Baldi}, {Bardelli}, {Biviano}, {Bonino}, {Borsato}, {Bozzo}, {Branchini}, {Brau-Nogue}, {Brescia}, {Brinchmann}, {Burigana}, {Camera}, {Capobianco}, {Carbone}, {Carretero}, {Castand er}, {Castellano}, {Cavuoti}, {Charles}, {Cledassou}, {Colodro-Conde}, {Congedo}, {Conselice}, {Conversi}, {Copin}, {Coupon}, {Courtois}, {Da Silva}, {Dupac}, {Fabbian}, {Farrens}, {Ferreira}, {Fosalba}, {Fourmanoit}, {Frailis}, {Fumana}, {Galeotta}, {Garilli}, {Gillard}, {Gillis}, {Giocoli}, {Graci{\'a}-Carpio}, {Grupp}, {Hormuth}, {Ili{\'c}}, {Israel}, {Jahnke}, {Keihanen}, {Kermiche}, {Kilbinger}, {Kirkpatrick}, {Kubik}, {Kunz}, {Kurki-Suonio}, {Laureijs}, {Le Mignant}, {Ligori}, {Lilje}, {Lloro}, {Maciaszek}, {Maiorano}, {Marggraf}, {Markovic}, {Martinet}, {Marulli}, {Massey}, {Mauri}, {Medinaceli}, {Mei}, {Mellier},
  {Meneghetti}, {Metcalf}, {Moresco}, {Moscardini}, {Munari}, {Neissner}, {Nichol}, {Niemi}, {Nutma}, {Padilla}, {Paltani}, {Pasian}, {Pettorino}, {Pires}, {Polenta}, {Raison}, {Renzi}, {Rhodes}, {Romelli}, {Roncarelli}, {Rossetti}, {Saglia}, {Sakr}, {S{\'a}nchez}, {Sapone}, {Scaramella}, {Schneider}, {Schrabback}, {Scottez}, {Secroun}, {Serrano}, {Sirignano}, {Sirri}, {Stanco}, {Starck}, {Sureau}, {Tallada-Cresp{\'\i}}, {Taylor}, {Tenti}, {Tereno}, {Toledo-Moreo}, {Torradeflot}, {Valenziano}, {Vannier}, {Vassallo}, {Zoubian}, \& {Zucca}}]{Paykari2020}
{Euclid Collaboration: Paykari}, P., {Kitching}, T., {Hoekstra}, H., {et~al.} 2020, \aap, 635, A139

\bibitem[{{Font-Ribera} {et~al.}(2014){Font-Ribera}, {McDonald}, {Mostek}, {Reid}, {Seo}, \& {Slosar}}]{2014JCAP...05..023F}
{Font-Ribera}, A., {McDonald}, P., {Mostek}, N., {et~al.} 2014, JCAP, 05, 023

\bibitem[{{Friedrich} {et~al.}(2021){Friedrich}, {Andrade-Oliveira}, {Camacho}, {Alves}, {Rosenfeld}, {Sanchez}, {Fang}, {Eifler}, {Krause}, {Chang}, {Omori}, {Amon}, {Baxter}, {Elvin-Poole}, {Huterer}, {Porredon}, {Prat}, {Terra}, {Troja}, {Alarcon}, {Bechtol}, {Bernstein}, {Buchs}, {Campos}, {Carnero Rosell}, {Carrasco Kind}, {Cawthon}, {Choi}, {Cordero}, {Crocce}, {Davis}, {DeRose}, {Diehl}, {Dodelson}, {Doux}, {Drlica-Wagner}, {Elsner}, {Everett}, {Fosalba}, {Gatti}, {Giannini}, {Gruen}, {Gruendl}, {Harrison}, {Hartley}, {Jain}, {Jarvis}, {MacCrann}, {McCullough}, {Muir}, {Myles}, {Pandey}, {Raveri}, {Roodman}, {Rodriguez-Monroy}, {Rykoff}, {Samuroff}, {S{\'a}nchez}, {Secco}, {Sevilla-Noarbe}, {Sheldon}, {Troxel}, {Weaverdyck}, {Yanny}, {Aguena}, {Avila}, {Bacon}, {Bertin}, {Bhargava}, {Brooks}, {Burke}, {Carretero}, {Costanzi}, {da Costa}, {Pereira}, {De Vicente}, {Desai}, {Evrard}, {Ferrero}, {Frieman}, {Garc{\'\i}a-Bellido}, {Gaztanaga}, {Gerdes}, {Giannantonio}, {Gschwend}, {Gutierrez}, {Hinton},
  {Hollowood}, {Honscheid}, {James}, {Kuehn}, {Lahav}, {Lima}, {Maia}, {Menanteau}, {Miquel}, {Morgan}, {Palmese}, {Paz-Chinch{\'o}n}, {Plazas}, {Sanchez}, {Scarpine}, {Serrano}, {Soares-Santos}, {Smith}, {Suchyta}, {Tarle}, {Thomas}, {To}, {Varga}, {Weller}, {Wilkinson}, {Wilkinson}, \& {DES Collaboration}}]{Friedrich:2020dqo}
{Friedrich}, O., {Andrade-Oliveira}, F., {Camacho}, H., {et~al.} 2021, \mnras, 508, 3125

\bibitem[{{Frusciante} {et~al.}(2024){Frusciante}, {Pace}, {Cardone}, {Casas}, {Tutusaus}, {Ballardini}, {Bellini}, {Benevento}, {Bose}, {Valageas}, {Bartolo}, {Brax}, {Ferreira}, {Finelli}, {Koyama}, {Legrand}, {Lombriser}, {Paoletti}, {Pietroni}, {Rozas-Fern{\'a}ndez}, {Sakr}, {Silvestri}, {Vernizzi}, {Winther}, {Aghanim}, {Amendola}, {Auricchio}, {Azzollini}, {Baldi}, {Bonino}, {Branchini}, {Brescia}, {Brinchmann}, {Camera}, {Capobianco}, {Carbone}, {Carretero}, {Castellano}, {Cavuoti}, {Cimatti}, {Cledassou}, {Congedo}, {Conversi}, {Copin}, {Corcione}, {Courbin}, {Cropper}, {Da Silva}, {Degaudenzi}, {Dinis}, {Dubath}, {Dupac}, {Dusini}, {Farrens}, {Ferriol}, {Fosalba}, {Frailis}, {Franceschi}, {Galeotta}, {Gillis}, {Giocoli}, {Grazian}, {Grupp}, {Guzzo}, {Haugan}, {Holmes}, {Hormuth}, {Hornstrup}, {Jahnke}, {Kermiche}, {Kiessling}, {Kilbinger}, {Kitching}, {Kunz}, {Kurki-Suonio}, {Ligori}, {Lilje}, {Lloro}, {Maiorano}, {Mansutti}, {Marggraf}, {Markovic}, {Marulli}, {Massey}, {Medinaceli}, {Meneghetti},
  {Meylan}, {Moresco}, {Moscardini}, {Munari}, {Niemi}, {Nightingale}, {Padilla}, {Paltani}, {Pasian}, {Pedersen}, {Percival}, {Pettorino}, {Polenta}, {Poncet}, {Popa}, {Raison}, {Rebolo}, {Renzi}, {Rhodes}, {Riccio}, {Romelli}, {Saglia}, {Sapone}, {Sartoris}, {Secroun}, {Seidel}, {Sirignano}, {Sirri}, {Stanco}, {Surace}, {Tallada-Cresp{\'\i}}, {Taylor}, {Tereno}, {Toledo-Moreo}, {Torradeflot}, {Valentijn}, {Valenziano}, {Vassallo}, {Verdoes Kleijn}, {Wang}, {Zacchei}, {Zamorani}, {Zoubian}, \& {Scottez}}]{Euclid:2023rjj}
{Frusciante}, N., {Pace}, F., {Cardone}, V.~F., {et~al.} 2024, \aap, 690, A133

\bibitem[{Giblin {et~al.}(2016)Giblin, Mertens, \& Starkman}]{Giblin:2016mjp}
Giblin, J.~T., Mertens, J.~B., \& Starkman, G.~D. 2016, ApJ, 833, 247

\bibitem[{{Guzzo} {et~al.}(2008){Guzzo}, {Pierleoni}, {Meneux}, {Branchini}, {Le F{\`e}vre}, {Marinoni}, {Garilli}, {Blaizot}, {De Lucia}, {Pollo}, {McCracken}, {Bottini}, {Le Brun}, {Maccagni}, {Picat}, {Scaramella}, {Scodeggio}, {Tresse}, {Vettolani}, {Zanichelli}, {Adami}, {Arnouts}, {Bardelli}, {Bolzonella}, {Bongiorno}, {Cappi}, {Charlot}, {Ciliegi}, {Contini}, {Cucciati}, {de la Torre}, {Dolag}, {Foucaud}, {Franzetti}, {Gavignaud}, {Ilbert}, {Iovino}, {Lamareille}, {Marano}, {Mazure}, {Memeo}, {Merighi}, {Moscardini}, {Paltani}, {Pell{\`o}}, {Perez-Montero}, {Pozzetti}, {Radovich}, {Vergani}, {Zamorani}, \& {Zucca}}]{Guzzo:2008ac}
{Guzzo}, L., {Pierleoni}, M., {Meneux}, B., {et~al.} 2008, \nat, 451, 541

\bibitem[{Harnois-Deraps {et~al.}(2019)Harnois-Deraps, Giblin, \& Joachimi}]{Harnois-Deraps:2019rsd}
Harnois-Deraps, J., Giblin, B., \& Joachimi, B. 2019, A\&A, 631, A160

\bibitem[{Heavens {et~al.}(2011)Heavens, Jimenez, \& Maartens}]{Heavens:2011mr}
Heavens, A.~F., Jimenez, R., \& Maartens, R. 2011, JCAP, 09, 035

\bibitem[{Huterer {et~al.}(2017)Huterer, Shafer, Scolnic, \& Schmidt}]{Huterer:2016uyq}
Huterer, D., Shafer, D., Scolnic, D., \& Schmidt, F. 2017, JCAP, 05, 015

\bibitem[{Ishak(2019)}]{Ishak:2018his}
Ishak, M. 2019, Living Reviews in Relativity, 22, 1

\bibitem[{Jiao {et~al.}(2023)Jiao, Borghi, Moresco, \& Zhang}]{Jiao:2022aep}
Jiao, K., Borghi, N., Moresco, M., \& Zhang, T.-J. 2023, ApJS, 265, 48

\bibitem[{Joyce {et~al.}(2016)Joyce, Lombriser, \& Schmidt}]{Joyce:2016vqv}
Joyce, A., Lombriser, L., \& Schmidt, F. 2016, Ann. Rev. Nucl. Part. Sci., 66, 95

\bibitem[{{Kammerer} {et~al.}(2025){Kammerer}, {Bartlett}, {Kronberger}, {Desmond}, \& {Ferreira}}]{Kammerer:2025dbi}
{Kammerer}, L., {Bartlett}, D.~J., {Kronberger}, G., {Desmond}, H., \& {Ferreira}, P.~G. 2025, arXiv:2506.08783

\bibitem[{Kazantzidis \& Perivolaropoulos(2021)}]{kazantzidis2021sigma}
Kazantzidis, L. \& Perivolaropoulos, L. 2021, in Modified Gravity and Cosmology: An Update by the CANTATA Network (Springer), 507--537

\bibitem[{{Laureijs} {et~al.}(2011){Laureijs}, {Amiaux}, {Arduini}, {Augu{\`e}res}, {Brinchmann}, {Cole}, {Cropper}, {Dabin}, {Duvet}, {Ealet}, {Garilli}, {Gondoin}, {Guzzo}, {Hoar}, {Hoekstra}, {Holmes}, {Kitching}, {Maciaszek}, {Mellier}, {Pasian}, {Percival}, {Rhodes}, {Saavedra Criado}, {Sauvage}, {Scaramella}, {Valenziano}, {Warren}, {Bender}, {Castander}, {Cimatti}, {Le F{\`e}vre}, {Kurki-Suonio}, {Levi}, {Lilje}, {Meylan}, {Nichol}, {Pedersen}, {Popa}, {Rebolo Lopez}, {Rix}, {Rottgering}, {Zeilinger}, {Grupp}, {Hudelot}, {Massey}, {Meneghetti}, {Miller}, {Paltani}, {Paulin-Henriksson}, {Pires}, {Saxton}, {Schrabback}, {Seidel}, {Walsh}, {Aghanim}, {Amendola}, {Bartlett}, {Baccigalupi}, {Beaulieu}, {Benabed}, {Cuby}, {Elbaz}, {Fosalba}, {Gavazzi}, {Helmi}, {Hook}, {Irwin}, {Kneib}, {Kunz}, {Mannucci}, {Moscardini}, {Tao}, {Teyssier}, {Weller}, {Zamorani}, {Zapatero Osorio}, {Boulade}, {Foumond}, {Di Giorgio}, {Guttridge}, {James}, {Kemp}, {Martignac}, {Spencer}, {Walton}, {Bl{\"u}mchen}, {Bonoli},
  {Bortoletto}, {Cerna}, {Corcione}, {Fabron}, {Jahnke}, {Ligori}, {Madrid}, {Martin}, {Morgante}, {Pamplona}, {Prieto}, {Riva}, {Toledo}, {Trifoglio}, {Zerbi}, {Abdalla}, {Douspis}, {Grenet}, {Borgani}, {Bouwens}, {Courbin}, {Delouis}, {Dubath}, {Fontana}, {Frailis}, {Grazian}, {Koppenh{\"o}fer}, {Mansutti}, {Melchior}, {Mignoli}, {Mohr}, {Neissner}, {Noddle}, {Poncet}, {Scodeggio}, {Serrano}, {Shane}, {Starck}, {Surace}, {Taylor}, {Verdoes-Kleijn}, {Vuerli}, {Williams}, {Zacchei}, {Altieri}, {Escudero Sanz}, {Kohley}, {Oosterbroek}, {Astier}, {Bacon}, {Bardelli}, {Baugh}, {Bellagamba}, {Benoist}, {Bianchi}, {Biviano}, {Branchini}, {Carbone}, {Cardone}, {Clements}, {Colombi}, {Conselice}, {Cresci}, {Deacon}, {Dunlop}, {Fedeli}, {Fontanot}, {Franzetti}, {Giocoli}, {Garcia-Bellido}, {Gow}, {Heavens}, {Hewett}, {Heymans}, {Holland}, {Huang}, {Ilbert}, {Joachimi}, {Jennins}, {Kerins}, {Kiessling}, {Kirk}, {Kotak}, {Krause}, {Lahav}, {van Leeuwen}, {Lesgourgues}, {Lombardi}, {Magliocchetti}, {Maguire},
  {Majerotto}, {Maoli}, {Marulli}, {Maurogordato}, {McCracken}, {McLure}, {Melchiorri}, {Merson}, {Moresco}, {Nonino}, {Norberg}, {Peacock}, {Pello}, {Penny}, {Pettorino}, {Di Porto}, {Pozzetti}, {Quercellini}, {Radovich}, {Rassat}, {Roche}, {Ronayette}, {Rossetti}, {Sartoris}, {Schneider}, {Semboloni}, {Serjeant}, {Simpson}, {Skordis}, {Smadja}, {Smartt}, {Spano}, {Spiro}, {Sullivan}, {Tilquin}, {Trotta}, {Verde}, {Wang}, {Williger}, {Zhao}, {Zoubian}, \& {Zucca}}]{Laureijs:2011gra}
{Laureijs}, R., {Amiaux}, J., {Arduini}, S., {et~al.} 2011, arXiv:1110.3193

\bibitem[{{Laurent} {et~al.}(2016){Laurent}, {Le Goff}, {Burtin}, {Hamilton}, {Hogg}, {Myers}, {Ntelis}, {P{\^a}ris}, {Rich}, {Aubourg}, {Bautista}, {Delubac}, {du Mas des Bourboux}, {Eftekharzadeh}, {Palanque Delabrouille}, {Petitjean}, {Rossi}, {Schneider}, \& {Yeche}}]{Laurent:2016eqo}
{Laurent}, P., {Le Goff}, J.-M., {Burtin}, E., {et~al.} 2016, JCAP, 11, 060

\bibitem[{Linder(2005)}]{Linder:2005in}
Linder, E.~V. 2005, Phys. Rev. D, 72, 043529

\bibitem[{Ma \& Bertschinger(1995)}]{Ma:1995ey}
Ma, C.-P. \& Bertschinger, E. 1995, ApJ, 455, 7

\bibitem[{{Maciaszek} {et~al.}(2022){Maciaszek}, {Ealet}, {Gillard}, {Jahnke}, {Barbier}, {Prieto}, {Bon}, {Bonnefoi}, {Caillat}, {Carle}, {Costille}, {Ducret}, {Fabron}, {Foulon}, {Gimenez}, {Grassi}, {Jaquet}, {Le Mignant}, {Martin}, {Pamplona}, {Sanchez}, {Cl{\'e}mens}, {Caillat}, {Niclas}, {Secroun}, {Kubik}, {Ferriol}, {Berthe}, {Barri{\`e}re}, {Fontignie}, {Valenziano}, {Auricchio}, {Battaglia}, {De Rosa}, {Farinelli}, {Franceschi}, {Medinaceli}, {Morgante}, {Sortino}, {Trifoglio}, {Corcione}, {Capobianco}, {Ligori}, {Dusini}, {Borsato}, {Dal Corso}, {Laudisio}, {Sirignano}, {Stanco}, {Ventura}, {Patrizii}, {Chiarusi}, {Fornari}, {Giacomini}, {Margiotta}, {Mauri}, {Pasqualini}, {Sirri}, {Spurio}, {Tenti}, {Travaglini}, {Bonoli}, {Bortoletto}, {Balestra}, {Dalessandro}, {Grupp}, {Penka}, {Steinwagner}, {Hormuth}, {Schirmer}, {Seidel}, {Padilla}, {Casas}, {Lloro}, {Toledo-Moreo}, {Gomez}, {Colodro-Conde}, {Liz{\'a}n}, {Diaz}, {Lilje}, {Andersen}, {Andersen}, {S{\o}rensen}, {Hornstrup}, {Jessen}, {Thizy},
  {Holmes}, {Pniel}, {Jhabvala}, {Pravdo}, {Seiffert}, {Waczynski}, {Laureij}, {Racca}, {Salvignol}, {Boenke}, {Strada}, \& {Mellier}}]{NISP_paper_2022}
{Maciaszek}, T., {Ealet}, A., {Gillard}, W., {et~al.} 2022, in Society of Photo-Optical Instrumentation Engineers (SPIE) Conference Series, Vol. 12180, Space Telescopes and Instrumentation 2022: Optical, Infrared, and Millimeter Wave, ed. L.~E. {Coyle}, S.~{Matsuura}, \& M.~D. {Perrin}, 121801K

\bibitem[{Marra \& Sapone(2018)}]{Marra:2017pst}
Marra, V. \& Sapone, D. 2018, Phys. Rev. D, 97, 083510

\bibitem[{{Martinelli} {et~al.}(2020){Martinelli}, {Martins}, {Nesseris}, {Sapone}, {Tutusaus}, {Avgoustidis}, {Camera}, {Carbone}, {Casas}, {Ili{\'c}}, {Sakr}, {Yankelevich}, {Auricchio}, {Balestra}, {Bodendorf}, {Bonino}, {Branchini}, {Brescia}, {Brinchmann}, {Capobianco}, {Carretero}, {Castellano}, {Cavuoti}, {Cledassou}, {Congedo}, {Conversi}, {Corcione}, {Dubath}, {Ealet}, {Frailis}, {Franceschi}, {Fumana}, {Garilli}, {Gillis}, {Giocoli}, {Grupp}, {Haugan}, {Holmes}, {Hormuth}, {Jahnke}, {Kermiche}, {Kilbinger}, {Kitching}, {Kubik}, {Kunz}, {Kurki-Suonio}, {Ligori}, {Lilje}, {Lloro}, {Marggraf}, {Markovic}, {Massey}, {Mei}, {Meneghetti}, {Meylan}, {Moscardini}, {Niemi}, {Padilla}, {Paltani}, {Pasian}, {Pettorino}, {Pires}, {Polenta}, {Poncet}, {Popa}, {Pozzetti}, {Raison}, {Rhodes}, {Roncarelli}, {Saglia}, {Schneider}, {Secroun}, {Serrano}, {Sirignano}, {Sirri}, {Sureau}, {Taylor}, {Tereno}, {Toledo-Moreo}, {Valenziano}, {Vassallo}, {Wang}, {Welikala}, {Weller}, \& {Zacchei}}]{Martinelli:2020hud}
{Martinelli}, M., {Martins}, C.~J.~A.~P., {Nesseris}, S., {et~al.} 2020, \aap, 644, A80

\bibitem[{{Martinelli} {et~al.}(2021){Martinelli}, {Martins}, {Nesseris}, {Tutusaus}, {Blanchard}, {Camera}, {Carbone}, {Casas}, {Pettorino}, {Sakr}, {Yankelevich}, {Sapone}, {Amara}, {Auricchio}, {Bodendorf}, {Bonino}, {Branchini}, {Capobianco}, {Carretero}, {Castellano}, {Cavuoti}, {Cimatti}, {Cledassou}, {Corcione}, {Costille}, {Degaudenzi}, {Douspis}, {Dubath}, {Dusini}, {Ealet}, {Ferriol}, {Frailis}, {Franceschi}, {Garilli}, {Giocoli}, {Grazian}, {Grupp}, {Haugan}, {Holmes}, {Hormuth}, {Jahnke}, {Kiessling}, {K{\"u}mmel}, {Kunz}, {Kurki-Suonio}, {Ligori}, {Lilje}, {Lloro}, {Mansutti}, {Marggraf}, {Markovic}, {Massey}, {Meneghetti}, {Meylan}, {Moscardini}, {Niemi}, {Padilla}, {Paltani}, {Pasian}, {Pedersen}, {Pires}, {Poncet}, {Popa}, {Raison}, {Rebolo}, {Rhodes}, {Roncarelli}, {Rossetti}, {Saglia}, {Secroun}, {Seidel}, {Serrano}, {Sirignano}, {Sirri}, {Starck}, {Tavagnacco}, {Taylor}, {Tereno}, {Toledo-Moreo}, {Valenziano}, {Wang}, {Zamorani}, {Zoubian}, {Baldi}, {Brescia}, {Congedo}, {Conversi},
  {Copin}, {Fabbian}, {Farinelli}, {Medinaceli}, {Mei}, {Polenta}, {Romelli}, \& {Vassallo}}]{Euclid:2021cfn}
{Martinelli}, M., {Martins}, C.~J.~A.~P., {Nesseris}, S., {et~al.} 2021, \aap, 654, A148

\bibitem[{Moresco(2015)}]{Moresco:2015cya}
Moresco, M. 2015, MNRAS, 450, L16

\bibitem[{{Moresco} {et~al.}(2022){Moresco}, {Amati}, {Amendola}, {Birrer}, {Blakeslee}, {Cantiello}, {Cimatti}, {Darling}, {Della Valle}, {Fishbach}, {Grillo}, {Hamaus}, {Holz}, {Izzo}, {Jimenez}, {Lusso}, {Meneghetti}, {Piedipalumbo}, {Pisani}, {Pourtsidou}, {Pozzetti}, {Quartin}, {Risaliti}, {Rosati}, \& {Verde}}]{Moresco:2022phi}
{Moresco}, M., {Amati}, L., {Amendola}, L., {et~al.} 2022, Living Reviews in Relativity, 25, 6

\bibitem[{{Moresco} {et~al.}(2012){Moresco}, {Cimatti}, {Jimenez}, {Pozzetti}, {Zamorani}, {Bolzonella}, {Dunlop}, {Lamareille}, {Mignoli}, {Pearce}, {Rosati}, {Stern}, {Verde}, {Zucca}, {Carollo}, {Contini}, {Kneib}, {Le F{\`e}vre}, {Lilly}, {Mainieri}, {Renzini}, {Scodeggio}, {Balestra}, {Gobat}, {McLure}, {Bardelli}, {Bongiorno}, {Caputi}, {Cucciati}, {de la Torre}, {de Ravel}, {Franzetti}, {Garilli}, {Iovino}, {Kampczyk}, {Knobel}, {Kova{\v{c}}}, {Le Borgne}, {Le Brun}, {Maier}, {Pell{\'o}}, {Peng}, {Perez-Montero}, {Presotto}, {Silverman}, {Tanaka}, {Tasca}, {Tresse}, {Vergani}, {Almaini}, {Barnes}, {Bordoloi}, {Bradshaw}, {Cappi}, {Chuter}, {Cirasuolo}, {Coppa}, {Diener}, {Foucaud}, {Hartley}, {Kamionkowski}, {Koekemoer}, {L{\'o}pez-Sanjuan}, {McCracken}, {Nair}, {Oesch}, {Stanford}, \& {Welikala}}]{Moresco:2012jh}
{Moresco}, M., {Cimatti}, A., {Jimenez}, R., {et~al.} 2012, JCAP, 08, 006

\bibitem[{Moresco {et~al.}(2016)Moresco, Pozzetti, Cimatti, Jimenez, Maraston, Verde, Thomas, Citro, Tojeiro, \& Wilkinson}]{Moresco:2016mzx}
Moresco, M., Pozzetti, L., Cimatti, A., {et~al.} 2016, JCAP, 05, 014

\bibitem[{Nadolny {et~al.}(2021)Nadolny, Durrer, Kunz, \& Padmanabhan}]{Nadolny:2021hti}
Nadolny, T., Durrer, R., Kunz, M., \& Padmanabhan, H. 2021, JCAP, 11, 009

\bibitem[{Nesseris(2009)}]{Nesseris:2008mq}
Nesseris, S. 2009, Phys. Rev. D, 79, 044015

\bibitem[{Nesseris {et~al.}(2011)Nesseris, Blake, Davis, \& Parkinson}]{Nesseris:2011pc}
Nesseris, S., Blake, C., Davis, T., \& Parkinson, D. 2011, JCAP, 07, 037

\bibitem[{Nesseris \& Garc\'ia-Bellido(2012)}]{Nesseris:2012tt}
Nesseris, S. \& Garc\'ia-Bellido, J. 2012, JCAP, 11, 033

\bibitem[{Nesseris \& Garc\'\i{}a-Bellido(2013)}]{Nesseris:2013bia}
Nesseris, S. \& Garc\'\i{}a-Bellido, J. 2013, Phys. Rev. D, 88, 063521

\bibitem[{Nesseris \& Mazumdar(2009)}]{Nesseris:2009jf}
Nesseris, S. \& Mazumdar, A. 2009, Phys. Rev. D, 79, 104006

\bibitem[{Nesseris {et~al.}(2017)Nesseris, Pantazis, \& Perivolaropoulos}]{Nesseris:2017vor}
Nesseris, S., Pantazis, G., \& Perivolaropoulos, L. 2017, Phys. Rev. D, 96, 023542

\bibitem[{Nesseris \& Sapone(2015{\natexlab{a}})}]{Nesseris:2015fqa}
Nesseris, S. \& Sapone, D. 2015{\natexlab{a}}, Phys. Rev. D, 92, 023013

\bibitem[{Nesseris \& Sapone(2015{\natexlab{b}})}]{Nesseris:2014mfa}
Nesseris, S. \& Sapone, D. 2015{\natexlab{b}}, Int. J. Mod. Phys. D, 24, 1550045

\bibitem[{Nesseris {et~al.}(2015)Nesseris, Sapone, \& Garc\'\i{}a-Bellido}]{Nesseris:2014qca}
Nesseris, S., Sapone, D., \& Garc\'\i{}a-Bellido, J. 2015, Phys. Rev. D, 91, 023004

\bibitem[{Nesseris \& Shafieloo(2010)}]{Nesseris:2010ep}
Nesseris, S. \& Shafieloo, A. 2010, MNRAS, 408, 1879

\bibitem[{Nesseris {et~al.}(2022)}]{Euclid:2021frk}
Nesseris, S. {et~al.} 2022, A\&A, 660, A67

\bibitem[{Ocampo {et~al.}(2025)Ocampo, Alestas, Nesseris, \& Sapone}]{Ocampo:2024fvx}
Ocampo, I., Alestas, G., Nesseris, S., \& Sapone, D. 2025, Phys. Rev. Lett., 134, 041002

\bibitem[{{Okumura} {et~al.}(2016){Okumura}, {Hikage}, {Totani}, {Tonegawa}, {Okada}, {Glazebrook}, {Blake}, {Ferreira}, {More}, {Taruya}, {Tsujikawa}, {Akiyama}, {Dalton}, {Goto}, {Ishikawa}, {Iwamuro}, {Matsubara}, {Nishimichi}, {Ohta}, {Shimizu}, {Takahashi}, {Takato}, {Tamura}, {Yabe}, \& {Yoshida}}]{Okumura:2015lvp}
{Okumura}, T., {Hikage}, C., {Totani}, T., {et~al.} 2016, \pasj, 68, 38

\bibitem[{Orjuela-Quintana {et~al.}(2024)Orjuela-Quintana, Sapone, \& Nesseris}]{Orjuela-Quintana:2024hha}
Orjuela-Quintana, J.~B., Sapone, D., \& Nesseris, S. 2024, arXiv:2407.16640

\bibitem[{{Peebles}(2020)}]{2020coce.book.....P}
{Peebles}, P.~J.~E. 2020, {Cosmology's Century: An Inside History of our Modern Understanding of the Universe} (Princeton University Press)

\bibitem[{Percival(2005)}]{Percival:2005vm}
Percival, W.~J. 2005, A\&A, 443, 819

\bibitem[{Perivolaropoulos \& Skara(2022)}]{Perivolaropoulos:2021jda}
Perivolaropoulos, L. \& Skara, F. 2022, New Astron. Rev., 95, 101659

\bibitem[{Pezzotta {et~al.}(2017)}]{Pezzotta:2016gbo}
Pezzotta, A. {et~al.} 2017, A\&A, 604, A33

\bibitem[{Pitjeva {et~al.}(2021)Pitjeva, Pitjev, Pavlov, \& Turygin}]{Pitjeva:2021hnc}
Pitjeva, E.~V., Pitjev, N.~P., Pavlov, D.~A., \& Turygin, C.~C. 2021, A\&A, 647, A141

\bibitem[{{Planck Collaboration: Aghanim} {et~al.}(2020){Planck Collaboration: Aghanim}, {Akrami}, {Ashdown}, {Aumont}, {Baccigalupi}, {Ballardini}, {Banday}, {Barreiro}, {Bartolo}, {Basak}, {Battye}, {Benabed}, {Bernard}, {Bersanelli}, {Bielewicz}, {Bock}, {Bond}, {Borrill}, {Bouchet}, {Boulanger}, {Bucher}, {Burigana}, {Butler}, {Calabrese}, {Cardoso}, {Carron}, {Challinor}, {Chiang}, {Chluba}, {Colombo}, {Combet}, {Contreras}, {Crill}, {Cuttaia}, {de Bernardis}, {de Zotti}, {Delabrouille}, {Delouis}, {Di Valentino}, {Diego}, {Dor{\'e}}, {Douspis}, {Ducout}, {Dupac}, {Dusini}, {Efstathiou}, {Elsner}, {En{\ss}lin}, {Eriksen}, {Fantaye}, {Farhang}, {Fergusson}, {Fernandez-Cobos}, {Finelli}, {Forastieri}, {Frailis}, {Fraisse}, {Franceschi}, {Frolov}, {Galeotta}, {Galli}, {Ganga}, {G{\'e}nova-Santos}, {Gerbino}, {Ghosh}, {Gonz{\'a}lez-Nuevo}, {G{\'o}rski}, {Gratton}, {Gruppuso}, {Gudmundsson}, {Hamann}, {Handley}, {Hansen}, {Herranz}, {Hildebrandt}, {Hivon}, {Huang}, {Jaffe}, {Jones}, {Karakci}, {Keih{\"a}nen},
  {Keskitalo}, {Kiiveri}, {Kim}, {Kisner}, {Knox}, {Krachmalnicoff}, {Kunz}, {Kurki-Suonio}, {Lagache}, {Lamarre}, {Lasenby}, {Lattanzi}, {Lawrence}, {Le Jeune}, {Lemos}, {Lesgourgues}, {Levrier}, {Lewis}, {Liguori}, {Lilje}, {Lilley}, {Lindholm}, {L{\'o}pez-Caniego}, {Lubin}, {Ma}, {Mac{\'\i}as-P{\'e}rez}, {Maggio}, {Maino}, {Mandolesi}, {Mangilli}, {Marcos-Caballero}, {Maris}, {Martin}, {Martinelli}, {Mart{\'\i}nez-Gonz{\'a}lez}, {Matarrese}, {Mauri}, {McEwen}, {Meinhold}, {Melchiorri}, {Mennella}, {Migliaccio}, {Millea}, {Mitra}, {Miville-Desch{\^e}nes}, {Molinari}, {Montier}, {Morgante}, {Moss}, {Natoli}, {N{\o}rgaard-Nielsen}, {Pagano}, {Paoletti}, {Partridge}, {Patanchon}, {Peiris}, {Perrotta}, {Pettorino}, {Piacentini}, {Polastri}, {Polenta}, {Puget}, {Rachen}, {Reinecke}, {Remazeilles}, {Renzi}, {Rocha}, {Rosset}, {Roudier}, {Rubi{\~n}o-Mart{\'\i}n}, {Ruiz-Granados}, {Salvati}, {Sandri}, {Savelainen}, {Scott}, {Shellard}, {Sirignano}, {Sirri}, {Spencer}, {Sunyaev}, {Suur-Uski}, {Tauber}, {Tavagnacco},
  {Tenti}, {Toffolatti}, {Tomasi}, {Trombetti}, {Valenziano}, {Valiviita}, {Van Tent}, {Vibert}, {Vielva}, {Villa}, {Vittorio}, {Wandelt}, {Wehus}, {White}, {White}, {Zacchei}, \& {Zonca}}]{Aghanim:2018eyx}
{Planck Collaboration: Aghanim}, N., {Akrami}, Y., {Ashdown}, M., {et~al.} 2020, \aap, 641, A6

\bibitem[{{Pozzetti} {et~al.}(2016){Pozzetti}, {Hirata}, {Geach}, {Cimatti}, {Baugh}, {Cucciati}, {Merson}, {Norberg}, \& {Shi}}]{Pozzetti:2016cch}
{Pozzetti}, L., {Hirata}, C.~M., {Geach}, J.~E., {et~al.} 2016, \aap, 590, A3

\bibitem[{Qin {et~al.}(2019)Qin, Howlett, \& Staveley-Smith}]{Qin:2019axr}
Qin, F., Howlett, C., \& Staveley-Smith, L. 2019, MNRAS, 487, 5235

\bibitem[{Racca {et~al.}(2016)Racca, Laureijs, Stagnaro, Salvignol, Alvarez, Criado, Venancio, Short, Strada, B{\"o}nke, Colombo, Calvi, Maiorano, Piersanti, Prezelus, Rosato, Pinel, Rozemeijer, Lesna, Musi, Sias, Anselmi, Cazaubiel, Vaillon, Mellier, Amiaux, Berth{\'e}, Sauvage, Azzollini, Cropper, Pottinger, Jahnke, Ealet, Maciaszek, Pasian, Zacchei, Scaramella, Hoar, Kohley, Vavrek, Rudolph, \& Schmidt}]{Racca:2016qpi}
Racca, G.~D., Laureijs, R., Stagnaro, L., {et~al.} 2016, in Space Telescopes and Instrumentation 2016: Optical, Infrared, and Millimeter Wave, ed. H.~A. MacEwen, G.~G. Fazio, M.~Lystrup, N.~Batalha, N.~Siegler, \& E.~C. Tong, Vol. 9904, International Society for Optics and Photonics (SPIE), 99040O

\bibitem[{Ratsimbazafy {et~al.}(2017)Ratsimbazafy, Loubser, Crawford, Cress, Bassett, Nichol, \& V\"ais\"anen}]{Ratsimbazafy:2017vga}
Ratsimbazafy, A.~L., Loubser, S.~I., Crawford, S.~M., {et~al.} 2017, MNRAS, 467, 3239

\bibitem[{{Ruiz-Zapatero} {et~al.}(2021){Ruiz-Zapatero}, {St{\"o}lzner}, {Joachimi}, {Asgari}, {Bilicki}, {Dvornik}, {Giblin}, {Heymans}, {Hildebrandt}, {Kannawadi}, {Kuijken}, {Tr{\"o}ster}, {van den Busch}, \& {Wright}}]{2021A&A...655A..11R}
{Ruiz-Zapatero}, J., {St{\"o}lzner}, B., {Joachimi}, B., {et~al.} 2021, A\&A, 655, A11

\bibitem[{Sagredo {et~al.}(2018)Sagredo, Nesseris, \& Sapone}]{Sagredo:2018ahx}
Sagredo, B., Nesseris, S., \& Sapone, D. 2018, Phys. Rev. D, 98, 083543

\bibitem[{Sakr {et~al.}(2018)Sakr, Ili\'c, Blanchard, Bittar, \& Farah}]{Sakr:2018new}
Sakr, Z., Ili\'c, S., Blanchard, A., Bittar, J., \& Farah, W. 2018, A\&A, 620, A78

\bibitem[{Sapone {et~al.}(2010)Sapone, Kunz, \& Amendola}]{Sapone:2010uy}
Sapone, D., Kunz, M., \& Amendola, L. 2010, Phys. Rev. D, 82, 103535

\bibitem[{Sapone \& Majerotto(2012)}]{Sapone:2012nh}
Sapone, D. \& Majerotto, E. 2012, Phys. Rev. D, 85, 123529

\bibitem[{Sapone {et~al.}(2014)Sapone, Majerotto, \& Nesseris}]{Sapone:2014nna}
Sapone, D., Majerotto, E., \& Nesseris, S. 2014, Phys. Rev. D, 90, 023012

\bibitem[{Scrimgeour {et~al.}(2012)}]{Scrimgeour:2012wt}
Scrimgeour, M. {et~al.} 2012, MNRAS, 425, 116

\bibitem[{Shafieloo {et~al.}(2009)Shafieloo, Sahni, \& Starobinsky}]{Shafieloo:2009ti}
Shafieloo, A., Sahni, V., \& Starobinsky, A.~A. 2009, Phys. Rev. D, 80, 101301

\bibitem[{Silveira \& Waga(1994)}]{Silveira:1994yq}
Silveira, V. \& Waga, I. 1994, Phys. Rev. D, 50, 4890

\bibitem[{Song \& Percival(2009)}]{Song:2008qt}
Song, Y.-S. \& Percival, W.~J. 2009, JCAP, 10, 004

\bibitem[{Stern {et~al.}(2010)Stern, Jimenez, Verde, Kamionkowski, \& Stanford}]{Stern:2009ep}
Stern, D., Jimenez, R., Verde, L., Kamionkowski, M., \& Stanford, S.~A. 2010, JCAP, 02, 008

\bibitem[{Sui {et~al.}(2025)Sui, Bartlett, Pandey, Desmond, Ferreira, \& Wandelt}]{Sui:2024wob}
Sui, C., Bartlett, D.~J., Pandey, S., {et~al.} 2025, A\&A, 698, A1

\bibitem[{Tsujikawa(2007)}]{Tsujikawa:2007gd}
Tsujikawa, S. 2007, Phys. Rev. D, 76, 023514

\bibitem[{Turnbull {et~al.}(2012)Turnbull, Hudson, Feldman, Hicken, Kirshner, \& Watkins}]{Turnbull:2011ty}
Turnbull, S.~J., Hudson, M.~J., Feldman, H.~A., {et~al.} 2012, MNRAS, 420, 447

\bibitem[{Valkenburg {et~al.}(2014)Valkenburg, Marra, \& Clarkson}]{Valkenburg:2012td}
Valkenburg, W., Marra, V., \& Clarkson, C. 2014, MNRAS, 438, L6

\bibitem[{van Dokkum {et~al.}(2000)van Dokkum, Franx, Fabricant, Illingworth, \& Kelson}]{vanDokkum:2000rs}
van Dokkum, P.~G., Franx, M., Fabricant, D., Illingworth, G.~D., \& Kelson, D.~D. 2000, ApJ, 541, 95

\bibitem[{Viljoen {et~al.}(2020)Viljoen, Fonseca, \& Maartens}]{Viljoen:2020efi}
Viljoen, J.-A., Fonseca, J., \& Maartens, R. 2020, JCAP, 09, 054

\bibitem[{Wang \& Steinhardt(1998)}]{Wang:1998gt}
Wang, L.-M. \& Steinhardt, P.~J. 1998, ApJ, 508, 483

\bibitem[{Wang(2008)}]{Wang:2007ht}
Wang, Y. 2008, JCAP, 05, 021

\bibitem[{Wang {et~al.}(2013)Wang, Chuang, \& Hirata}]{Wang:2012bx}
Wang, Y., Chuang, C.-H., \& Hirata, C.~M. 2013, \mnras, 430, 2446

\bibitem[{Zhang {et~al.}(2014)Zhang, Zhang, Yuan, Zhang, \& Sun}]{Zhang:2012mp}
Zhang, C., Zhang, H., Yuan, S., Zhang, T.-J., \& Sun, Y.-C. 2014, Res. Astron. Astrophys., 14, 1221

\bibitem[{Zheng {et~al.}(2024)Zheng, Sakr, \& Amendola}]{Zheng:2023yco}
Zheng, Z., Sakr, Z., \& Amendola, L. 2024, Phys. Lett. B, 853, 138647

\end{thebibliography}

\clearpage

\begin{appendix} 
\section{Data Compilations \label{sec:dataappdx}}
In \Cref{tab:data-fs8} and \Cref{tab:data-CC} we present the updated compilations of the currently available cosmic chronometer and $f\sigma_8$ data that were used in our analysis.

\begin{table}[!h]
\begin{center}
\setlength{\tabcolsep}{4pt}
\renewcommand{\arraystretch}{1.35}
\caption{The up-to-date compilation of currently available growth rate RSD data. \label{tab:data-fs8}}
\begin{tabular}{lcccc}
\hline
\hline
$z$ & $f \sigma_{8} \pm \sigma_{f \sigma_{8}}$ & $\Omega_\mathrm{m}$ & Ref. \\ 
\hline   
$0.17$ & $0.510 \pm 0.060$ & $0.3$ & \cite{Song:2008qt} \\
$0.02$ & $0.314 \pm 0.048$ & $0.266$ & \cite{Davis:2010sw} \\
  $0.02$ & $0.398 \pm 0.065$ & $0.3$ & \cite{Turnbull:2011ty} \\
  $0.44$ & $0.413 \pm 0.080$ & $0.27$ & \cite{Blake:2012pj} \\
  $0.60$ & $0.390 \pm 0.063$ & $0.27$ & \cite{Blake:2012pj} \\
  $0.73$ & $0.437 \pm 0.072$ & $0.27$ & \cite{Blake:2012pj} \\
  $0.18$ & $0.36 \pm 0.09$ & $0.27$ & \cite{Blake:2013nif} \\
  $0.38$ & $0.44 \pm 0.06$ & $0.27$ & \cite{Blake:2013nif} \\
  $1.4$ & $0.482\pm0.116$ & $0.27$ & \cite{Okumura:2015lvp} \\
  $0.02$ & $0.428_{-0.045}^{+0.048}$ & $0.3$ & \cite{Huterer:2016uyq} \\
  $0.6$ & $0.55 \pm 0.12$ & $0.3$ & \cite{Pezzotta:2016gbo} \\
  $0.86$ & $0.40 \pm 0.11$ & $0.3$ & \cite{Pezzotta:2016gbo} \\
 $0.03$&$0.404^{+0.082}_{-0.081}$ &$0.312$ & \cite{Qin:2019axr} \\
 $0.013$ & $0.46 \pm 0.06$ & $0.315$ & \cite{Avila:2021dqv} \\
 $0.15$ & $0.53 \pm 0.16$ & $0.31$ & \cite{eBOSS:2020yzd}  \\
 $0.38$ & $0.500 \pm 0.047$ & $0.31$ & \cite{eBOSS:2020yzd}  \\
 $0.51$ & $0.455 \pm 0.039$ & $0.31$ & \cite{eBOSS:2020yzd}  \\
 $0.70$ & $0.448 \pm 0.043$ & $0.31$ & \cite{eBOSS:2020yzd}  \\
 $0.85$ & $0.315 \pm 0.095$ & $0.31$ & \cite{eBOSS:2020yzd}  \\
 $1.48$ & $0.462 \pm 0.045$ & $0.31$ & \cite{eBOSS:2020yzd}  \\
\hline
\hline
\end{tabular}
\end{center}
\end{table}

\begin{table}
\begin{center}
\setlength{\tabcolsep}{4pt}
\renewcommand{\arraystretch}{1.35}
\caption{The up-to-date compilation of currently available cosmic chronometer data. All values of the Hubble parameter and its $68.3\%$ uncertainties are shown in units of km s$^{-1}$~Mpc$^{-1}$. \label{tab:data-CC}}
\begin{tabular}{lcccc}
\hline
\hline
$z$ & $H(z)$ $\pm \,\sigma_H$ & Ref. \\ 
\hline   
$0.09$    & $69 \pm 12$       & \cite{Stern:2009ep} \\
$0.17$    & $83 \pm 8$        & \cite{Stern:2009ep} \\
$0.27$    & $77 \pm 14$       & \cite{Stern:2009ep} \\
$0.40$    & $95 \pm 17$       & \cite{Stern:2009ep} \\
$0.48$    & $97 \pm 62$       & \cite{Stern:2009ep} \\
$0.88$    & $90 \pm 40$       & \cite{Stern:2009ep} \\
$0.90$    & $117 \pm 23$      & \cite{Stern:2009ep} \\
$1.30$    & $168 \pm 17$      & \cite{Stern:2009ep} \\
$1.43$    & $177 \pm 18$      & \cite{Stern:2009ep} \\
$1.53$    & $140 \pm 14$      & \cite{Stern:2009ep} \\
$1.75$    & $202 \pm 40$      & \cite{Stern:2009ep} \\
$0.44$    & $82.6 \pm 7.8$    & \cite{Blake:2012pj} \\
$0.60$    & $87.9 \pm 6.1$    & \cite{Blake:2012pj} \\
$0.73$    & $97.3 \pm 7.0$    & \cite{Blake:2012pj} \\
$0.179$   & $75 \pm 4$        & \cite{Moresco:2012jh} \\
$0.199$   & $75.0 \pm 5$      & \cite{Moresco:2012jh} \\
$0.352$   & $83.0 \pm 14$     & \cite{Moresco:2012jh} \\
$0.593$   & $104.0 \pm 13$    & \cite{Moresco:2012jh} \\
$0.68$    & $92.0 \pm 8$      & \cite{Moresco:2012jh} \\
$0.781$   & $105.0 \pm 12$    & \cite{Moresco:2012jh} \\
$0.875$   & $125.0 \pm 17$    & \cite{Moresco:2012jh} \\
$1.037$   & $154.0 \pm 20$    & \cite{Moresco:2012jh} \\
$0.35$    & $82.7 \pm 8.4$    & \cite{Chuang:2012qt}  \\
$0.07$    & $69.0 \pm 19.6$   & \cite{Zhang:2012mp} \\
$0.12$    & $68.6 \pm 26.2$   & \cite{Zhang:2012mp} \\
$0.20$    & $72.9 \pm 29.6$   & \cite{Zhang:2012mp} \\
$0.28$    & $88.8 \pm 36.6$   & \cite{Zhang:2012mp} \\
$0.57$    & $96.8 \pm 3.4$    & \cite{BOSS:2013rlg} \\
$2.34$    & $222.0 \pm 7.0$   & \cite{BOSS:2014hwf} \\
$1.363$   & $160.0 \pm 33.6$  & \cite{Moresco:2015cya} \\
$1.965$   & $186.5 \pm 50.4$  & \cite{Moresco:2015cya} \\
$0.3802$  & $83.0 \pm 13.5$   & \cite{Moresco:2016mzx} \\
$0.4004$   & $77.0 \pm 10.2 $  & \cite{Moresco:2016mzx} \\
$0.4247$   & $87.1 \pm 11.2$   & \cite{Moresco:2016mzx} \\
$0.4497$  & $92.8 \pm 12.9$   & \cite{Moresco:2016mzx} \\
$0.4783$   & $80.9 \pm 9.0$    & \cite{Moresco:2016mzx} \\
$0.47$     & $89 \pm 50$   & \cite{Ratsimbazafy:2017vga}  \\
$0.75$     & $98.8 \pm 33.6$   & \cite{Borghi:2021rft}  \\
$0.80$     & $113.1 \pm 20.73$  & \cite{Jiao:2022aep} \\
\hline
\hline
\end{tabular}
\end{center}
\end{table}

\end{appendix}

\end{document}